\documentclass[10pt]{article}
\pdfoutput=1
\usepackage{graphicx} 
\usepackage{amsmath} 
\usepackage{amsfonts}
\usepackage{amssymb}
\usepackage{subfigure}
\usepackage{multirow}

\title{\textbf{The Polycluster Theory for the Structure of Glasses: \\ 
Evidence from Low Temperature Physics}} 

\author{ Giancarlo Jug\cite{J} \\
Dipartimento di Scienza ed Alta Tecnologia and To.Sca.Lab \\
Universit\`a dell'Insubria,
Via Valleggio 11, 22100 Como, Italy \\
and INFN -- Sezione di Pavia, Italy }

\date{\today} 

\begin{document}

\maketitle 

\begin{abstract}
The problems of the intermediate-range atomic structure of glasses and of the 
mechanism for the glass transition are approached from the low-temperature end 
in terms of a scenario for the atomic organization that justifies the use of an 
extended tunneling model. The latter is crucial for the explanation of the magnetic 
and compositional effects discovered in non-metallic glasses in the Kelvin and 
milli-Kelvin temperature range. The model relies on the existence of multi-welled
local potentials for the effective tunneling particles that are a manifestation of 
a non-homogeneous atomic structure deriving from the established dynamical 
heterogeneities that characterize the supercooled liquid state. It is shown that the
extended tunneling model can successfully explain a range of experiments at low
temperatures, but the proposed non-homogeneous atomic structure scenario is 
then tested in the light of available high resolution electron microscopy imaging
of the structure of some glasses and of the behaviour near the glass transition.
\end{abstract}

\newpage

\section{Introduction}
The physics of glass-forming liquids, especially at higher temperature, continues 
to generate considerable research effort. These substances of extraordinary practical
and technological importance still present considerable scientific challenges in the
description of the glass-formation mechanism (nature of the glass transformation)
and of the nature of the atomic structure at intermediate- and long-range length
scales that characterize the solid. Standard X-ray and other scattering techniques fail 
in this respect to give a conclusive answer about the atomic structure of topologically 
disordered solids and the development of new investigation tools is desirable since 
there is no way to distinguish via scattering the structure of the liquid from that of 
the topologically disordered solid except through the vastly different relaxation time 
scales. Opinions as to why divergent time scales characterize the formation of the 
topologically disordered solid differ, however, and no agreement on a justification from 
structure for the mechanical properties below the glass transformation temperature 
$T_g$ has to date been reached. Why should indeed a microscopically liquid-looking 
assembly of interacting particles behave (mechanically) like a crystalline solid remains 
to date a true mystery. Owing to such difficulties much theoretical and computational 
research on the description of the topologically disordered solid 
and its properties takes its moves from the study of the corresponding liquid for which 
much has been understood thanks to equilibrium statistical mechanics. Though 
understandable, this approach would correspond to be wanting to understand the 
physical properties of a crystalline solid from the study of its melt, which of course 
would present some formidable challenges given the ergodicity-breaking phenomenon 
that characterizes crystallization. In this essay it is proposed that the glass transition is 
of a purely kinetic nature and the use of equilibrium statistical mechanics is at its very 
limit of applicability since ergodicy also gets broken, though perhaps not so sharply 
and completely as for crystals, through the onset of the glassy state. The idea is then 
to try to learn something about the structure below $T_g$ and the kinetic character of 
the glass transition starting from the phenomenology of the better understood 
supercooled liquid state (defined for temperatures $T$ in the range $T_g\le T\le T_c$, 
with $T_c$ the (equilibrium) crystallization temperature) and folding that knowledge 
in the study of the low-temperatures properties of laboratory amorphous solids. The 
return to low-temperatures is for glasses in a sense akin to starting to study the 
crystalline solid through X-ray and other (e.g. neutron-) scattering techniques at 
``zero'' temperatures assuming the atoms in fixed positions. However since at those 
temperatures the glass would look like a liquid in a static scattering experiment, one 
has to exploit other specific degrees of freedom of the cold, topologically disordered 
solid that are not present for the perfectly ordered crystals. These are
the so-called tunneling systems (TSs), local defects described in terms of new local 
degrees of freedom and that can be exploited -- much as the atomic nuclei in NMR 
research -- to probe the atomic structure and properties of the glassy state. The only
difficulty is that these TS probes are not entirely localized, each comprising several 
atoms as a rule, and have not been fully understood to date. However, progress in
their characterization and in the understanding of their microscopic nature is 
advancing also in view of some recent challenges posed by the discovery of puzzling
magnetic effects in non-magnetic glasses that cannot be attributed (despite some
interesting attempts) to trace paramagnetic impurities. Moreover, in the strategic
research for reliable solid-state qubits to be deployed for the fabrication of working
quantum computers (e.g. through use of Josephson-junction superconducting 
devices) the problems posed by the TSs (ubiquitous in the junctions) are paramount.
These new challenges foster enhanced research efforts that have culminated in the 
development of an extended tunneling model \cite{Jug2004} that relies on a new 
scenario for the intermediate-range atomic structure of glasses. The new model 
and structural scenario afford a reasonable - though not complete - explanation for 
the low-$T$ anomalies in glasses and in turns the structural scenario can be evolved 
at higher temperatures to formulate a possible mechanism for the onset of the glass 
transition from the supercooled state. Preliminary aspects of the structure that is 
expected are tested in this essay and seem to corroborate some known facts about 
the glassy state. In this new framework of ideas, research in the previously-exotic 
low-$T$ regime begins to give very useful hints about the onset of the glassy state 
from the supercooled liquid at much higher temperatures. The interplay of knowledge 
coming from the high-$T$ supercooled liquid state and from the low-$T$ cryogenic 
properties thus begins to provide a productive symbiosis for the understanding of this 
still mysterious, though ubiquitous, state of matter. 
  
At low temperatures (cryogenic) glasses are believed to be characterized by special
low-energy excitations (TSs) which are usually described through the use of 
double-welled potentials (DWPs) and of so-called two-level systems (2LSs) with 
energy asymmetry and tunneling barrier broadly distributed throughout the mass of 
the amorphous solid \cite{Phi1981,Esq1998}. While little is still known about the 
character (atomic, polyatomic or cluster-like) of the TSs, the general agreement 
is still that the intermediate-range atomic structure of glasses should be conveniently 
well described by Zachariasen's 1932 continuous random network (CRN) model 
\cite{Zac1932,War1934} (thus homogeneously disordered, just like for a liquid) and 
the 2LSs therefore result out of two slightly similar, probably localized atomic 
configurations. 
With this by now classic characterization, the 2LSs have been employed mainly in the 
1970s and 80s to explain with some success the anomalies in the physicsl properties 
of glasses at low and ultra-low temperatures, a research field thus far completely 
detached from the quest for the nature of the glass transition.
     
Yet, systematic deviations from the behaviour predicted by the standard tunneling 
model (STM) occasionally challenge (or have challenged) the validity of the model in 
the case of multi-component glasses with tunable content of the good crystal-forming 
(GCF) component (e.g. (SiO$_2$)$_{1-x}$(K$_2$O)$_{1-x}$ with changing $x$, the 
concentration of the alkali component, \cite{Mac1985,Jug2010}) and especially in 
glasses of the compositions type BaO-Al$_2$O$_3$-SiO$_2$ and then in the 
presence of a weak magnetic field \cite{Ens2002}. In such glasses (unfortunately
the mixed alkali-silicate glasses have not yet been investigated in a field, but the 
prediction is important magnetic effects there, and $x$-dependent also) a puzzling 
non-monotonous magnetic-field dependence in many properties has been unveiled in 
some physical properties \cite{Str2000,Woh2001,Nag2004}. The magnetic effect is 
typically (but not always) weak, but orders of magnitude larger than expected from 
basic thermodynamic-science considerations. The STM by itself is unable to account 
for the compositional and magnetic effects, therefore a suitable extension of the 
celebrated tunneling model for both situations has been proposed by the present 
Author \cite{Jug2004}. This so-called extended tunneling model (ETM) rests upon 
the existence (particularly in the multi-component glasses) of small regions of 
enhanced regularity (RERs) in the intermediate-range atomic structure of the 
somewhat incompletely-frozen (in fact) amorphous solid. A complete mathematical 
description and intimate physical justification of the ETM thus requires at the very 
least a partial (or complete) demise of the Zachariasen-Warren's vision of the 
intermediate-scale structure of glasses and amorphous solids in general.    

As it happens, an alternative to the homogeneously-disordered approach of 
Zachariasen-Warren has been proposed and developed in the former Soviet Union
and in some places in the West. Well before Zachariasen-Warren, Lebedev 
\cite{Leb1921} and his followers in Skt. Petersburg (but Randall in London, UK,
too \cite{Ran1930}) proposed that glasses should be made up of (initially, true) 
polycrystallites of sufficiently small size as to justify the X-ray peak rounding which 
was observed experimentally in diffractometers. Since the thermal and mechanical 
properties of glasses are reported not the same as those of polycrystalline solids, 
the concept of ``crystallite'' has evolved, meaning finally a failed micro-crystal of 
some sort. Also because finite clusters cannot be crystalline, strictly speaking. 
The evolution of the crystallite concept can be found in reviews by Porai-Koshits
(see, e.g., \cite{PK1990}) and the latest views rely on concepts such as 
``cybotactic groupings'', meaning atomic regions that can be rather extended and 
interpenetrating, but where the atomic ordering - though not complete due to 
finite size, thus highly defective - is better achieved than in the rest of the solid. 
A recent overview can be found in the works by Wright \cite{Wri2014}, and in the 
Russian literature recent ideas have been put forward also by Bakai 
\cite{Bak1994,Bak2013}. The latter Author envisages in fact better-ordered clusters 
being preferentially nucleated below $T_c$ through some kinetic mecanisms that 
have the clusters survive and grow at the expense of true micro-crystalline nuclei that 
have no time to grow. At the glass transition these ill-formed, but better-ordered 
clusters merge together and get to form a polycluster that is the macroscopic skeleton 
of the solid glass. Evidently, looser material is also present in a patchwork atomic
structure, characterized by spatial and (above $T_g$) temporal heterogeneities.
   
In this review we provide a scenario for the intermediate-range atomic structure 
of glasses, the cellular model, which is very much reminiscent of the polycluster 
theory of Bakai and a formulation within which the phenomenological assumptions 
for the ETM's mathematical framework (given in several papers by the present 
Author) become completely - or at least to a large extent - justified. The cellular 
or polycluster model provides for a definitely more realistic mathematical 
formulation of this framework in terms of a tetrahedric four-welled tunneling 
quasi-particle potential, as said, the simplified triangular three-welled version of 
which is nothing but a poor-man's, probably rather realistic, version affording a 
much speedier mathematical description. Within this cellular approch to the 
structure of glasses the most significant local tunneling potentials (probably in 
terms of number density) turn out to be the DWPs, this for a single (or very few) 
atomic particle(s), and the said tetrahedric four-welled potential (TFWP) for a 
correlated cluster of $N\gg 1$ charged real atomic particles. A reasonable and 
very useful simplification for the TFWP is thus the replacement of the $N$ 
interacting and tunneling atomic particles with a single fictive quasi-particle 
subject to a triangular tunneling potential (TWP) and carrying renormalized 
parameters (charge, magnetic threaded area, energy asymmetry as well as 
tunneling probability), quantum-mechanically moving about one face of the full 
tetrahedric potential. The renormalization gets fully justified by the proximity 
of (on average) four similarly quasi-ordered, close-packed atomic cells (RERs or 
better-ordered regions) and the reasonable assumption that most of the charged 
particles will avoid the interstice's tetrahedral potential's centre.
We present again, therefore, a brief mathematical description of the TWP and its 
quantum mechanics in those limits appropriate for practical applications. We 
then show the main results obtained for the description of the density of states 
(DOS) and the temperature and magnetic-field dependence for the specific heat of 
some specific glasses, then for the dielectric constant (real and imaginary parts) in 
the linear-response regime, and for the polarization echo - always in the presence 
of a magnetic field (limiting to results obtained for the multi-silicates).

With this scenario of the cellular, or polycluster structure of real glasses assimilated, 
the question of the description of the glass transition is one of the next challenges.
Having nucleation theory, and for better-ordered or Bakai clusters in mind the 
growth of such clusters with a cooling rate $\kappa$ can be considered and the 
temperature at which the polycluster appears is taken to be $T_g$. With some 
phenomenological assumptions in the light of standard Adam-Gibbs theory (where
the RERs, or Wright's better-ordered clusters, appear as CRCs or 
coherently-rearranging clusters) \cite{AG1965}, the dependence of $T_g$ from 
$\kappa$ is worked out and the known logarithmic dependence can be recovered. 

The paper is organized as follows. In Section 2 we introduce the cellular 
model for the structure of glasses, now a complete departure from Zachariasen's
continuous random-network model, and then examine the likely tunneling states 
that would emerge from such cellular picture, to conclude that only DWPs and 
TFWPs should be relevant for the physics of glasses below the glass transition 
temperature $T_g$. In Section 2 we also review the basic relevant quantum 
mechanics of the three-welled, poor man's version of the TFWP, a version that 
has been used to date to obtain a reasonable single explanation for all of the 
anomalies (and deviations from the STM predictions) due to composition changes 
and to the presence of a magnetic field. We also show how to evaluate the 
magnetic density of states (DOS) $g(E,B)$ and, in Section 3, we briefly review some 
of the magnetic-field dependent low-temperature physical properties that have 
been studied to date, such as the heat capacity $C_p(T,B)$ comparing with some 
of the published data for the multi-silicate glasses.  We then do the same for the 
dielectric constant, real part $\epsilon'$ and (just comment on the) imaginary part 
$\epsilon''$ as well, also showing some comparison with available data at low 
(kHz) frequency, and then we examine the application of the ETM to the explanation 
of some of the data for the polarization (or, electric) echoes in the silicates (only 
commenting about glycerol, for which case the ETM has been able to explain the 
so-called isotope effect, which is in fact a mere mass substitution effect). 
Section 4 contains a discussion on how the cellular model can be  implemented 
to gain information on the size of the better-ordered cells making up the polycluster
structure of amorphous solids. The typical values for the cell size extracted from
low-temperature experiment parameters are then compared with available high 
resolution electron microscopy (HREM) imaging of insulating glasses that show 
such structure. It is shown that estimates from low temperatures and from HREM 
imaging compare reasonably well. A brief derivation of the
model's prediction for the cooling rate dependence of $T_g$ is also provided and 
this Section 4 contains also our Conclusions.

\section{The Cellular Model for the Atomic Structure of Glasses and the 
Three-Welled Tunneling Potential}

Glass-forming liquids are an important class of materials for technology and 
ubiquitous applications, nevertheless the atomistic structure of glass in relation 
to its physical properties remains a mystery. As stated in the Introduction,
the widespread conception is that the atomic arrangement of a glass should be 
the same as that for a liquid, as is indeed implicit in Zachariasen's 1932 proposed 
continuous network picture \cite{Zac1932,War1934}, which has been widely 
adopted by scholars (at least in the West, see below). This picture differs from 
that of a liquid only in that a dynamical arrest has occurred, without specifying 
its ultimate origin. Relaxation times diverge ``near'' $T_g$, but why? In a 
spin-glass the ultimate origin of dynamical arrest is magnetic frustration (with 
or without disorder), but for ordinary structural glasses it remains mysterious 
and an important open issue \cite{Ber2011,Ang1988,Lub2007,Sim2009}. In the
opinion of the present Author, lack of justification for this arrest is the primary 
roadblock for the achievement of a theory of the glassy state encompassing all
aspects of glass physics in every possible temperature range. As mentioned in 
the Introduction, prior to Zachariasen's scheme, however, the Soviet scientist 
A.A. Lebedev had proposed, in 1921, the concept of ``crystallites" \cite{Leb1921}, 
small crystal-like (yet not truly ordered) regions jammed against each other in 
random orientation to contain altogether all (actually, almost all) atoms in the 
glassy substance. Later, Randall proposed that these be real micro-crystals and 
explained the rounded-up X-ray spectra from glasses in this manner \cite{Ran1930}. 
However, the density of glasses is typically some 10\% less than that of 
poly-crystalline aggregates and the thermal properties of glasses too cannot be 
explained by means of the Lebedev-Randall picture. Despite these observations, 
the East-West dychotomy continued to these days. The Zachariasen-Warren model 
of glass structure was for instance criticized by H\"agg \cite{Hag1935} in the West 
right in the early days of X-ray crystallography and a good review of the status 
quo of this controversy has been recently provided by Wright \cite{Wri2014} 
who concludes, from a re-analysis of X-ray and neutron-scattering data from 
many different covalent-bonded and network glasses that indeed so-called
{\em cybotactic groupings} (better-ordered regions) may well be present and 
frozen-in in most glasses, particularly if multi-component (or - one could add -
polymeric \cite{Koi2016}). 
The formation of {\em polyclusters}, instead of jammed crystallites, in most 
glasses is the latest scenario by the Eastern school \cite{Bak1994,Bak2013}, which
is based on observations and kinetic reasoning. As nicely set out by Bakai 
\cite{Bak1994}, the incipient crystals forming at and below $T_c$ (melting point) 
are in constant competition with kinetically swifter (for glass-forming liquids) 
embryo clusters (or crystallites) that can win thermodynamically and kinetically 
over crystals during a rapid enough quench forming a polycluster spanning the 
whole of the sample.  

On the experimental side, the concept of de-vitrification has been gradually taking 
sway with reports of metallic glasses \cite{Hwa2012} and also of monocomponent 
high-coordination covalent solids, like amorphous Si, forming {\em paracrystals} 
in their amorphous solids \cite{Tre2012}. Therefore, the stance will be taken in this 
new review that only the purest mono-component glasses may (perhaps) abide to 
the Zachariasen-Warren dogma of the continuous random-network model for a 
glass, whilst the vast majority of real glasses \cite{Phi1983} will be organized 
differently at the intermediate-range atomic structure level. Solid-like fluctuations 
of finite size are, in fact, likely to form already around and below the crystallization 
temperature $T_c$, the formed solid-like clusters continuously breaking up and 
reforming in the supercooled state between $T_c$ and $T_g$, which is indeed 
widely known to be characterised by the so-called dynamical heterogeneities (DH). 
The slower-particle regions of the DH are, in the present Author's view, to be 
identified with the solid-like (better-ordered) clusters and the faster-particle DH 
regions with the liquid-like (completely disordered) clusters that become thinner 
and thinner as $T_g$ is approached. In this vision, the lower temperature glassy
structure (now entirely solid-like) inherits the inhomogeneous DH structure of the 
supercooled state, the slower, solid-like better-ordered regions having grown to a 
limit size (e.g. a maximum average radius $\xi_0$) that is determined by the onset 
of the polycluster.
\begin{figure}[h]
\centering
   \subfigure[]{\includegraphics[scale=0.25] {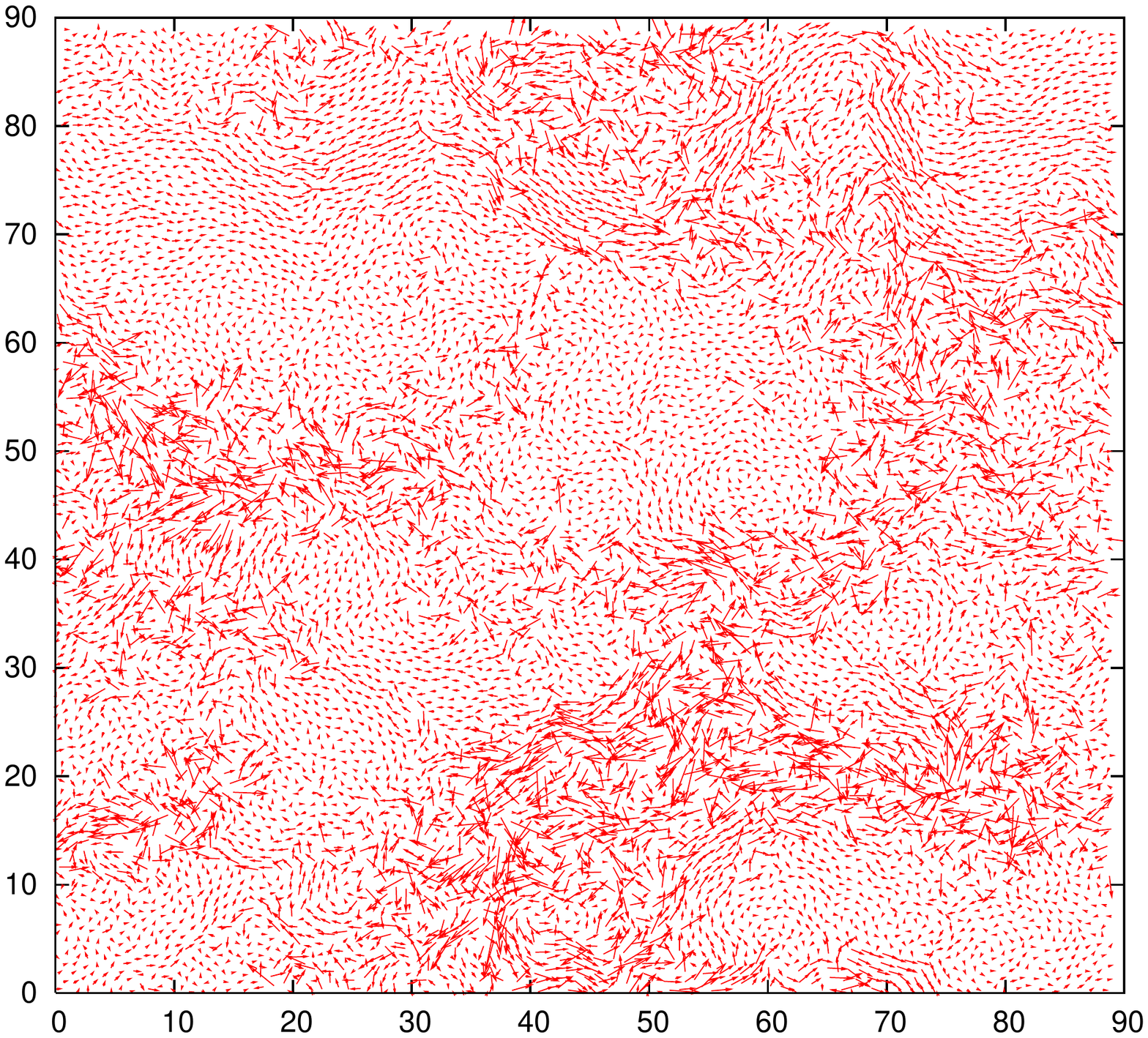} }
   \subfigure[]{\includegraphics[scale=0.36] {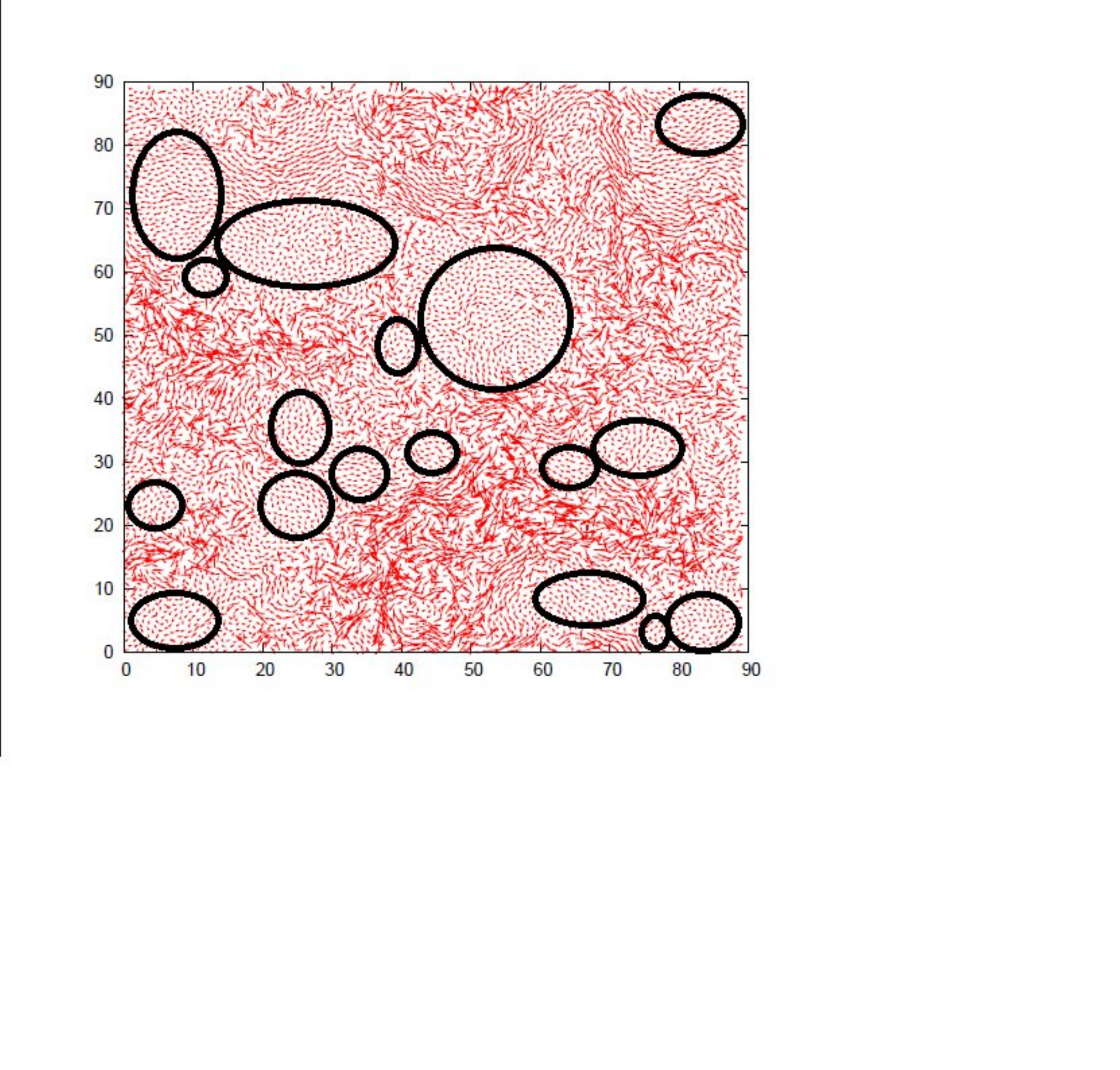} }
\vskip -1mm
\caption[2]{ (a) Spatial map of a single-particle displacements
in the simulation of a binary Lennard-Jones mixture in two dimensions. Arrows
show the displacement of each particle in a trajectory of length comparable to 
to the structural relaxation time. The map (courtesy of G. Biroli, from 
\cite{Ber2011}) reveals the existence of particles with different mobilities during
relaxation, also the existence of spatial correlations between these dynamical 
fluctuations. Faster and slower particle regions are clearly visible and a closer
inspection reveals that the slower particles form better-ordered regions, while
the faster particles form liquid-like regions. Note
the distinct clustering of slow particles (dynamical heterogeneities, DH). 
(b) The slower regions have been schematically highlighted to show their 
incipient cellular structure. At the temperature of the simulation ($T>T_g$)
the slower regions are continuously breaking up and reforming, the idea is that
as $T_g$ is approached their size grows to a finite limit value and their mutual 
hindering significally slows down their dynamics. The ringed regions of slower
moving particles become, below $T_g$, the RERs or cells characterizing glassy
intermediate-range atomic structure. }
\label{Fig1}
\end{figure} 
The structure now proposed is a cellular-type arrangement of better-ordered 
atomic regions (regions of enhanced ordering, RERs) that can have complicated, 
maybe fractal, but definitely compact shapes with a narrow size distribution and 
the interveening regions (let's call them interstitials) 
between them populated by still fast-moving particles (normally charged and
possibly dangling-bonded ions). The insterstitial regions between the RERs are 
here the equivalent of the concept of {\it cages}, that is often discussed in the 
glass structure's literature (see e.g. \cite{Ber2011}). We remark that a cellular-type 
structure was already embodied within the ``crystallite'' idea of Lebedev, but 
now no micro-crystals are claimed here to exists (except definitely in the 
ceramic-glasses case, to a good extent). The RERs are more like Wright's 
``cybotactic regions'' \cite{Wri2014} or Treacy's ``paracrystals'' \cite{Tre2012} 
(there for a-Si, not a quenched-melt solid) and issue from the DH picture for
the glass-forming supercooled liquid state above $T_g$ (see e.g. 
Fig.~\ref{Fig1}) \cite{Ber2011,Hur1995,Sil1999,Edi2000}. The DH picture does
indeed recognise the presence of regions of ``slower'' and ``faster'' particles,
and an inspection of the slower-particle regions of the supercooled liquid 
reveals that these are also better ordered (solid-like) whilst the 
faster-particle regions are rather much more liquid-like. DH are ubiquitous in 
perhaps all supercooled liquids \cite{Edi2000} and the claim (though still 
somewhat speculative) here is that the slower-particle regions will grow on 
approaching $T_g$, albeit only up to a finite size and will now be giving rise 
to the RERs in the frozen, glassy state below $T_g$. In fact, simulations in 
the frozen glassy state of the slower- and faster-particle regions do confirm
that a DH-like picture applies also below $T_g$ \cite{Vol2005} and with the 
slower-regions's size $\xi$ increasing as $T\to 0$. Earlier simulations (always for
model systems) \cite{Don1999} did point out the difficulty of simulating the DH 
picture below $T_g$ and came up with a picture of these slower-regions growing, 
and possibly diverging in size, on approaching $T_g$. In the present reviw, 
however, the stance will be taken that the slower-regions' average (even 
maximum) size growths in real systems, but does not diverge at $T_g$ or at any 
other characteristic temperature. Full experimental or numerical proofs of this 
fact are, unfortunately, still lacking and should be considered as a reasonable 
working hypothesis in this paper. Indeed, the absence of a diverging characteristic
length is, in a nutshell, the central enigma of the problem of the glass transition.
We remark in passing that a cellular structure (mosaic-like) for the glasses had 
been proposed by de Gennes \cite{deG2002} in the past and, in the context of 
the low-temperature anomalies, by Baltes \cite{Bal1973} who was able to explain 
the linear in $T$ anomaly in the heat capacity $C_p$ (but not those in the 
acoustic properties, though, which require the introduction of tunneling). A very 
similar picture is that of the polyclusters of Bakai \cite{Bak1994}, as was already 
mentioned, but in the present approach the thermal genesis of the cells, or RERs, 
is ascribed directly to the DH scenario already present above $T_g$.  In this 
approach the RERs, like grains or domains in the frozen structure below $T_g$, 
are the thermal-history continuation of the slower-particle regions of the DH 
above $T_g$. In Fig. \ref{mosaicstructure} we show how the RERs themselves 
(and not the single particles) can get to grow to a limit intermediate-maximum 
size $\xi_0$ and randomly close-pack together in the proximity of the 
glass transition temperature $T_g$, thus forming a highly correlated polycluster. 
The random close-packing process gets to be completed at a lower temperature
$T_K<T_g$ ($T_K$ possibly to be identified with the Kauzmann temperature) thus 
giving the glass transition more the character of a crossover.
\begin{figure}[h]
\centering
\vskip -60mm
{
   {\includegraphics[scale=0.60] {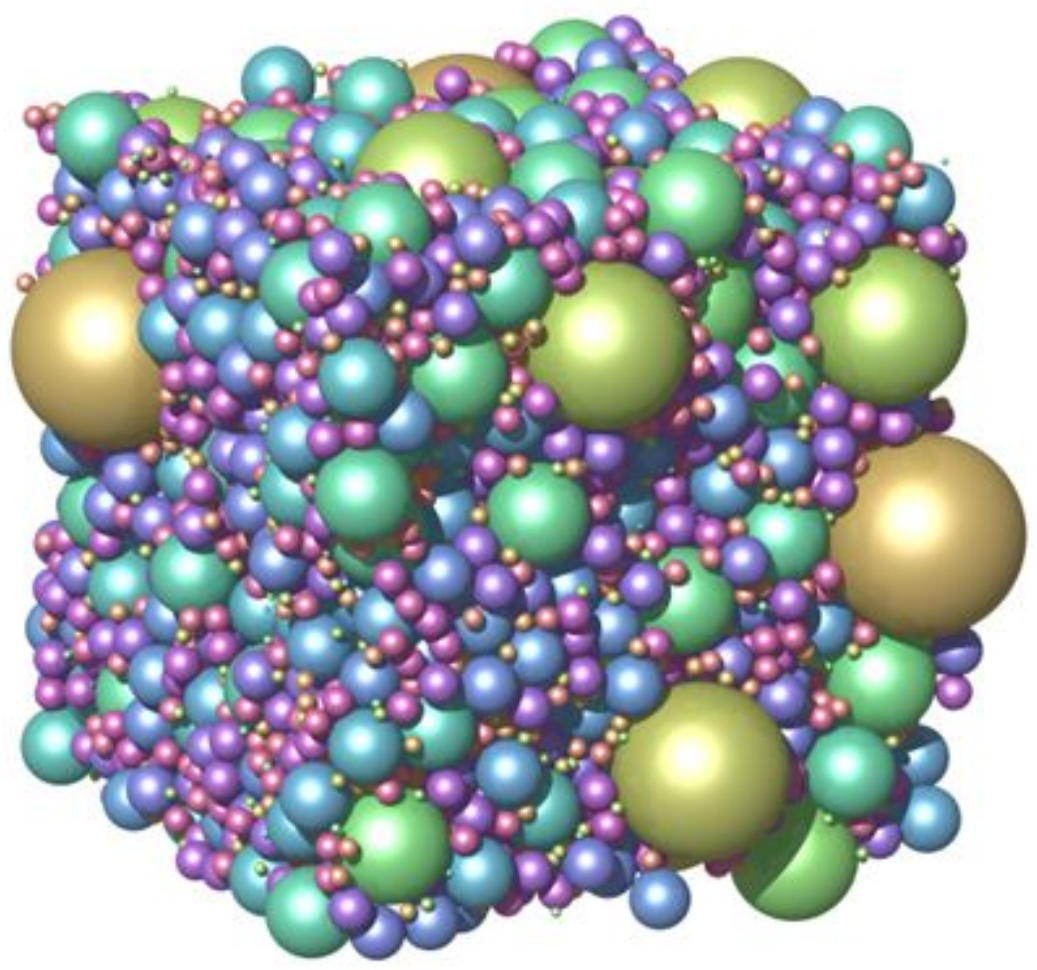} }
\vskip -60mm
}
\caption{Random close packing of compact RERs in the proximity of the glass
transition temperature $T_g$. The spheres represent regions of better-ordered 
atomic particles that have grown to a limit maximum size and hinder each other's
further growth. In the interstitial regions (voids between the spheres), not shown, 
are the remainder of the 
incipient glass particles: these are more liquid-like and mobile, while the RERs 
that are shown need not be spherical nor completely mutually exclusive, partial 
coalescence being allowed.  }
\label{mosaicstructure}
\end{figure}
As temperature drops further below $T_K$, the cells or RERs can slightly increase 
in size, consolidating their growth at the expense of the species present in the 
interestitials between the RERs. This idea of consolidation at lower temperatures is 
shown pictorially in Fig. \ref{lowTstructure}.
\begin{figure}[h]
\centering
{
   {\includegraphics[scale=0.30] {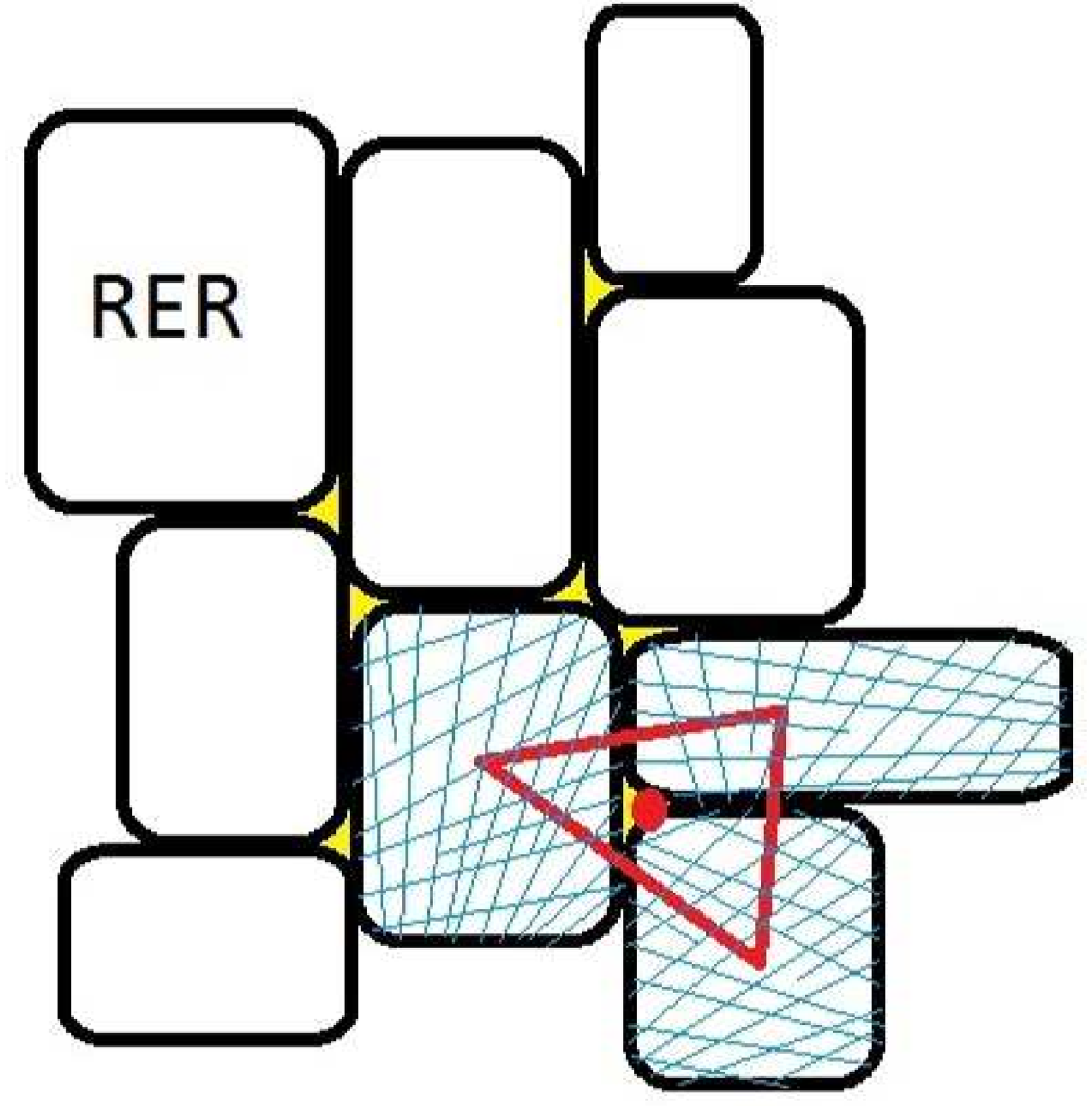} }
\vskip -3mm
}
\caption{ 2D cartoon of the consolidation of the RERs at the lower temperatures, 
where the better-ordered regions have grown at the expense of the particles in 
the interstitials. Ultimately this leads to a temperature-dependent number of 
liquid-like tunneling particles in each interstitial. }
\label{lowTstructure}
\end{figure}
The cells and interfaces between the cells will contain the bulk of the tunneling 
systems. These 2LSs are atomic tunneling states arising from the cells' own disorder, 
but most of the 2LSs should be located at the meeting point between two cells (two 
RERs). In the interstitial spaces between cells the remnant faster particles 
of the DH at $T>T_g$ give rise for $T<T_g$ to regions where a large number $N$ 
(on average, per interstitial) of charged atomistic tunneling particles are constrained 
to move in a coherent fashion due to the high Coulomb repulsion forces between 
them. Fig. \ref{cellstructure} shows in a
schematic way how the atomic/ionic matter can get organised below $T_g$ in a
real glass. 
\begin{figure}[h]
\centering
   \subfigure[]{\includegraphics[scale=0.50] {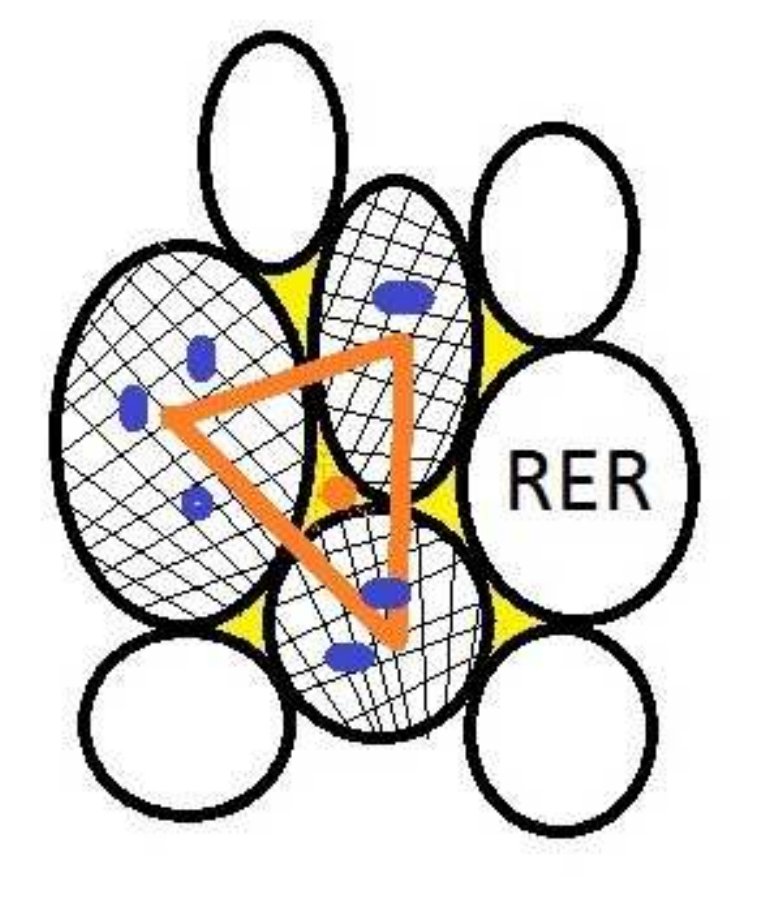} }
   \subfigure[]{\includegraphics[scale=0.25] {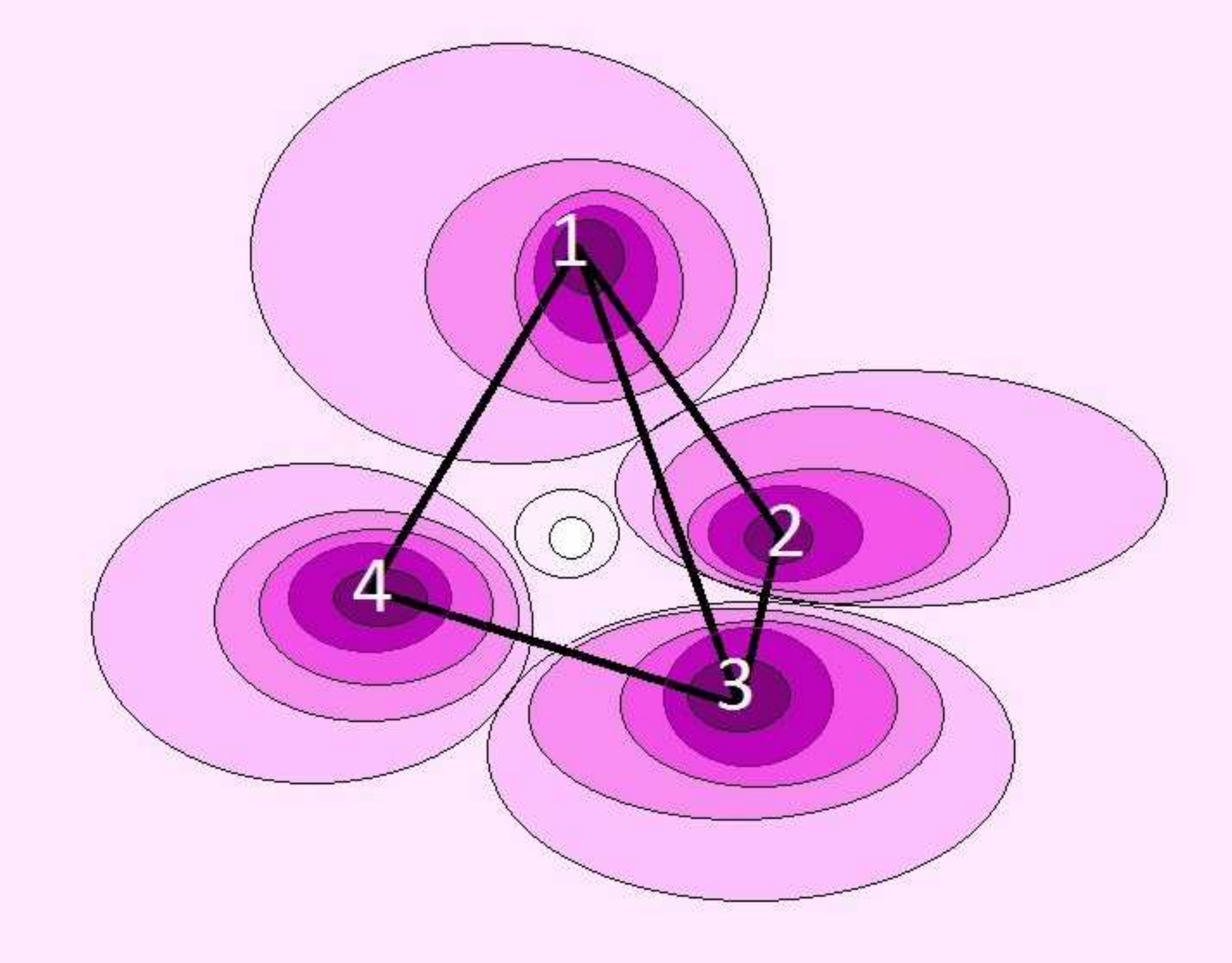} } 
\vskip -3mm
\caption[2]{ (a) A 2D cartoon of the cellular structure of an amorphous 
(melt-quenched) solid just below $T_g$. The RERs (black-circled blobs, an 
oversimplified schematics for fractal-like, but compact objects) have grown 
to completely fill the space and enclose atomistic tunneling states of the 
2LS type (blue blobs). At the same time, in the RER interstitials (yellow 
regions, connecting to each other) the trapped, charged and faster particles 
of the DH existing above $T_g$ (now probably charged dangling bonds), 
give rise to coherently-tunneling large groups of ions here represented by a 
single, fictitious, quasi-particle (orange dot) subjected to an effective 
tunneling potential having typically four natural wells in distorted-tetrahedral 
configuration (b) for close-packed RERs. (b) The tetrahedral four-welled
potential (TFWP) in a 3D representation with a colour-coded potential intensity
(dark=deepest, light=highest). }
\label{cellstructure}
\end{figure}     
Since the charged ions (dangling bonds, most likely) should act as a 
coherently tunneling ensemble, it seems natural to simplify the description of the 
physics at the lower temperatures using only phonons, propagating in the 
collection of cells now jammed against each other, and remnant ergodic 
localized degrees of freedom acting as TSs. These TSs will be of two types: the 
2LSs within the cells and at their points of contact (owing to inherent disorder in 
the cells' atomic arrangement) and effective quasi-particles sitting in the 
close-packed cells' interstices and now representing the collective motion of the 
coherently-tunneling ions which are trapped in each interstice 
(Fig. \ref{cellstructure}). 
\begin{figure}[h]
\centering
   \subfigure[]{\includegraphics[scale=0.35] {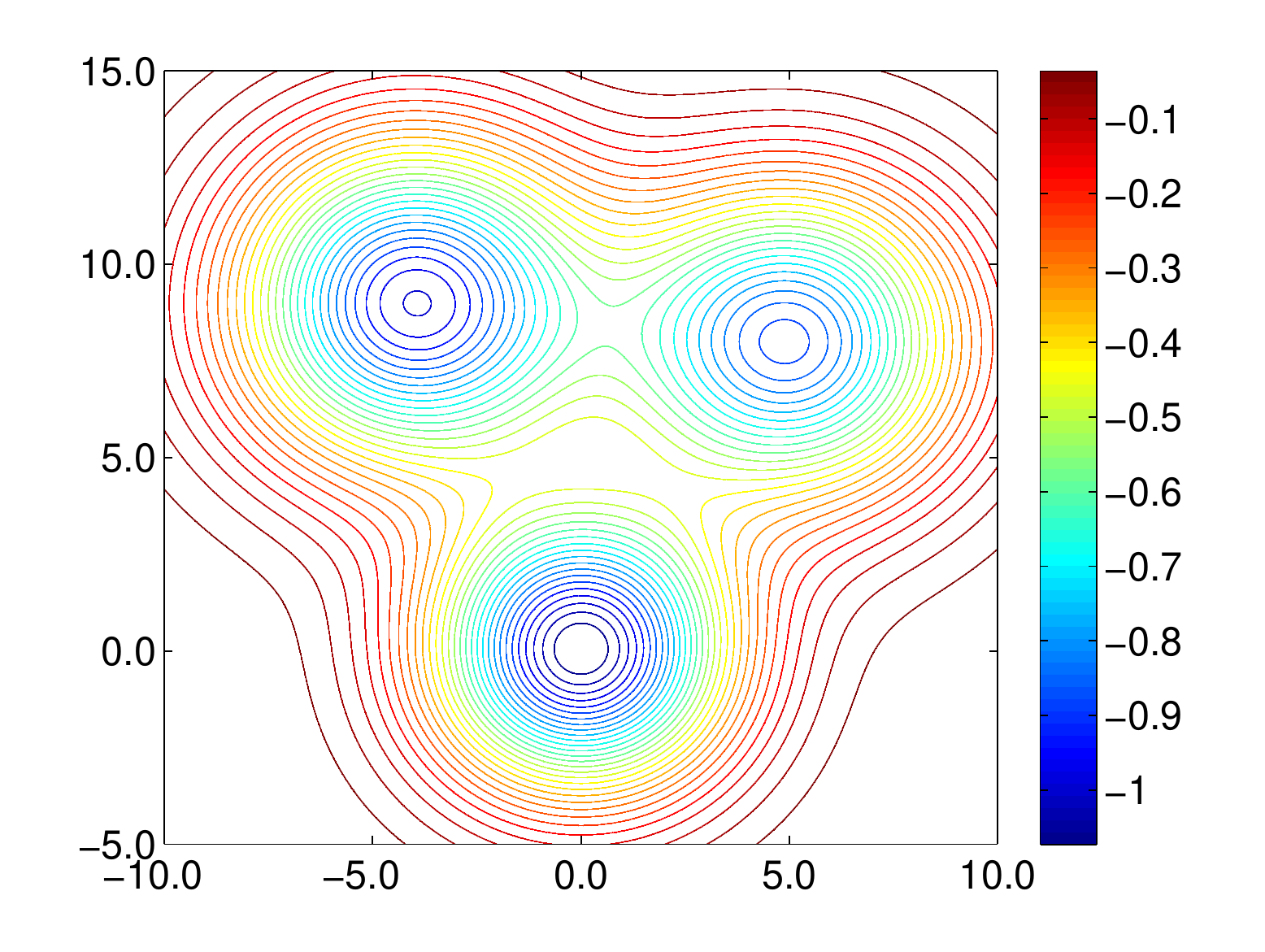} }
   \subfigure[]{\includegraphics[scale=0.35] {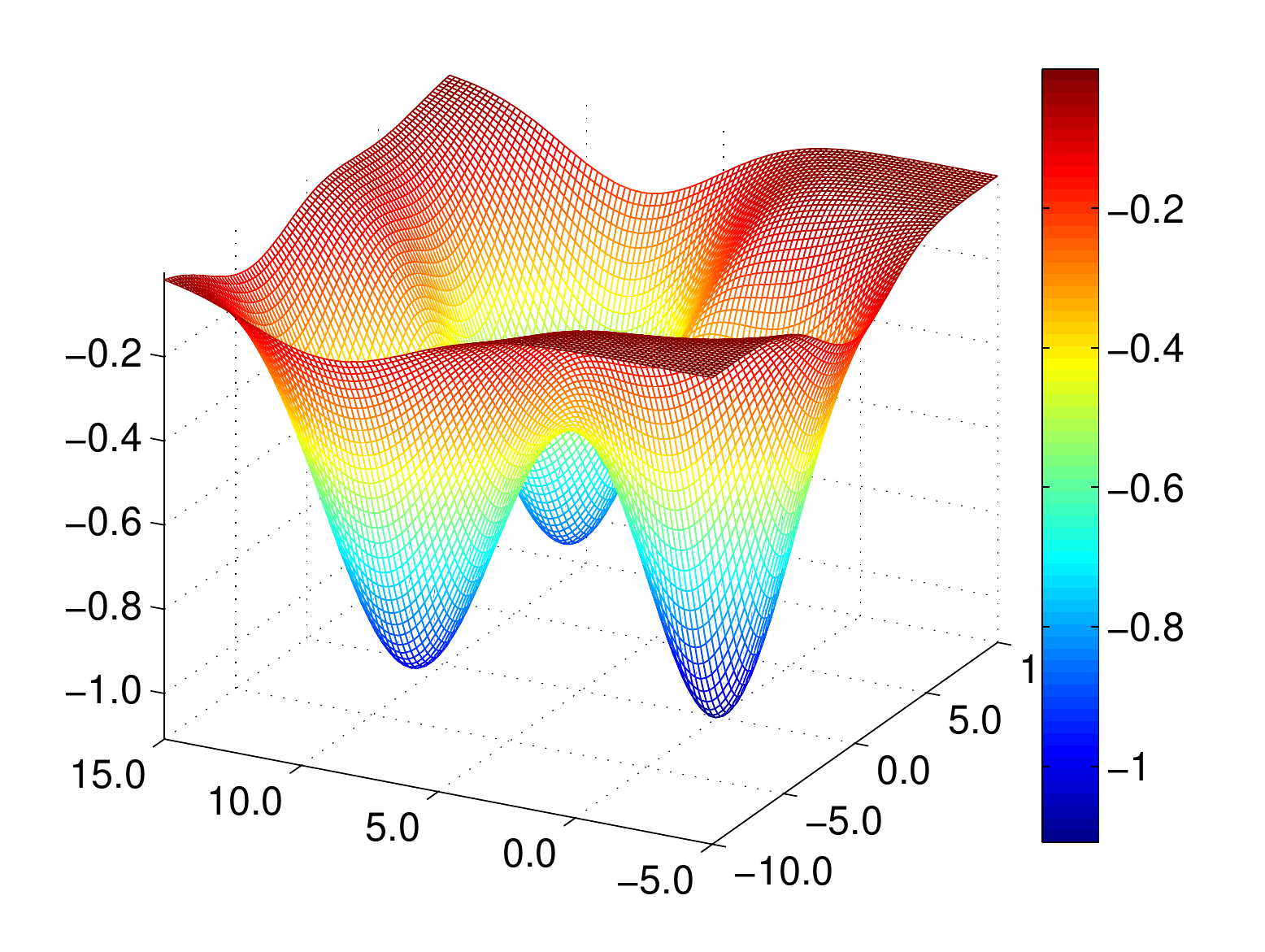} }
\vskip -3mm
\caption[2]{ (a) Contour-plot of a possible realization of the 2D effective 
three-welled potential (TWP) very likely felt by the quasi-particle of those 
charged real particles dangling from, or being trapped by, a group of three 
RERs on each one face of the tetrahedral configuration for an interstice 
formed by close-packed RERs well below $T_g$ (and $T_K$). (b) 3D 
attempted visualization of the same example of a TWP potential. }
\label{threewells}
\end{figure}
The quasi-particle will be subjected to an effective potential of distorted
tetrahedral topology characterized by four wells for each RER interstice, with
minima inside each RER or - depending on the material - at the RER meeting
points and a high barrier in the interstice's centre. {\it De facto} this 3D interstitial
TFWP potential can be replaced by four local 2D potentials for the four 
quasi-particles describing the coherent tunneling of the faster-moving 
particles sitting near each face of the distorted tetrahedron, close to a group of 
three (on average) quasi-ordered cells (Fig. \ref{threewells}). Because of the 
better-ordering implicit in this model of the glassy intermediate-range atomic
structure, and in each cell, the three wells of each effective local 2D 
potential for the tunneling quasi-particles (four per interstice, on average)
should be near-degenerate in terms of their ground-state energy asymmetries:
$E_1 \simeq E_2 \simeq E_3 \simeq 0$.
With this, still qualitative, picture in mind we now turn to the mathematical 
description of the physics of the remnant and still ergodic degrees of freedom
(phonons in the cells' network - likely the origin of the Boson peak - 2LSs and 
ATSs (anomalous tunneling systems, four in each interstice)). For our model 
of a real glass, by construction the 2LSs will be much more numerous than the 
ATSs.   

In this approach \cite{Jug2004} the relevant degrees of freedom, beside the 
phonons, are generalised dilute gases of independent 2LSs described by the STM 
and of fictitious quasi-particles tunneling in TWPs.
The formulation of the STM (the 2LS model) for the low temperature properties 
of glasses is very well known \cite{Phi1981,Esq1998}. One assumes a collection 
of DWPs distributed in the substance and represented each by a 2$\times$2 
effective Hamiltonian of the form, in the potential-well (or real-space) representation:
\begin{equation}
{\cal H}_{2LS}=-\frac{1}{2}\left( \begin{array}{cc}
\Delta & \Delta_0 \cr
\Delta_0 & -\Delta \end{array} \right).
\label{2lstunneling}
\end{equation}
Here the two parameters $\Delta$ (the energy asymmetry) and $\Delta_0$ 
(twice the tunneling parameter) are typically characterized by a probability
distribution that views $\Delta$ and $\ln(\Delta_0)$ (the latter linked to
the particular DWP energy barrier) broadly (in fact, uniformly) distributed 
throughout the topologically disordered solid~\cite{Phi1987}:
\begin{equation}
{\cal P}_{2LS}(\Delta,\Delta_0)=\frac{\bar{P}}{\Delta_0}
\label{2lsdistribution}
\end{equation}
where some cutoffs must be introduced when needed and where $\bar{P}$ is 
an elusive material-dependent parameter, like the cutoffs. In fact, Eq. 
(\ref{2lsdistribution}) embodies entirely the Zachariasen-Warren hypothesis for 
the intermediate atomic structure of a glass, assuming broadly-distributed the
energy asymmetry $\Delta=E_2-E_1$, hence the single-well ground state 
energies $E_1$ and $E_2$ themselves, as well as the potential barrier height 
$V_0$ appearing typically in relations such as:
\begin{eqnarray}
&&\Delta_0\simeq\hbar\Omega e^{-\frac{d}{\hbar}\sqrt{2mV_0}} \\
&& \Delta_0=\frac{\hbar\Omega}{2} \left[3-\sqrt{\frac{8V_0}{\pi\hbar\Omega}}
\right]e^{-2\frac{V_0}{\hbar\Omega}} \label{overlap}
\end{eqnarray}   
Here, the first relation is the generic WKB result for an arbitrarily shaped DWP 
(where $m$ is the particle's mass, $\Omega$ its single-well harmonic frequency 
(or tunneling attempt frequency) and $d$ the tunneling distance) and the second 
formula refers instead to a symmetric ($\Delta=0$) DWP made up by two 
superimposed parabolic wells. In fact, in the end the distribution (\ref{2lsdistribution})
refers to the combination of parameters $\Delta_0/\hbar\Omega$. The
energies of the two levels $|0>$ and $|1>$ are then given by 
${\cal E}_{0,1}=\pm\frac{1}{2}\sqrt{\Delta^2+\Delta_0^2}$, and so on 
\cite{Phi1981,Phi1987}.

The tunneling Hamiltonian of a particle in a TWP is also easily written down, in 
the same low-$T$ spirit as for a 2LS, as a generalization of the above matrix 
formulation to the case of three wells~\cite{Jug2004}:
\begin{equation}
{\cal H}_{3LS}=\left( \begin{array}{ccc}
E_1 & D_0 & D_0 \cr
D_0 & E_2 & D_0 \cr
D_0 & D_0 & E_3 \end{array} \right)
\label{3lstunneling}
\end{equation}
where $E_1, E_2, E_3$ are now the energy asymmetries between the wells and 
where $D_0$ is the most relevant tunneling amplitude (through saddles of the 
glassy potential energy landscape, or PEL, in fact). No disorder in $D_0$ is 
considered, for simplicity, within each single DWP.
This 3LS Hamiltonian has the important advantage of readily allowing for the 
inclusion of a magnetic field $B>0$, when this is coupling orbitally with a 
tunneling ``particle'' (in fact, a quasi-particle) having charge $q$ ($q$ being 
some multiple of the electron's charge $e$)~\cite{Jug2004}:
\begin{equation}
{\cal H}_{3LS}(B)=\left( \begin{array}{ccc}
E_1 & D_0e^{i\varphi/3} & D_0e^{-i\varphi/3} \cr
D_0e^{-i\varphi/3} & E_2 & D_0e^{i\varphi/3} \cr
D_0e^{i\varphi/3} & D_0e^{-i\varphi/3} & E_3 \end{array} \right)
\label{3lsmagtunneling}
\end{equation}
where $\varphi/3$ is the so-called Peierls phase for the tunneling particle 
through a saddle and in the field, and $\varphi$ is the Aharonov-Bohm (A-B) 
phase for a tunneling loop (triangular shaped) and is given by the usual formula:
\begin{equation}
\varphi=2\pi\frac{\Phi}{\Phi_0}, \qquad \Phi_0=\frac{h}{\vert q\vert}
\label{ABphase}
\end{equation}
$\Phi_0$ being the appropriate magnetic flux quantum ($h$ is Planck's constant) 
and $\Phi={\bf B}\cdot{\bf S}_{\triangle}$ the magnetic flux threading the area 
$S_{\triangle}$ and formed by the tunneling paths of the particle in this now 
simplified (poor man's, yet as we have seen realistic) model. The three energy
asymmetries $E_1, E_2, E_3$ typically enter through their natural combination
$D\equiv\sqrt{E_1^2+E_2^2+E_3^2}$. 

For $n_w$=3 wells an exact solution for the $k$=0, 1, 2 eigenvalues of the
multi-welled tunneling Hamiltonian Eq. (\ref{3lsmagtunneling}) is still 
possible (but not in the $n_w$=4 TFPW case):
\begin{eqnarray}
&&{\cal E}_k = 2D_0\sqrt{ 1-\frac{\sum_{i\not=j}E_iE_j}{6D_0^2} } ~
\cos\bigg( \frac{1}{3}\theta+\theta_k \bigg)
\label{3ls} \\
&&\cos\theta = \left( \cos\varphi+\frac{E_1E_2E_3}{2D_0^3} \right)
\left( 1-\frac{\sum_{i\not=j}E_iE_j}{6D_0^2} \right)^{-3/2}
\nonumber
\label{solution}
\end{eqnarray}
$\theta_k=0,+\frac{2}{3}\pi,-\frac{2}{3}\pi$ an index distinguishing the three
eigenstates. In the physically relevant limit, which we now consider, and in which 
$\varphi\to 0$ (weak fields) and
$D=\sqrt{E_1^2+E_2^2+E_3^2}\to 0$ (near-degenerate distribution), and 
always at low temperatures, we can approximate (in a now simplified calculation) 
the $n_w=3$ - eigenstate system with an {\em effective 2LS} having
its energy gap $\Delta{\cal E}={\cal E}_1-{\cal E}_0$ widening with increasing 
$\varphi$ provided $D_0>0$ (see below):
\begin{equation}
\lim\Delta{\cal E}\simeq
\frac{2}{\sqrt{3}}\sqrt{ D_0^2\varphi^2+\frac{1}{2}(E_1^2+E_2^2+E_3^2) }
\to \sqrt{D_0^2\varphi^2+D^2}
\label{3gap}
\end{equation}
(a trivial rescaling of $D_0$ and of the $E_i$ has been applied).
One can easily convince oneself that if such TWP is used and with the standard 
parameter distribution, Eq.~(\ref{2lsdistribution}) (with $D, D_0$ replacing 
$\Delta, \Delta_0$) for the description of the TS, one would then obtain 
essentially the very same physics as for the STM's 2LS-description. In other 
words, there would be no need to complicate the popular, minimal 2LS-description 
in order to study glasses at low temperatures, unless structural inhomogeneities of 
the RER-type and a magnetic field are present. Without the RERs, hence no 
distribution of the type (\ref{atsdistribution}) below, the phase interference from 
separate tunneling paths is only likely to give rise to an exceedingly weak quantum 
effect. Hence, it will be those TSs nesting between the RERs that will give rise to 
an enhanced quantum phase interference and these TSs can be minimally 
described -- most appropriately and conveniently -- through Hamiltonian 
(\ref{3lsmagtunneling}) and with a distribution of asymmetries
thus modified in order to favour near-degeneracy ~\cite{Jug2004}:
\begin{equation}
{\cal P}_{3LS}^*(E_1,E_2,E_3;D_0)=\frac{P^{\ast}}{D_0(E_1^2+E_2^2+E_3^2)}
\label{atsdistribution}
\end{equation}
$P^*$ being a dimensionless (which is pleasing) material parameter. We remark 
that the incipient ``crystallinity'' (better-ordering, in fact) of the RERs calls for 
near-degeneracy in $E_1, E_2, E_3$ simultaneously and not in a single one
of them, whence the correlated form of (\ref{atsdistribution}). We now have
basically three-level systems (3LSs) with energy levels
${\cal E}_0<{\cal E}_1\ll {\cal E}_2$, periodic in $\varphi$. The typical ATS 
spectrum, with $D_0>0$ (see below), is shown in Fig. \ref{spectrum} as a
function of $\varphi$ and one can now see that the third and highest level 
${\cal E}_2$ can be safely neglected for most low-field applications. Other
descriptions, with TFWPs or modified three-dimensional DWPs are possible for 
the TSs nested in the RERs and yet they lead to the same physics as that from 
Eqs.~(\ref{3lsmagtunneling}) and (\ref{atsdistribution}) above 
(which describe what we like to call the anomalous tunneling
systems, or ATSs, nesting in the interstitials between the RERs).
\begin{figure}[h]
\centering
{
   {\includegraphics[scale=0.50] {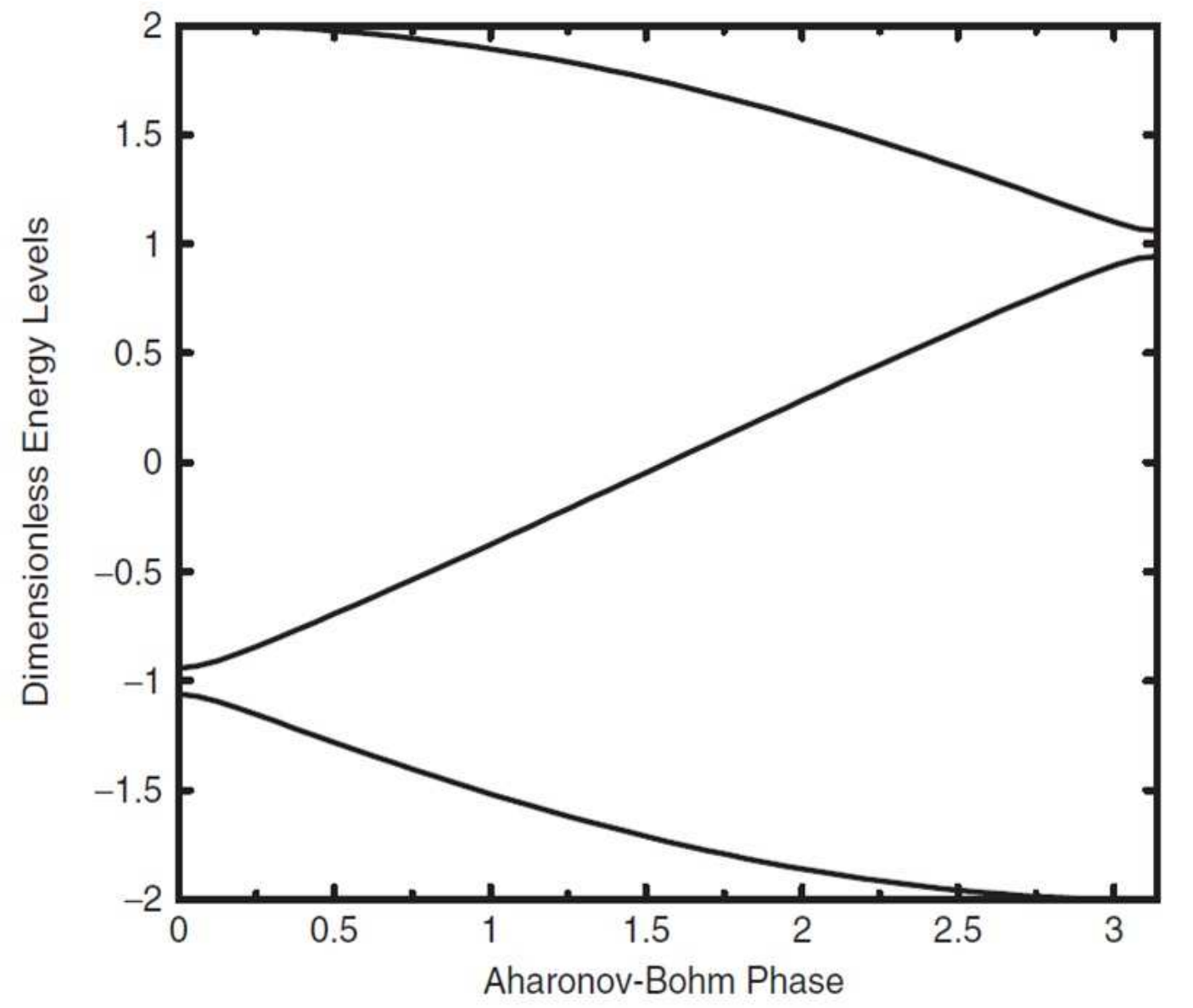} }
\vskip -3mm
}
\caption{ The energy spectrum (for $D_0=1$ units) of the ATS (3LS, TWP) model, 
in the physically appropriate limits of a weak magnetic field and near-degeneracy 
due to the embedding within RER's interstitials. On the horizontal axis is the 
A-B phase $\varphi\propto B$. }
\label{spectrum}
\end{figure}
The final and a most important consideration for the construction of a suitable 
mathematical model of such complex systems is that the TSs appear to be rather
diluted defects in the glass (indeed their concentration is of the order of magnitude 
of that for trace, paramagnetic (e.g. Fe), impurities, as we shall see). Hence
the tunneling ``particles'' are, de facto, embedded in a medium otherwise 
characterized only by fairly simple acoustic-phonon degrees of freedom. This 
embedding, however, does mean that the rest of the material takes its part in the 
making of the tunneling potential for the TS's ``particle'', which itself is not 
moving quantum-mechanically in a true vacuum. Sussmann~\cite{Sus1962}
has shown that this should lead to local trapping potentials that (for the case of
triangular and tetrahedral perfect symmetry, like in our limit no-disorder case) 
must be characterized by a degenerate ground state. This should mean that, as 
a consequence of the TS embedding, our poor man's model, 
Eq. (\ref{3lsmagtunneling}), for the ATSs should be chosen with a strictly 
positive tunneling parameter~\cite{Jug2004}:
\begin{equation}
D_0>0
\label{degeneracy}
\end{equation}
where of course perfect degeneracy gets always removed through weak disorder in 
the asymmetries. Intrinsic near-degeneracy of (\ref{atsdistribution}) of course
implies that the model should be used in its $D/D_0\ll 1$ limit, and that in turn 
reduces the ATSs to effective magnetic-field dependent 2LSs, greatly simplifying 
the mathematical analysis when the limit $\varphi\to 0$, which we always take for
relatively weak magnetic fields, is used. The ETM was first proposed in \cite{Jug2004}, 
and consists basically in a collection of independent, non-interacting 2LSs which are
described by the STM and also 3LS-TWPs, which are described by 
Eqs.~(\ref{3lsmagtunneling}) and (\ref{atsdistribution}) above, in the mentioned 
$D/D_0\ll 1$ and $\varphi\to 0$ limits. The 3LSs are meant to be nested in the 
interstitials between the close-packed RERs while the magnetic-field insensitive 2LSs 
are distributed in the remaining homogeneously-disordered granular matrix of RERs 
and at their touch points or interfaces \cite{Jug2013}.
\begin{figure}[htbp]
  \centering
  \includegraphics[scale=0.25] {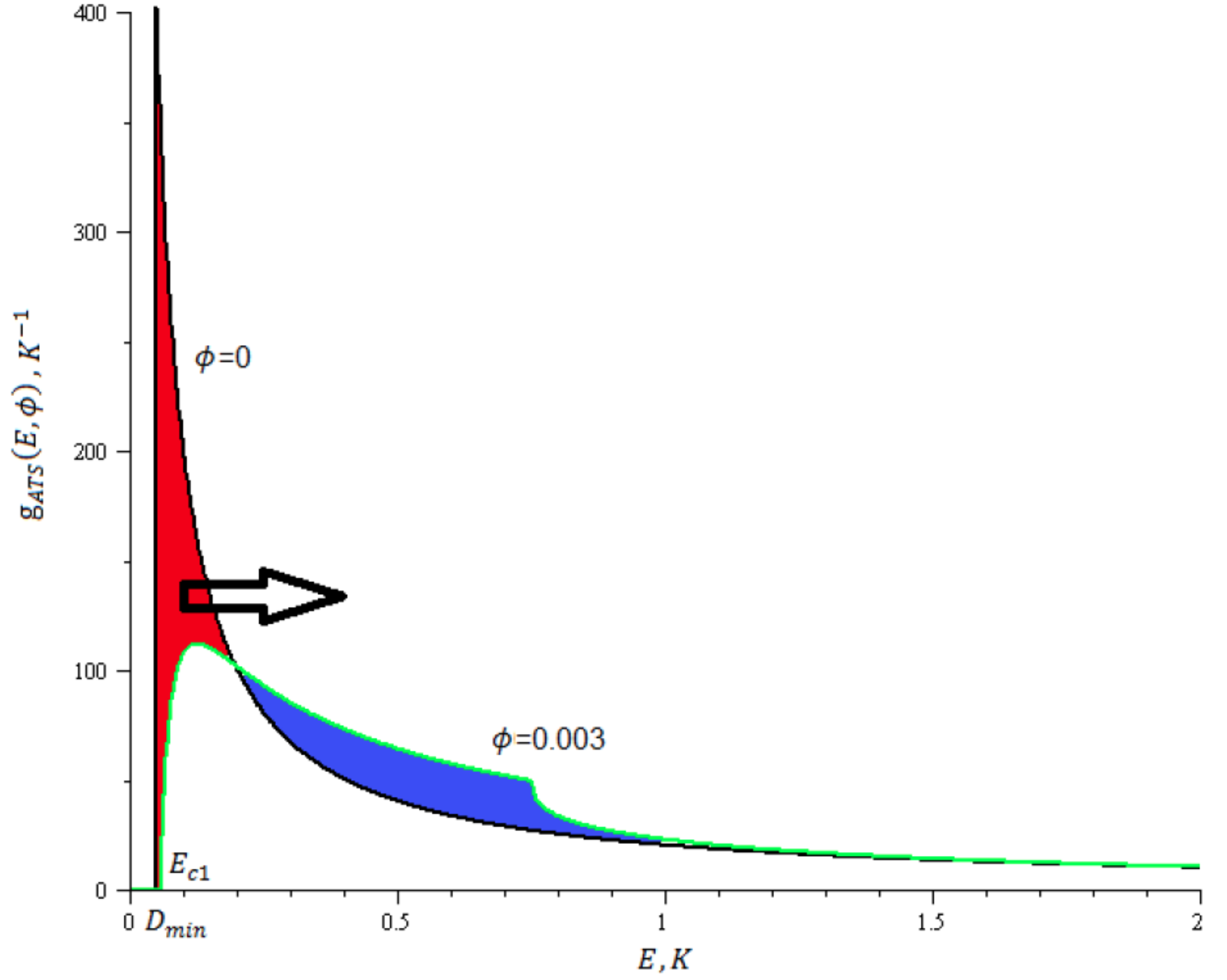}
\caption{ The magnetic-sensitive contribution of the density of states (DOS) as 
the function of the energy gap $E$ and selected Aharon-Bohm (A-B) phases 
$\varphi$ (proportional to the magnetic field $B$) (here $n_{ATS}P^{\ast}$ has 
been set to 1). The rapid transfer of quantum states to higher energy when a 
very weak magnetic field $B$ is switched on is the basic explanation for the origin 
of the magnetic effects. }  
  \label{dosfigphi}
\end{figure}
In Fig.~\ref{dosfigphi} we illustrate the behaviour of the magnetic part of the
density of states (DOS) for this model as a function of the ATS gap energy, $E$, 
for different values of the A-B phase $\varphi$. This figure demonstrates the 
physical origin of the magnetic effects: the number of quantum states being 
conserved, they get to be very rapidly shifted towards high values of the energy 
when a magnetic field, even very weak, is turned on.
\begin{figure}[htbp]
  \centering
  \includegraphics[scale=0.50] {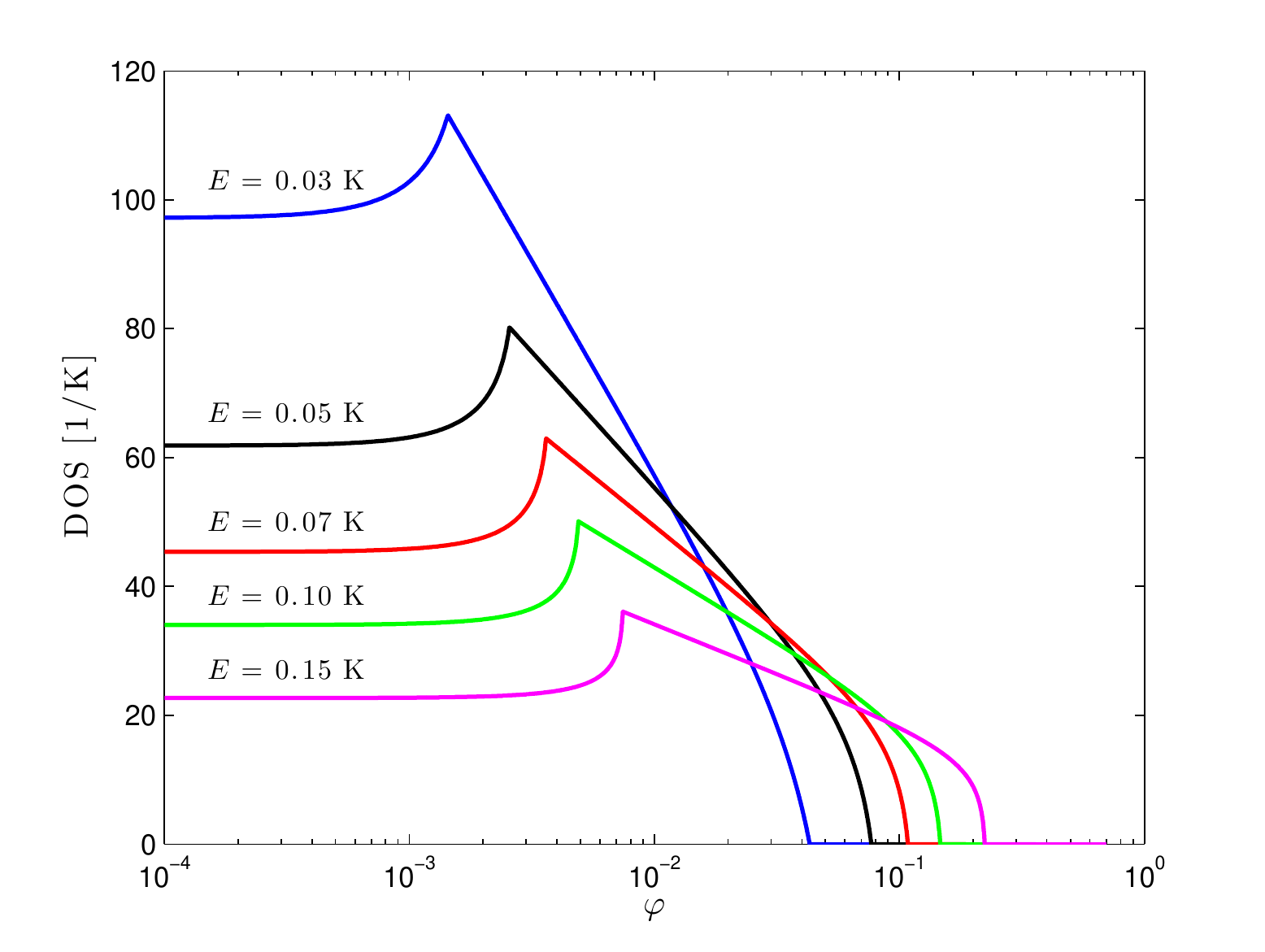}
\caption{ The magnetic-sensitive part of the density of states
(DOS) as a function of the A-B phase $\varphi$ (proportional to the magnetic field 
$B$, upon averaging) and different energies (again $n_{ATS}P^{\ast}$ has been 
set equal to 1). The shape of this DOS contribution (coming from the TWPs with 
a parameter distribution (\ref{atsdistribution}) which is favouring near-degeneracy) is 
the ultimate source of all the magnetic effects (see also Fig. \ref{alltogether}). 
The cusp is an artifact of the effective 2LS approximation~\cite{Jug2004}, albeit
also of the existence of upper and lower bounds for $D_0$ due to the disordered 
nature of the RER glassy atomic structure.} 
  \label{dosfigE}
\end{figure}
The ETM, proposed in this form by the present Author in 2004,  has been able to 
explain so far the magnetic effects in the heat capacity~\cite{Jug2004}, in the 
real~\cite{Jug2009} and imaginary~\cite{Jug2014} parts of the dielectric constant 
and in the polarization echo amplitude~\cite{Jug2014} data reported to date for
a variety of glasses at low temperatures. The ETM has also shed much new light  
into the composition-dependent anomalies~\cite{Jug2010,Jug2013}.
The new physics provided by the magnetic-field dependent TS DOS, as mentioned,
comes from a term due to the near-degenerate TWPs~\cite{Jug2004} and that gets 
to be added up to the constant DOS from the STM 2LSs (having constant density 
$n_{2LS})$:
\begin{eqnarray}
g_{tot}(E,B)&=&n_{2LS}\bar{P}
+n_{ATS}\frac{P^{\ast}}{E}f_{ATS}(E,B)\theta(E-E_{c1}) \nonumber \\ 
&=&g_{2LS}(E)+g_{ATS}(E,B) 
\label{dos}
\end{eqnarray}
where $n_{ATS}$ is the ATSs' concentration, while $f_{ATS}$ is a 
magnetic-field dependent dimensionless function already described in previous 
papers~\cite{Jug2004} and $E_{c1}$ is a material and weakly $B$-dependent 
energy cutoff:
\begin{eqnarray}
g_{ATS}(E,\varphi)=\int \Pi_i dE_i \delta(\Sigma_j E_j) \int d D_0
{\cal P}^*_{3LS}(\{E_k\},D_0) \delta(E-\Delta{\cal E}) \cr
=\begin{cases}
\frac{2\pi P^*}{E}\ln\left( \frac{D_{0max}}{D_{0min}}
\sqrt{ \frac{E^2-D_{0min}^2\varphi^2}{E^2-D_{0max}^2\varphi^2} }
\right) \qquad &{\rm if} \quad E>E_{c2}  \cr
\frac{2\pi P^*}{E}\ln \frac{ \sqrt{ (E^2-D_{0min}^2\varphi^2)
(E^2-D_{min}^2) } }{ D_{0min}D_{min}\varphi }
&{\rm if} \quad E_{c1}\leq E \leq E_{c2}   \cr
0 &{\rm if} \quad E<E_{c1}. \cr
\end{cases}
\label{3dos}
\end{eqnarray}
Moreover, after an irrelevant renormalization of parameters: 
$E_{c1}=\sqrt{D_{min}^2+D_{0min}^2\varphi^2}$, 
$E_{c2}=\sqrt{D_{min}^2+D_{0max}^2\varphi^2}$, $D_{min}$, $D_{0min}$ and
$D_{0max}$ being suitable (material parameters) cutoffs. A more microscopic model
does not exists, at present. The gross $1/E$ dependence of the ATS DOS is one
consequence of the chosen tunneling parameters' distribution, 
Eq.~(\ref{atsdistribution}), and it gives rise to a peak in $g_{tot}$ near $E_{c2}$ 
that is rapidly eroded away as soon as a (weak, depending on material parameters) 
magnetic field is switched on. The form and evolution of the magnetic part 
of the DOS is shown in Fig.~\ref{dosfigE} for some chosen parameters and as a 
function of $\varphi\propto B$ for different values of the ATS energy gap $E$.
This behaviour of the DOS with $B$ is essentially the underlying mechanism
for all of the experimentally observed magnetic-field effects in the studied cold
glasses within this model: the measured physical properties turn out to be 
convolutions of this DOS (with some appropriate $B$-independent functions) and 
in turn they reproduce the shape of this DOS as functions of $B$ (see also 
Fig. \ref{alltogether}). We remark that already at $B=0$ the TS total DOS 
$g_{tot}(E)$ that is here advocated, Eq. (\ref{dos}) above (see also 
Fig. \ref{dosfigphi}, has the correct qualitative form that was solicited by Yu and 
Leggett \cite{CYu1988} in order to achieve a better qualitative explanation for 
many experiments at low temperatures in the real glasses. Namely, a broad constant 
(due to the majority 2LSs), surmounted by a triangular-shaped contribution in a 
range of energy $E$ (this part, coming from the minority ATSs). This shape is now 
being to a large extent justified by our model and indeed explains experimental data  
rather well. 

\section{The Magnetic Field Effetcs (Multi-silicate Glasses only)}
We now come to the main theme for this article, the crucial piece of new 
experimental evidence for the cellular, RER-based scenario for glass structure. 
As mentioned in the Introduction, a large number of magnetic-field dependent
effects have been reported in the late 1990s and early 2000s for several insulating
glasses and that could not be ascribed to the (ubiquitous) trace paramgnetic 
impurities. We refer the reader to a review article of this Author's work so far \cite{JBK2016}, here we just review the main results that have been obtained 
through the proposed cellular approach and ETM low-temperature model in order 
to explain all such magnetic effects.

\subsection{The Magnetic Field Dependent Specific Heat}
Though unrecognized by the experimentalists, the heat capacity measured in 
some multi-component silicate glasses (pure, mono-component silica does not 
display such effect) has shown a strong dependence on the applied magnetic field. 
Such dependence was the first to be explained by the present approach 
\cite{Jug2004,Bon2015}. The total TS heat capacity is easily evaluated for this 
model (ETM) and obtained from the above calculated DOS
\begin{equation}
C_{pTS}(T,B)=\int_0^{\infty}dE~g_{tot}(E,B)C_{p0}(E,T)
\label{convolution}
\end{equation}
where
\begin{equation}
C_{p0}(E,T)=k_B\left( \frac{E}{2k_{B}T} \right)^2\cosh^{-2}\left(
\frac{E}{2k_{B}T} \right)
\end{equation}
is the heat capacity contribution from each single TS having energy gap $E$ and
where $g_{tot}(E,B)$ is given by Eq. (\ref{dos}).
We have made use the resulting expression to the available data~\cite{Sie2001} that 
display a magnetic effect in the heat capacity in the case of two multi-component 
silicate glasses: commercial borosilicate Duran (from Schott GmbH) and 
barium-allumo-silicate (AlBaSiO, or BAS in short) glass (from Heraeus GmbH), in 
order to show that the ATS-ETM model works well for the magnetic-field dependent 
$C_p$. In order to fit the data, the standard phonon ($T^3$) contribution as well as 
the Langevin-paramagnetism contribution from the trace iron impurities [mostly 
Fe$^{2+}$ as it turns out] need to be added to the said tunneling contributions. 

\begin{figure}[htbp]
\centering
  \subfigure[]{\includegraphics[scale=0.50] {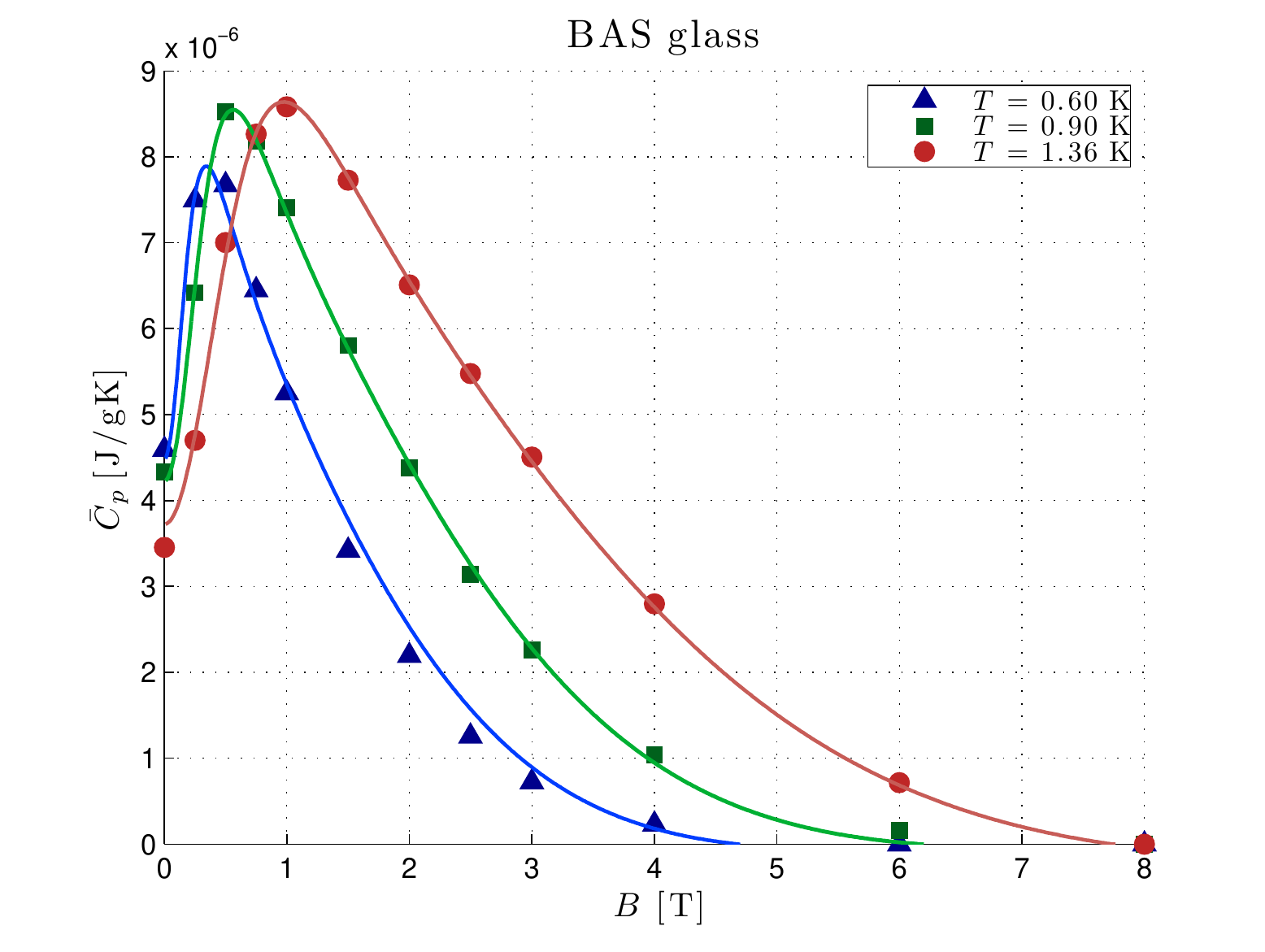} \label{c_tot_albasi}}
  \subfigure[]{\includegraphics[scale=0.50] {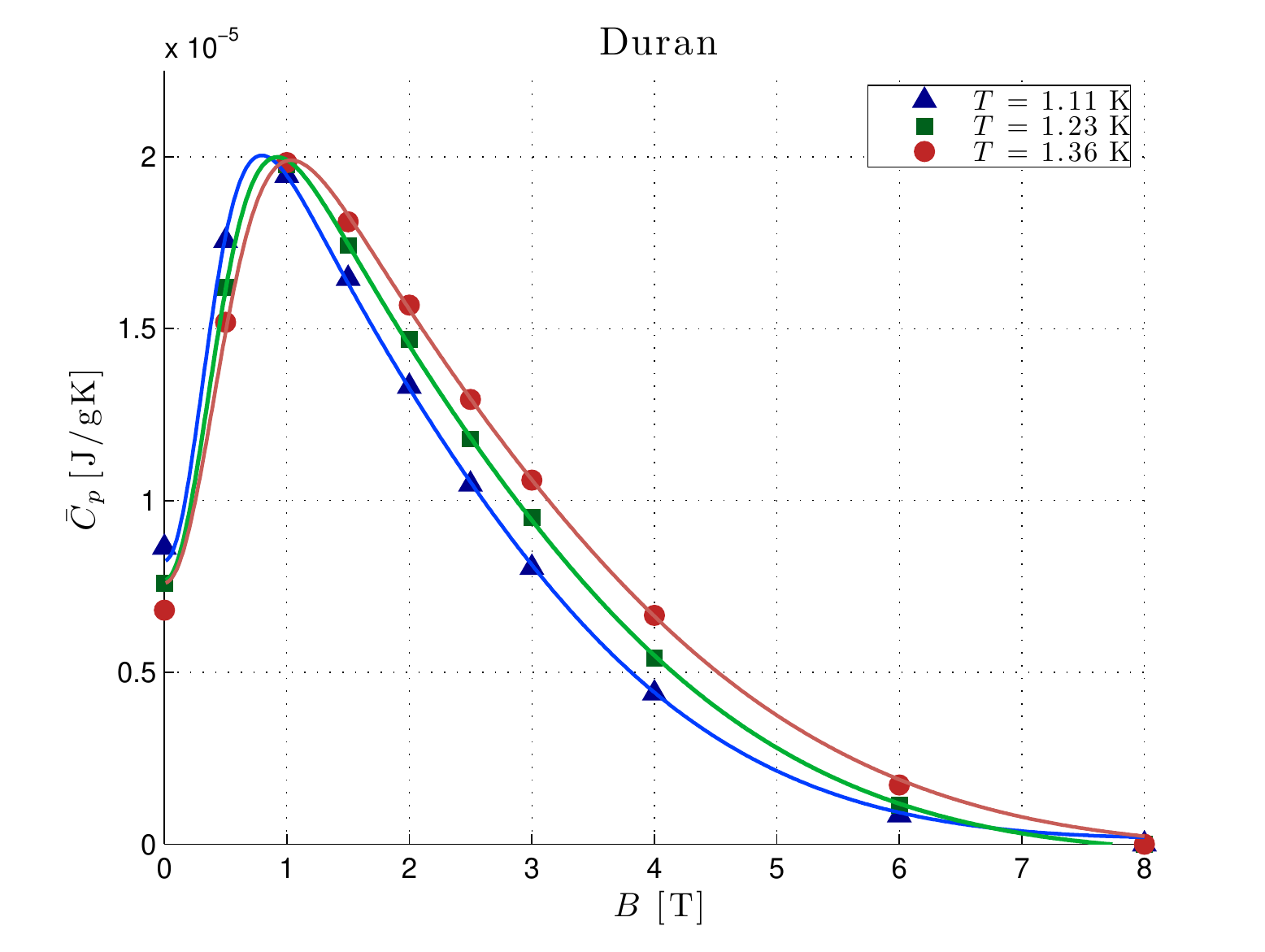} \label{c_tot_duran} }
\caption{ The heat capacity best fit for the a)~BAS (or AlBaSiO)
and b)~Duran glasses. The continuous lines are the predictions from the ETM. Data
from Ref. \cite{Sie2001}.}
\label{heat_capa_fit}
\end{figure}

The best fit for the available data is reported in Fig.~\ref{heat_capa_fit}(a) in the
case of BAS glass.
\begin{table}[htbp]
\begin{center}
\begin{tabular}{|c|c|c|}
\hline
BAS glass & Concentration $\mathrm{[g^{-1}]}$ & Concentration [ppm] \\
\hline
\hline
\textbf{$n_{Fe^{2+}}$} & 1.06$\times10^{17}$ & 14.23 \\
\textbf{$n_{Fe^{3+}}$} & 5.00$\times10^{16}$ &  6.69\\
\textbf{$P^{\ast}n_{ATS}$} & 5.19$\times10^{16}$ & - \\
\hline
\end{tabular}
\caption{Extracted parameters (from the specific heat data \cite{Sie2001}) for the
concentrations of ATSs and Fe-impurities for the BAS glass.}
\label{tab_imp_extr}
\end{center}
\end{table}
\begin{table}[htbp]
\begin{center}
\begin{tabular}{|c|c|c|c|}
\hline
Temperature [K] & $D_{min}$ [K] & $D_{0min}\vert\frac{q}{e}\vert S$ [K$\AA^2$] & $D_{0max}\vert\frac{q}{e}\vert S$ [K$\AA^2$]\\
\hline
\hline
0.60 & 0.49 & 4.77$\times 10^{4}$ & 3.09$\times 10^{5}$ \\
0.90 & 0.53 & 5.07$\times 10^{4}$ & 2.90$\times 10^{5}$ \\
1.36 & 0.55 & 5.95$\times 10^{4}$ & 2.61$\times 10^{5}$ \\
\hline
\end{tabular}
\caption{Extracted tunneling parameters (from the $C_p$ data \cite{Sie2001}) for the 
BAS glass.}
\label{tab_d0_ALBASI_extr}
\end{center}
\end{table}
The concentrations of the ATSs and Fe-impurities extracted from the best fit of
the heat capacity data \cite{Sie2001} as a function of $B$, for Duran, are reported in
Table~\ref{tab_imp_extr_dur}; having fixed the concentrations, it is then possible
to extract the remaining parameters for Duran (Table~\ref{tab_d0_duran_extr}). The
fit of the available data \cite{Sie2001} is reported in Fig.~\ref{heat_capa_fit}(b) for 
Duran.
\begin{table}[htbp]
\begin{center}
\begin{tabular}{|c|c|c|}
\hline
Duran & Concentration $\mathrm{[g^{-1}]}$ & Concentration [ppm] \\
\hline
\hline
\textbf{$n_{Fe^{2+}}$} & 3.21$\times10^{17}$ &  33.01 \\
\textbf{$n_{Fe^{3+}}$} & 2.11$\times10^{17}$ &  21.63 \\
\textbf{$P^{\ast}n_{ATS}$} & 8.88$\times10^{16}$ & - \\
\hline
\end{tabular}
\caption{Extracted parameters (from the heat capacity data \cite{Sie2001}) for the
concentration of ATSs and Fe-impurities for Duran.}
\label{tab_imp_extr_dur}
\end{center}
\end{table}
\begin{table}[htbp]
\begin{center}
\begin{tabular}{|c|c|c|c|}
\hline
Temperature [K] & $D_{min}$ [K] & $D_{0min}\vert\frac{q}{e}\vert S$ [K$\AA^2$] & $D_{0max}\vert\frac{q}{e}\vert S$ [K$\AA^2$]\\
\hline
\hline
1.11 & 0.34 & 4.99$\times 10^{4}$ & 2.68$\times 10^{5}$ \\
1.23 & 0.32 & 5.30$\times 10^{4}$ & 2.50$\times 10^{5}$ \\
1.36 & 0.32 & 5.54$\times 10^{4}$ & 2.46$\times 10^{5}$ \\
\hline
\end{tabular}
\caption{Extracted tunneling parameters (from the $C_p$~data \cite{Sie2001}) 
for Duran.}
\label{tab_d0_duran_extr}
\end{center}
\end{table}
The results of our repeated \cite{Jug2004} $C_p$ analysis definitely indicate that 
the magnetic-field senstive ATSs give a significant contribution to the low 
temperature heat capacity data. The explanation of these data \cite{Jug2004} was the 
first test of the ETM and gave the first indication that the cellular structure of glass has 
a manifestation in experiments outside the traditional  X-ray scattering domain of 
investigation. 

\subsection{The Magnetic Field Dependent Dielectric Constant}
As it turns out, historically the magnetic effect on the dielectric properties of the
multi-silicate glasses was the first to be discovered \cite{Str2000}. Only the present 
approach - at the time of writing - is able to explain the data, qualitatively 
\cite{Jug2009} and then quantitatively \cite{Jug2014}.
We consider the contribution to the dielectric constant $\epsilon(\omega)$ 
from the TWPs (or ATSs) that are sitting in the interstices between the RERs. One 
can then treat each ATS again as an {\it effective} 2LS having first energy gap 
$\Delta{\cal E}={\cal E}_1-{\cal E}_0=E=\sqrt{D^2+D_0^2\varphi^2}$ for 
relatively weak fields. Within this simplified picture, the linear-response, quasi-static 
resonant and relaxational contributions to the polarizability tensor $\alpha_{\mu\nu}$ 
can be extracted according to the general 2LS approach described in various papers 
\cite{Car1994}, as well as in the review in Ref. \cite{JBK2016}, to obtain
\begin{equation}
\alpha^{RES}_{\mu\nu}=\int_0^{\infty}\frac{dE}{2E}
{\cal G}_{\mu\nu}\left ( \left\{ \frac{E_i}{E} \right\};{{\bf p}_i}\right )
\tanh\big( \frac{E}{2k_{B}T} \big)\delta(E-\Delta{\cal E})
\label{respolariz}
\end{equation}
as well as
\begin{equation}
\alpha^{REL}_{\mu\nu}=\frac{1}{4k_{B}T}\int_0^{\infty} dE \left( \sum_{i,j=1}^{3}
\frac{E_iE_j}{E^2}p_{i\mu}p_{j\nu} \right )
\cosh^{-2}\left( \frac{E}{2k_{B}T} \right) \delta(E-\Delta{\cal E})
\label{relpolariz}
\end{equation}
where
\begin{equation}
{\cal G}_{\mu\nu}\left ( \left\{ \frac{E_i}{E} \right\} ;{{\bf p}_i}\right )=
\sum_{i=1}^{3}p_{i\mu}p_{i\nu}-\sum_{i,j}\frac{E_iE_j}{E^2}p_{i\mu}p_{j\nu}
\label{geomcorr}
\end{equation}
is a disorder correlator that makes use of the single-wells' electric dipoles 
${\bf p}_i=q{\bf a}_i$ ($i=1,2,3$). 
This expression assumes vanishing electric fields and no TS-TS interactions yet, an
approximation which indeed does not realistically apply to experiments at the lowest 
temperatures. To keep the theory treatment simple, however, one can still make use 
of Eq. (\ref{respolariz}) and of the analogous one for the relaxational contribution to 
the polarizability.  Interactions will play a role, though from estimates this should 
only happen around the 1 to 10 mK range. Eq. (\ref{respolariz}) must be averaged 
over the random energies' distribution (\ref{atsdistribution})
(denoted by $[\dots]_{av}$, responsible for the high sensitivity to weak fields) and
also over the dipoles' orientations and strengths (denoted by $\overline{(\dots)}$). 
For a collection of ATS with $n_w>2$ wells this averaging, as it turns out,  presents 
serious difficulties and one must resort to the reasonable decoupling:
\begin{equation}
\overline{{\cal G}_{\mu\nu}\delta(E-\Delta{\cal E})}\simeq
\overline{{\cal G}_{\mu\nu}}\cdot\overline{\delta(E-\Delta{\cal E})},
\label{decoupl}
\end{equation}
where now $\overline{[\delta(E-\Delta{\cal E})]_{av}}=g_{ATS}(E,B)$ is the
fully-averaged DOS. To calculate $\overline{{\cal G}_{\mu\nu}}$, one can begin by
envisaging a fully-isotropic distribution of planar $n_w$-polygons to then obtain:
\begin{equation}
\overline{{\cal G}_{\mu\nu}}=\frac{1}{3}\left ( \frac{n_w}{n_w-1} \right )
\overline{p_i^2}\frac{(n_w-2)E^2+D_0^2\varphi^2}{E^2}\delta_{\mu\nu}.
\label{isoglass}
\end{equation}
The second term in the numerator of Eq. (\ref{isoglass}) gives rise, one discovers, 
to a peak in $\delta\epsilon/\epsilon$ as a function of $B$ at very low fields, while 
the first term (which is present only if $n_w>2$) produces a {\it negative} 
contribution to $\delta\epsilon/\epsilon$ at larger $B$ which can dominate over 
the enhancement term for all values of $B$ when $D_{0max}\gg D_{0min}$, 
where $D_{0min}$, $D_{0max}$ correspond to material-dependent cutoffs in 
the distribution of ATS energy barriers ($V_{0max}$, $V{0min}$ respectively).
Observations in Duran and BK7 indeed show a significant depression of
$\epsilon'(B)$ for weak fields \cite{Woh2001}, thus presenting direct evidence for
the existence of ATSs carrying $n_w>2$ in the multi-silicate glasses. Performing
the averaging $[\dots]_{av}$ one then gets some analytical expressions for the
polarizability. The uniform average over orientation angles $\theta$ must be
performed numerically (although it was checked that a very good approximation 
consists in the replacement $\varphi^2 \to \frac{1}{3}\varphi^2$ for averaged 
expressions, which corresponds to the replacement 
$\overline{\cos^2\theta}\to\frac{1}{3}$).

\begin{figure}[htp]
  \centering
 \includegraphics[scale=0.25] {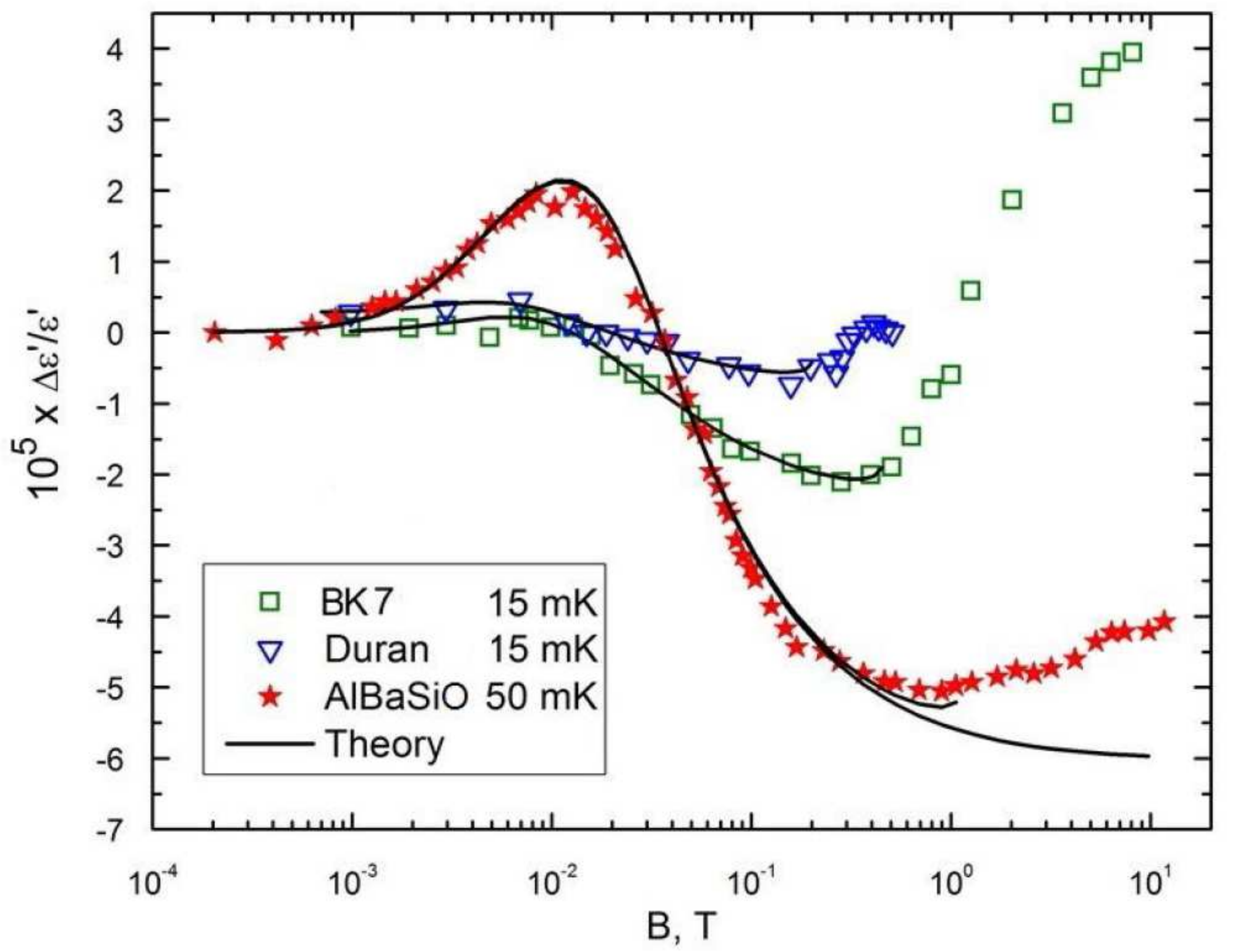}
 \caption[]{ Relative dielectric constant variation versus the magnetic field for AlBaSiO (or BAS) \cite{Woh2001}b, BK7 \cite{Woh2001} and Duran \cite{Woh2001}b glasses. With best-fit parameters given in Table~\ref{tab71}, the continuous curves are the results of 
our theory in the ``weak field'' approximation with (and - for AlBaSiO - also without) 
higher order correction. From \cite{Pal2011}.}
\label{fig71}
\end{figure}

\begin{figure}[htp]
  \centering
 \includegraphics[scale=0.25] {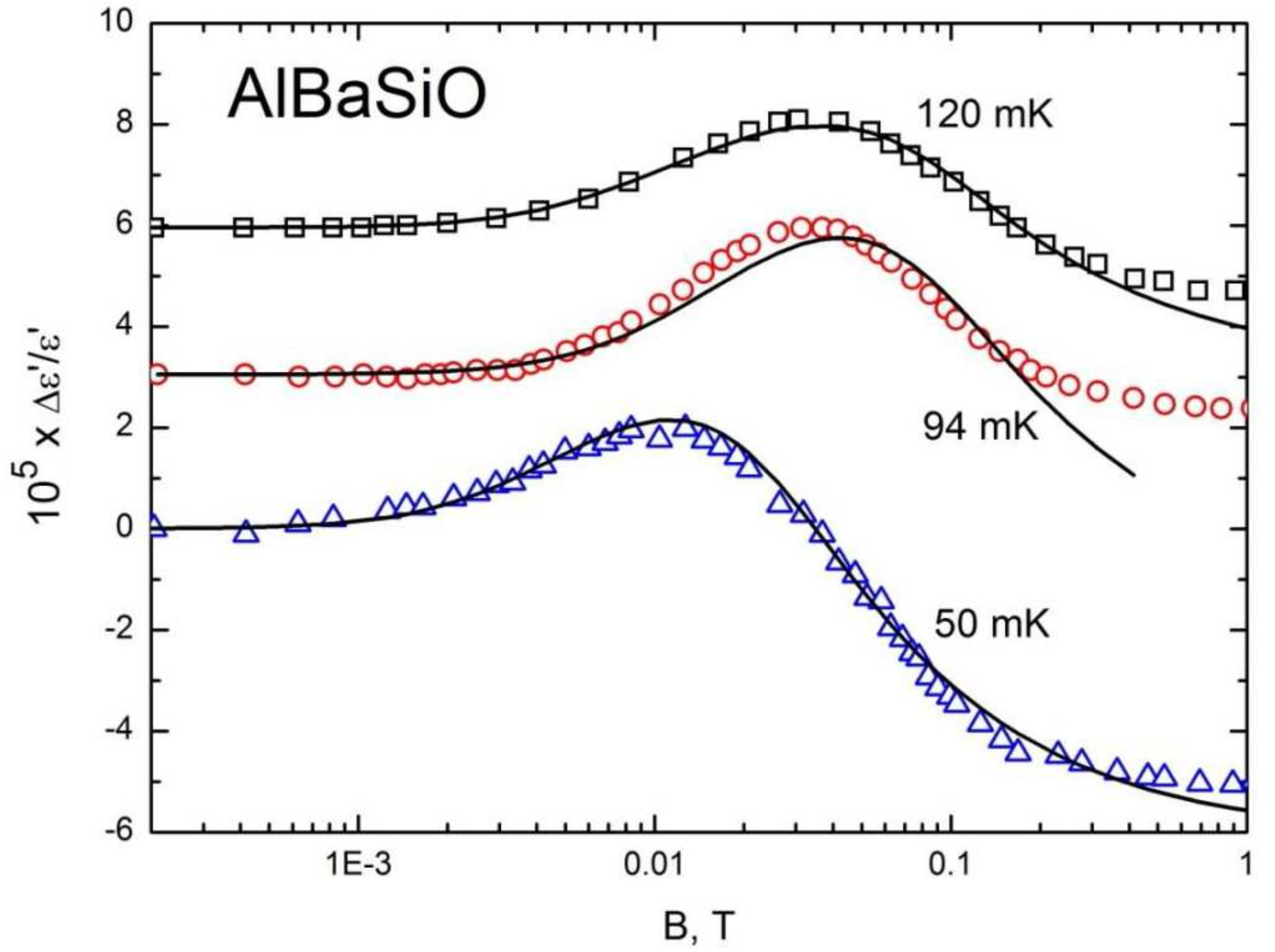}
 \caption[]{ Relative dielectric constant variation versus the magnetic field and temperature for AlBaSiO (or BAS) glass \cite{Woh2001}b. With fitting parameters as in Table~\ref{tab71}, the continuous curves are the result of our theory in the 
``weak field'' approximation. From \cite{Pal2011}.}
\label{fig72}
\end{figure}

\begin{figure}[htp]
  \centering
 \includegraphics[scale=0.25] {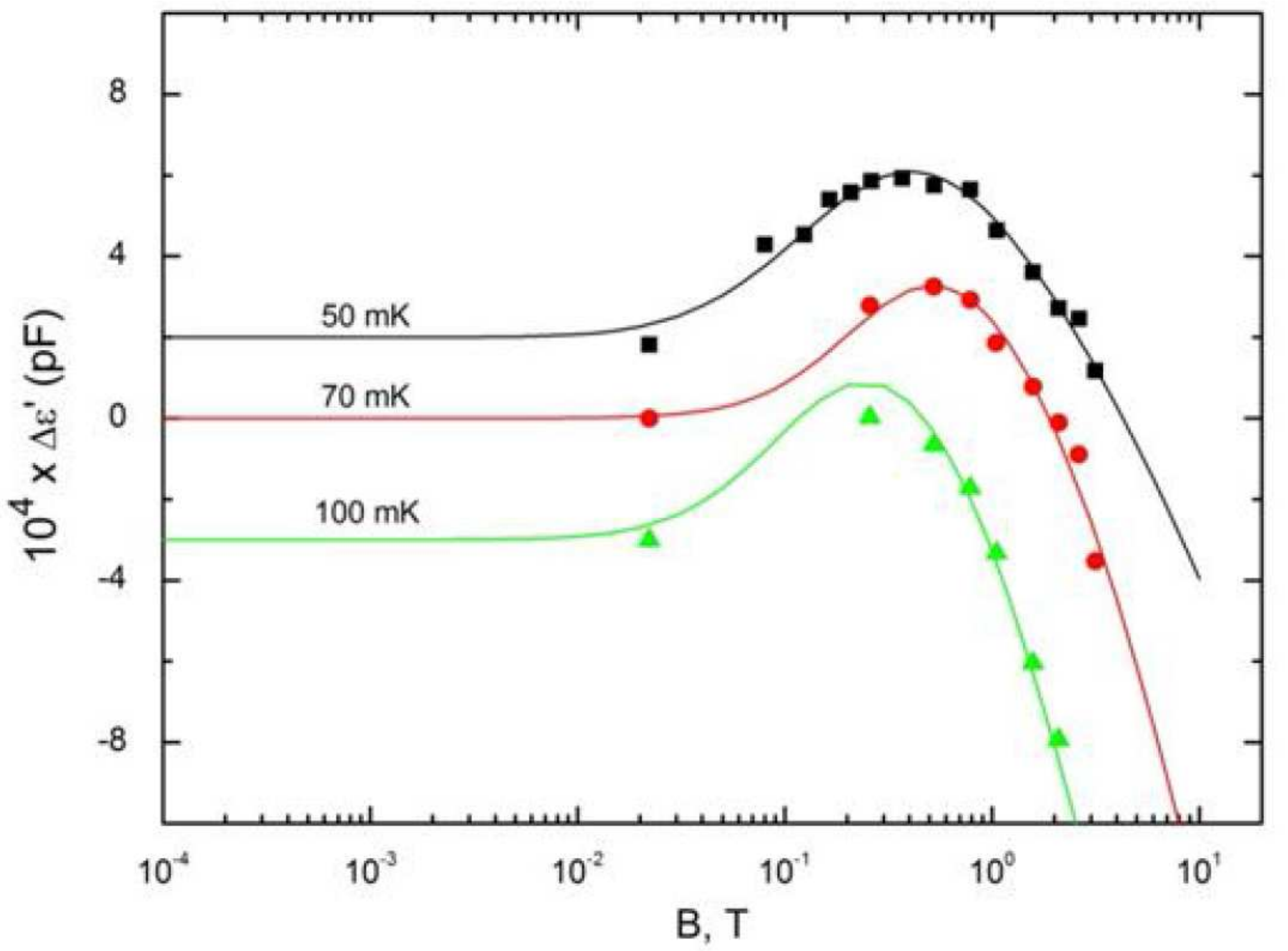}
 \caption[]{ Dielectric constant variation as a function of the magnetic field and temperature for the (non nuclear-quadrupole moments containing) SiO${}_{2+}$\textit{${}_{x}$}C\textit{${}_{y}$}H\textit{${}_{z}$} glass \cite{LeC2002}. Fitting parameters as in Table~\ref{tab72}. 
From \cite{Pal2011}.}
\label{fig73}
\end{figure}

Details of the evaluation of the dielectric constant can be found in \cite{JBK2016} 
and the resulting expressions appear to describe well most experimental data and 
for different glasses, as is shown in Figs.~\ref{fig71}-\ref{fig73} using the fitting 
parameters presented in the Tables~\ref{tab71},~\ref{tab72}.

For the sake of clarity, the data and theoretical curves in Figs. \ref{fig72} and 
\ref{fig73} have been shifted apart vertically. The quantity $x_{ATS}$ now refers to
volume concentrations of ATSs, linked to mass concentrations $n_{ATS}$
through use of the solid's density $\rho$: $x_{ATS}=n_{ATS}\rho$.
\begin{table}[htbp]
\begin{center}
\begin{tabular}{ |c|c|c|c|c|}
\hline
Material and Temperature & ${\pi x_{ATS}P}^*\overline{p^2_1}/\epsilon_r\epsilon_0$ & $D_{min}$,K & $D_{0min}\left|\frac{q}{e}\right|S_{\Delta }$,KÅ$^{2}$ & $D_{0max}\left|\frac{q}{e}\right|S_{\Delta }$,KÅ$^{2}$  \\
\hline
BK7 15 mK & $0.089\cdot {10}^{-5}$ & $0.03$ & $1.668\cdot {10}^5$ & $4.576\cdot {10}^5$  \\
\hline
Duran 15 mK & $0.052\cdot {10}^{-5}$ & 0.021 & $2.457\cdot {10}^5$ & $4.151\cdot {10}^5$ \\
\hline
AlBaSiO 50 mK & $0.89\cdot {10}^{-5}$ & $0.015$ & $2.440\cdot {10}^5$ & $3.080\cdot {10}^5$ \\
\hline
AlBaSiO 94 mK & $3.75\cdot {10}^{-5}$ & 0.025 & $1.225\cdot {10}^5$ & $1.589\cdot {10}^5$ \\
\hline
AlBaSiO 120 mK & $3.09\cdot {10}^{-5}$ & 0.0227 & $1.767\cdot {10}^5$ & $2.248\cdot {10}^5$ \\
 \hline
\end{tabular}
\caption{Fitting parameters extracted for the dielectric constant in a magnetic field and
for three different types of glasses.}
 \label{tab71}
\end{center}
\end{table}

\begin{table}[htbp]
\begin{center}
\begin{tabular}{ |c|c|c|c|c|}
\hline
Temperature & ${\pi x_{ATS}P}^*\overline{p^2_1}/\epsilon_r\epsilon_0$ & $D_{min}$\textit{${}_{,}$, }K & $D_{0min}\left|\frac{q}{e}\right|S_{\Delta }$\textit{, }KÅ${}^{2}$ & $D_{0max}\left|\frac{q}{e}\right|S_{\Delta }$\textit{, }KÅ${}^{2}$ \\
\hline
50 mK & $4.38\cdot {10}^{-5}$ & $0.015$ & $0.076\cdot {10}^3$ & $3.047\cdot {10}^4$ \\
\hline
70 mK & $12.22\cdot {10}^{-5}$ & 0.0486 & $0.600\cdot {10}^3$ & $2.662\cdot {10}^4$ \\
\hline
100 mK & $13.63\cdot {10}^{-5}$ & 0.0486 & $3.035\cdot {10}^3$ & $7.616\cdot {10}^4$ \\
\hline
\end{tabular}
\caption{Fitting parameters for the (non nuclear-quadrupole moments containing) SiO${}_{2+}$\textit{${}_{x}$}C\textit{${}_{y}$}H\textit{${}_{z}$} glass for different temperatures.}
 \label{tab72}
\end{center}
\end{table}


Analogous results have been obtained for the explanation of the dielectric loss 
data in a magnetic field, the theory and fits of the available experimental data can be
found in \cite{JBK2016}. The tunneling parameters extracted from the data fitting 
compare well with those given above, so that in fact a single model with a single
set of material parameters per specimen could be used in all these studies (except
that the fit of all the data sets would not be so good).
Here we only like to remark that for the loss the theory must also describe the 
magnetic-field dependence of a phenomenological relaxation time, a new parameter 
entering in the description of dielectric dissipation.
This relaxation time for ATSs at low temperature and in a magnetic field is now 
found to be given by the following expression (derived, yet no detailed published, by 
the present Author) \cite{JBK2016}:
\begin{equation}
{\tau }^{-1}_{ATS}={\tau }^{-1}\left(E,\varphi\right)=\frac{E^3\left(D^2_0\varphi^2+\frac{5}{6}D^2\right)}{{\Gamma {\rm \ tanh} \left(\frac{E}{2k_{B}T}\right)\ }}=\frac{E^3\left(E^2-\frac{1}{6}D^2\right)}{{\Gamma {\rm \ tanh} \left(\frac{E}{2k_{B}T}\right)\ }}={\tau }^{-1}\left(E,D\right)
\label{eq85}
\end{equation}
where, as usual, the A-B phase $\varphi$ is directly proportional to the magnetic 
field $B$. It appears, therefore, that the total dielectric relaxation time, 
obtained through its inverse:
\begin{equation}
\frac{1}{{\tau }_{tot}}=\frac{1}{{\tau }_{2LS}}+\frac{1}{{\tau }_{ATS}(\varphi)}
\label{eq86}
\end{equation}
must diminish in a non-trivial manner as the magnetic field is switched on. 
This very interesting prediction of the present theory appears to be confirmed 
explicitly, albeit only qualitatively, in laboratory experiments and for some special 
multi-silicate glasses so far only via the work of a Russian group at 
liquid-He temperatures \cite{Smo1979}. A systematic study of the 
magnetic-field dependence of ${\tau }_{tot}$ in the multi-component glasses is 
lamentably still lacking.

\subsection{The Magnetic Field Dependent Polarization Echo Amplitude}
The experimental detection of electric and phonon echoes in glasses is one strongly 
convincing argument for the 2LSs' existence. Echoes in glasses are similar (yet with
important differences) to other echo phenomena such as nuclear spin echo, photon 
echo and so on. But only at very low temperatures the relaxation of the TSs 
becomes so slow that coherent phenomena like polarization echoes become 
observable in the insulating glasses \cite{Phi1987}.

\begin{figure}[h]
\centering
   \subfigure[]{\includegraphics[scale=0.90] {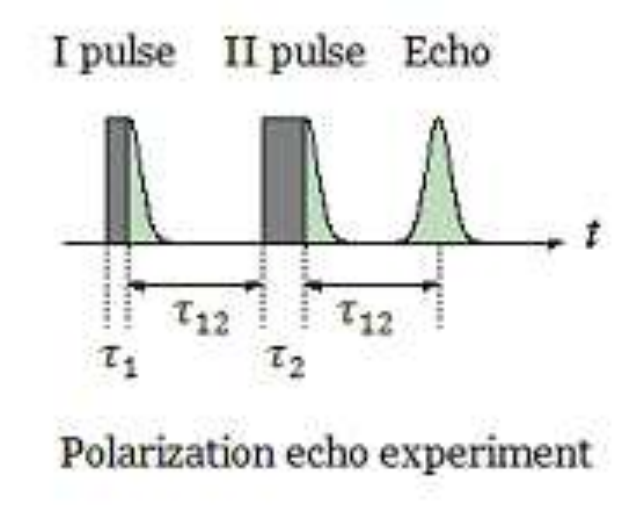}}   
    \subfigure[]{\includegraphics[scale=0.70] {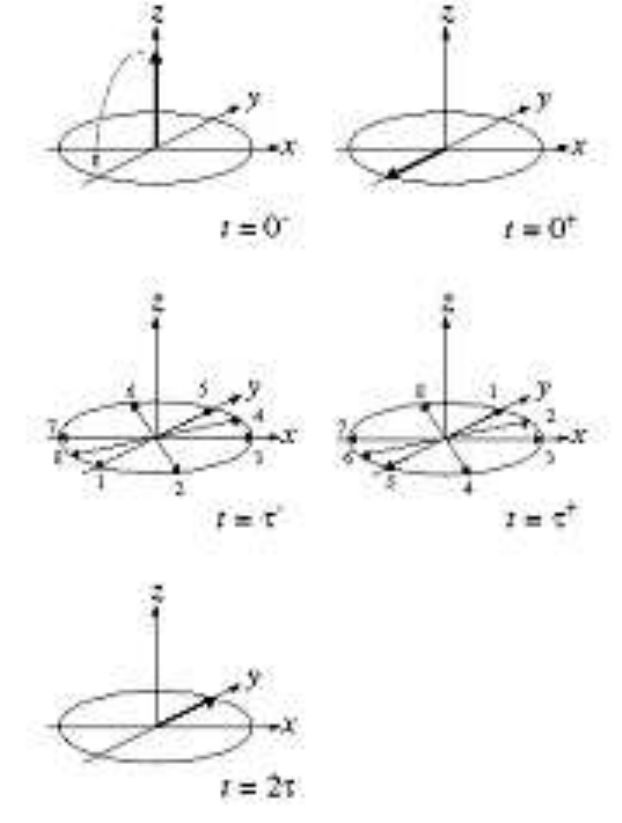}}
\caption{ Schematics of the two-pulse polarization echo experiment. 
Hahn's vectorial interpretation on the right hand side is for NMR's spin-echo 
experiment.}
\label{image1-2e}
\end{figure}

The essence of the effect is as follows (see Fig. \ref{image1-2e}). A glass sample 
placed in a reentrant resonating cavity (``Topfkreisresonator'') is subjected to two 
very short ac electromagnetic pulses at the nominal frequency of about 1 GHz 
and separated by a time interval $\tau_{12}$. The durations $\tau_1$ and 
$\tau_2$ of the pulses must be much shorter that all relaxation process times in 
the observed system. The macroscopic polarization produced by the first pulse 
then vanishes rapidly, due to the broad distribution of parameters of the TSs 
in glasses. This phenomenon is similar to the well-known free-induction decay 
that is observed in nuclear magnetic resonance (NMR) experiments. The ``phase'' 
(energy-level populations) of each 2LS develops freely between the two exciting 
pulses. The second pulse causes an effective ``time reversal'' for the 
development of the phase of the 2LSs. The initial macroscopic polarization of the 
glass is then recovered at a time $\tau_{12}$, roughly, after the second pulse. 
Because the thermal relaxation processes and (see below) spectral diffusion are 
strongly temperature dependent, polarization echoes in glasses can be observed, 
in practice, only at very low temperatures: typically below 100 mK. The echo 
amplitude is clearly proportional to the number of 2LSs that are in or near 
resonance with the exciting microwave pulse and that do not loose their phase 
coherence during the time $2\tau_{\rm 12}$ \cite{Phi1987}.
It should be stressed that, because of the wide distribution for the parameters 
of the two-level systems in glasses, the theoretical description of polarization echoes 
in glasses is considerably more complicated than in the case of nuclear spin systems. 
In analogy to the two-pulse echo in NMR experiments this phenomenon is referred to 
as the spontaneous electric echo.

Polarization echo phenomena can help us understand more about the 
microscopic structure of TSs in general in glasses and give different kinds of 
information. The analysis of these experiments is similar to that for the 
analogous NMR case, except that the TS problem is complicated by three new 
factors. First, the elastic (or electric) dipoles are not aligned with respect to the 
driving field(s) and a calculation of the echo signal involves averaging over 
their orientations. 
Second, for a given pumping frequency $\omega$ there exists a distribution of 
induced moments (elastic or electric) and relaxation times, which must be included 
in the theoretical analysis. Finally, in electric echo experiments the local field as  
seen by the TSs is not equal to the externally applied field, and a local-field 
correction factor must be used when evaluating absolute values of the polarization
\cite{Phi1987}.

In the polarization echo experiments, done typically at radio frequencies and at 
very low temperature (about 10 to 100 mK), it has been established that the 
TSs in glasses couple directly to the magnetic field $B$ \cite{Lud2003,Lud2002}. 
This time, the amplitude of two-pulse echoes in the BAS, Duran and BK7 glasses 
was found to dependent strongly on the applied magnetic field showing a 
non-monotonic (perhaps oscillatory) field variation and for all glass types. Since 
the very beginning \cite{Nag2004,Wur2002}, such behavior was attributed to the 
existence of nuclear electric quadrupole moments (NEQM) carried by some of the 
tunneling species (having nuclear spin $I>\frac{1}{2}$) interacting via their 
nuclear magnetic dipole with the magnetic field and also with gradients of the 
internal microscopic electric field. The NEQM model is based on the consideration 
that the levels of tunneling particles with non-zero nuclear quadrupole moment 
experience a nuclear quadrupole splitting, which is different in the ground state 
and in the excited state of a tunneling 2LS. The magnetic field then causes an 
additional Zeeman splitting of these levels, giving rise to interference effects. 
In turn, these two different nuclear intractions, though very weak, are thought to
be causing the non-monotonic magnetic field variation of the echo amplitude.

\begin{figure}[h!]
\centering
   \subfigure[]{\includegraphics[scale=0.25] {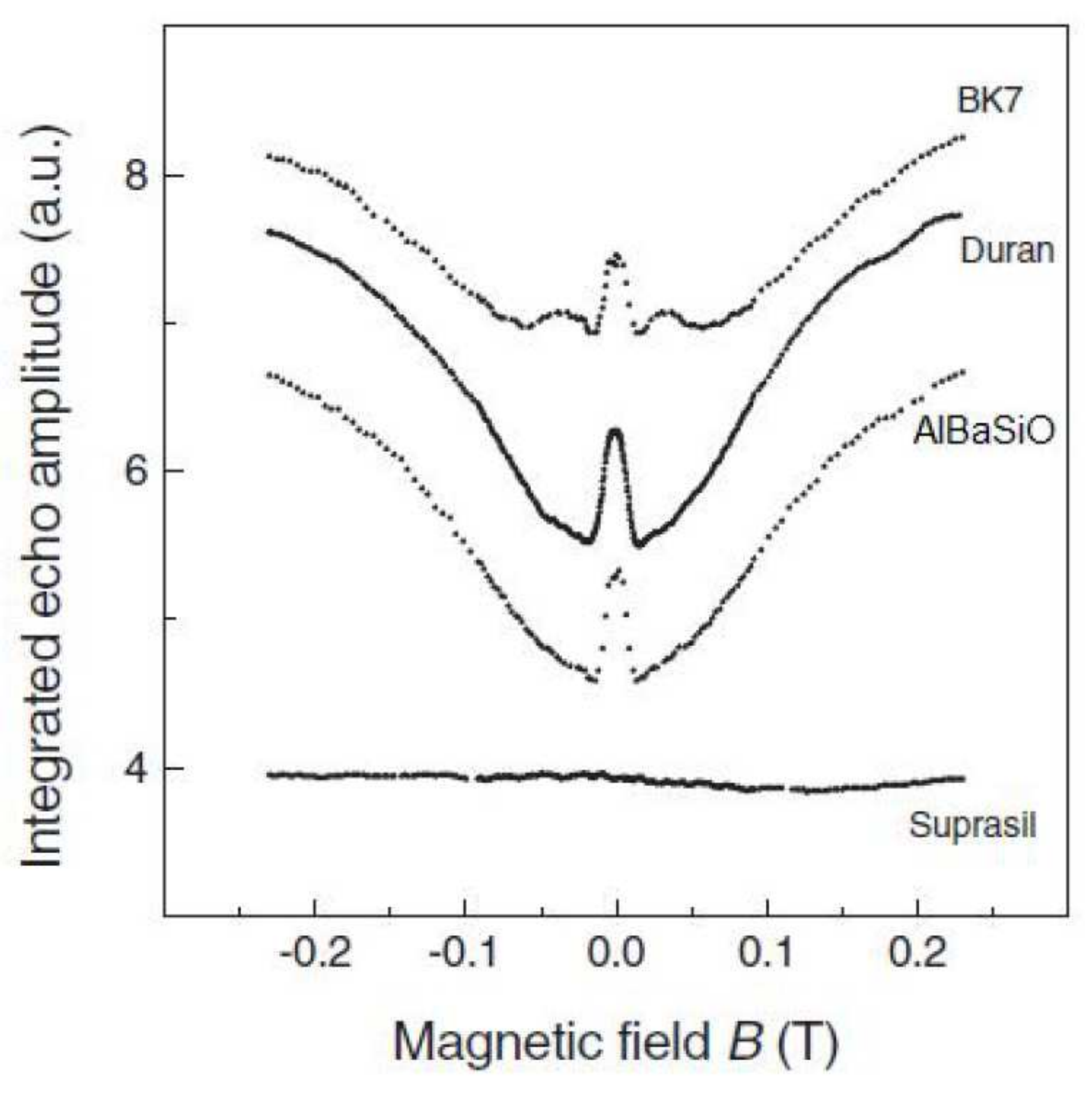}}
   \subfigure[]{\includegraphics[scale=0.255] {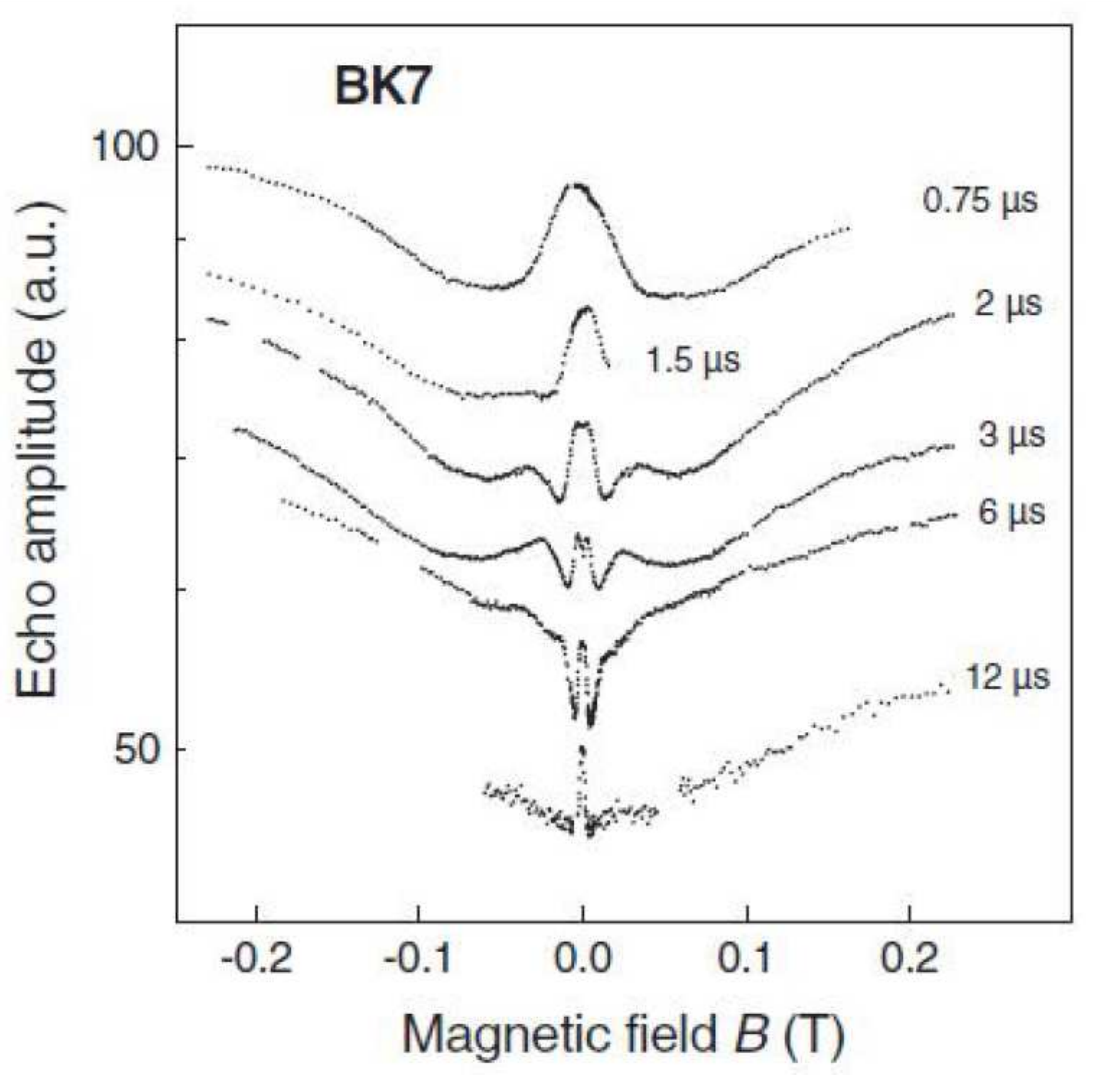}}
\caption{Data from \cite{Lud2003} a) The integrated echo amplitude plotted as a 
function of the magnetic field for different silicate glasses: BK7, Duran, AlBaSiO 
(BAS) and Suprasil I. All data were taken at $T$=12 mK, $\tau_{12}$=2 $\mu$s, 
and nominally 1 GHz, except for Duran, where the delay time was $\tau_{12}$=1.7 
$\mu$s. b) The amplitude of two-pulse echoes in BK7 glass vs. the magnetic field 
for different values of the waiting time $\tau_{12}$ between pulses. All data 
sets were taken at 4.6 GHz and 12 mK except that for $\tau_{12}$=2$\mu$s 
which was taken at 0.9 GHz (this behaviour, right panel, remains unexplained by 
all existing theories and a real challenge). }
\label{image3-4e}
\end{figure}

The amplitude (in fact, integrated amplitude) of two-pulse polarization echoes of 
four types of silicate glasses is now shown in Fig.~\ref{image3-4e}(a) and as a 
function of the magnetic field \cite{Lud2002}. In contrast to various other 
low-temperature properties of glasses, the influence of the magnetic field on 
the amplitude of spontaneous echoes is manifestly not universal on the qualitative
level already. BK7 and Duran show similar effects, and yet the concentration of 
magnetic impurities differs by at least a factor of 20 (a clear indication, that such
impurities are irrelevant for such effects). The most remarkable result of these 
measurements is the fact that Suprasil I (very pure $a$-SiO${}_2$) shows no 
detectable magnetic-field effect. While Duran, BAS and BK7 contain nuclei 
with non-zero nuclear quadrupole moment, Suprasi I is virtually free of such 
nuclei, except for a tiny \% of $^{17}$O. This fact has been used to provide the 
justification for the nuclear quadrupole theory. The variation of the echo amplitude 
with the applied magnetic field is similar for Duran, BK7 and BAS, but not at all 
qualitatively identical. All three samples exhibit a marked principal maximum at 
very weak fields, $B\sim$ 10 mT, but only BK7 has a relevant second maximum 
and some hint to an oscillation as a function of $B$. Unexplained by the NEQM 
theory, the high fields the amplitude of the echo rises well above its value at zero 
magnetic field and seemingly saturates (yet, see Figs.~\ref{heat_capa_fit} and
\ref{alltogether} below, this is very similar to what happens to the inverted heat 
capacity, $-C_p$, as a function of $B$). Again unexplained by NEQM theory, there
is a piece of linear dependence on $B$ at intermediate fields. 

In Figure~\ref{image3-4e}(b) the amplitude of spontaneous echoes for the BK7 
glass is reproduced as a function of the applied magnetic field and for different 
delay times $\tau_{12}$ between the exciting pulses. We can see some obvious 
qualitative differences for different values of $\tau_{\rm 12}$ and that a second 
maximum (the ``oscillation'') is not always present. These findings necessitate a 
good theory for spectral diffusion \cite{Bla1977} in real glasses (containing both 
2LSs and ATSs)  and this theory remains to date unaccomplished. 

One of the most remarkable facts about the experiments on echoes from glasses 
in a weak magnetic field is that strong magnetic response is not confined to the 
inorganic, multi-silicate glasses. The amplitude of spontaneous echoes in 
part-deuterated and in undeuterated amorphous glycerol as a function of the weak 
magnetic field $B$ has been demontrated to show completely different behaviour 
\cite{Nag2004}. In the case of ordinary glycerol there is some, very small change 
of the echo amplitude with $B$. However, for partially deuterated glycerol the 
change is considerably more noticeable, of different shapes and duration. These 
experiments {\it seemingly} provide proof that the magnetic effects are of nuclear 
origin, since the two amorphous glycerol samples differ only in the content of 
nuclei carrying a NEQM. As it turns out, a comparative analysis of experiments in 
different isotope-concentration samples hints to the fact that the effect does not 
scale at all with NEQM concentration. The explanation of the magnetic effects in 
amorphous glycerol echo experiments goes beyond the aims of this review and 
an explanation based on the ETM can be found in the review \cite{JBK2016} and
first published in \cite{Jug2014}. The vitreous-glycerol experiments show, however,
that the tunneling species are H$^+$ and/or D$^+$ in these samples: correctly, 
the lighter atomic species. Then, in the multi-silicates the tunneling particles ought 
to be overwhelmingly O$^{2-}$, hardly the NEQM-carrying B$^{3+}$, Na$^+$, 
K$^+$, or Al$^{3+}$ as claimed by NEQM-theory supporters. The status of the
NEQM-theory remains unclear.

We will discuss in this essay principally the echo experiments on the silicates. A 
theory for the echo signal from a collection of 2LSs can be obtained - and from first 
principles - from a lengthy but straightforward Schr\"odinger equation (or density
matrix) treatment in which high-frequency modes are neglected and the
phonon-damping is treated in a phenomenological way \cite{Pal2011}. In the more
rigorous way, we have obtained for the echo signal an expression which confirms 
and in fact improves on the theory of 2LSs' electric echos by \cite{Gur1990} [some 
details can also be found in \cite{JBK2016}]. It is then possible to extend this 
polarization echo's calculation to the case of the ATS-ETM describing glasses in a 
magnetic field \cite{Pal2011,Jug2014}. The point of view is taken that a 
background of ordinary 2LSs - insensitive to the magnetic field - also exists in the 
glass, but is not needed in order to explain the data as a function of the magnetic 
field.

One begins with a collection of 3LSs (now $n_w=3$ is not just computationally 
convenient, but physically correct as explained in Section 2), but with each single 
ATS Hamiltonian written in the (diagonal) energy representation:
\begin{equation}H'=SHS^{-1}=\left( \begin{array}{ccc}
{\cal E }_0 & 0 & 0 \\
0 & {\cal E }_1 & 0 \\
0 & 0 & {\cal E }_2 \end{array}
\right)+S\left( \begin{array}{ccc}
-{{\mathbf p}}_1\cdot {\mathbf F} & 0 & 0 \\
0 & -{{\mathbf p}}_2\cdot {\mathbf F} & 0 \\
0 & 0 & -{{\mathbf p}}_3\cdot {\mathbf F} \end{array}
\right)S^{-1}
\label{3lselectric}
\end{equation}
Here the transformation matrix $S=S(\varphi)$ is magnetic-field dependent, the 
${\cal E }_i$\textit{ }are the (\textit{B}-dependent) ATS energy levels and the 
${{\mathbf p}}_i$ are the single wells' electric dipoles. As in the treatment of 
Gurevich \textit{et al.} \cite{Gur1990} there is also a phonon bath and this will be 
treated - as usual - phenomenologically and resulting in a standard phonon-damping 
exponential. The second term in Eq.~(\ref{3lselectric}) causes irrelevant 
energy-level shifts and also produces an extra matrix term $\Delta H'(t)=(A_{ij})$ 
of which the only relevant element is of the form
\begin{equation}
A_{01}=A^*_{10}=\sum^3_{k=1}{-{{\mathbf p}}_k\cdot {{\mathbf F}}_0{\rm \ }{{\rm S}}_{{\rm 0k}}{\rm (}\varphi{\rm )}{{\rm S}}^{{\rm *}}_{{\rm 1k}}{\rm (}\varphi{\rm )}}\ {\cos  \omega t\ }
\end{equation}
These $A_{ij}$ cause quantum transitions between the ATS levels $(|0>, |1>, |2>)$ 
when the electric-field pulses are applied. In the weak magnetic field limit (now 
most appropriate for these experiments) and in the usual approximation $D\ll D_0$  
(which is always consistent with our best fits to the data), one discovers that the 
second excited level $|2>$ remains unperturbed and one can make use of the 
effective 2LS approximation (where, however, the ground-state single-well 
wavefunctions mix). One can then repeat the Schr\"odinger equation (or 
density-matrix) calculation carried out for the 2LS case, introducing though the 
complex Rabi frequency:
\begin{equation}
{\Omega }_0= \frac{A_{01}}{\hbar}
\end{equation}
The evolution of the generic ATS during, or in the absence, of pulses then gets to 
be followed exactly in much the same way as before, except that in order to 
simplify the formalism it is convenient to introduce right from the beginning the 
orientationally-averaged Rabi frequency (which is now a real quantity):
\begin{equation}
{\Omega }_R=\sqrt{\overline{{|{\Omega }_0|}^2}}
\end{equation}
the bar denoting averaging over the 3LS base-triangle's orientation. Replacing 
${\Omega }_0$ with ${\Omega }_R$ before carrying out this averaging of the 
sample's polarization is here our main approximation, allowing for a considerably 
simplified treatment and leading to the following magnetic-field dependent Rabi
frequency:
\begin{equation}
{\Omega }_R=\frac{{{\rm p}}_{{\rm 1}}{{\rm F}}_0}{\hbar }\sqrt{\frac{D^2_0\varphi^2+\frac{5}{6}D^2}{{6E}^2}}
\end{equation}
Here, ${\bf p}_1$ is the single-well (orientation-averaged) electric dipole and 
$E=\hbar\omega_0=\sqrt{D^2+D^2_0\varphi^2}$ is the usual magnetic-field 
dependent lower energy gap, in the weak field and near-degenerate approximations. 
The above form for $\Omega_0$ of course treats incorrectly the ATSs that have 
${\bf F}_0$ roughly orthogonal to the ATS base triangle; fortunately these have 
$\Omega_0\approx 0$ and in fact do not contribute to the echo signal.

Proceeding as in our own derivation of the ordinary 2LSs echo \cite{Pal2011},  one 
finds that there is indeed a magnetic contribution to the polarization of the sample 
from the generic ATS and (partly averaged) given by:
\begin{equation}
\begin{split}
\Delta {\wp }_\parallel(t)\cong& -\frac{\hbar }{{{\rm F}}_0}\tanh  \bigg( \frac{E}{2k_B T} \bigg) e^{-\frac{\overline{\gamma }}{2}t}\frac{\Omega^4_R}{\Omega^3_G} \\
&{\rm Im}~\left\{ {\sin}^2 (\frac{{\Omega }_G{\tau }_2}{2})\bigg[{\sin ({\Omega }_G{\tau }_1) }-2i\frac{{\omega }_0-\omega }{{\Omega }_R}{{\rm sin}}^2(\frac{{\Omega }_G{\tau }_1}{2})\bigg] \right\} e^{i\Phi(t)-i\int^t_0{\Delta \omega (t')s(t')dt'}} \\
\end{split}
\label{eq930}
\end{equation}
Here, ${\frac{\overline{\gamma }}{2}=\tau }^{-1}$ is the magnetic ATS relaxation 
rate due to phonons and given by Eq. (\ref{eq85}), so the generalized Rabi frequency 
is still given by 
${\Omega }_G{\rm =}\sqrt{{\Omega }^{{\rm 2}}_R{\rm +}{{\rm (}{\omega }_0{\rm -}\omega {\rm )}}^{{\rm 2}}}$ 
and, moreover:
\begin{equation}
\Phi\left(t\right)={\omega }_0\left(t-2{\overline{\tau }}_{12}\right)+\omega \Delta \tau 
\end{equation}
is what we find to be the appropriate time argument. From this, it is obvious that the 
time at which all the ATSs (regardless of energy gap $E=\hbar {\omega }_0$ value) 
get to be refocused is $t=2{\overline{\tau }}_{12}$ and this determines the correct
echo's peak position (when the echo signal has a reasonable shape, this not being 
always the case \cite{Lud2003}). 
The measured echo amplitude's contribution from the magnetic ATS is then
(also allowing for an arbitrary amplification factor $A_0$):
\begin{equation}
\begin{split}
\Delta A\left(\varphi\right)=&A_0\frac{d}{{\varepsilon }_0{\varepsilon }_r}x_{ATS}2\pi P^*\int^{\infty }_0{dE}\int{\frac{dD}{D}}\int{\frac{dD_0}{D_0}\Theta (}D,D_0)\\
&\times \delta \left(E-\sqrt{D^2+D^2_0\varphi^2}\right)\Delta {\wp }_\parallel\left(2{\overline{\tau }}_{12}\right)
\end{split}
\label{eq932}
\end{equation}
where now $d$ is the sample's thickness, $\Theta (D,D_0)$ is a Dirichlet's
theta-function restriction for the integration domain (see \cite{Pal2011}) and 
where a final orientational-averaging with respect to the angle (defining the A-B 
phase $\varphi$, see Eq. (\ref{ABphase}))
$\beta =\widehat{{\mathbf B}{{\mathbf S}}_{\triangle }}$ is in order. 
One deals with the delta-function's constraint and the energy parameters' integrations 
in the usual way to arrive at, after a lengthy calculation:
\begin{equation}
\begin{split}
\Delta A(\varphi)\cong & {-A}_0\frac{d}{{\varepsilon }_0{\varepsilon }_r}x_{ATS}\frac{4\pi {\hbar }^2P^*}{{{\rm F}}_0}{\cos (\omega \Delta \tau) }\\
& \times \int^{E_{c_2}}_{E_{c_1}}{\frac{dE}{E}}\int^{D_2(\varphi)}_{D_{min}}{\frac{dD}{D}{\tanh (\frac{E}{2k_{B}T})}\frac{E^2}{E^2-D^2}}\\
& \cdot e^{-w2{\overline{\tau }}_{12}}{\Omega }^{2}_R\sigma (E)\left[{\rm S}({\theta }_{{1}},{\theta }_{{2}}){\tan (\omega \Delta \tau)}{\rm +C}({\theta }_{{1}},{\theta }_{{2}})\right] \\  
&+\int^{\infty }_{E_{c_2}}{\frac{dE}{E}}\int^{D_2(\varphi)}_{D_1(\varphi)}{\frac{dD}{D}}({\rm same~integrand~as~above}\dots ) \\
\end{split}
\label{eq933}
\end{equation}
where we have defined the special functions:
\begin{eqnarray}
&&\sigma \left({\rm E}\right)=\frac{{\Omega }^{{\rm 2}}_{{\rm R}}}{{{\rm 2}\hbar\Omega }^{{\rm 3}}_{{\rm G}}}{\rm =}\frac{{\Omega }^{{\rm 2}}_R}{{\rm 2}\hbar {\left({\Omega }^{{\rm 2}}_R{\rm +}{{\rm (}{\omega }_0-\omega {\rm )}}^{{\rm 2}}\right)}^{{\rm 3/2}}}  \nonumber \\
&&{\rm S}\left({\theta }_{{\rm 1}},{\theta }_{{\rm 2}}\right)=\sin \left(\Omega_G\tau_1\right) \sin^2\left(\Omega_G\tau_2/2\right) \\
&&{\rm C}\left(\theta_1,\theta_2\right)=-2\frac{\omega_0-\omega}{\Omega_R}\sin ^2\left(\Omega_G \tau_1/2\right)\sin ^2\left(\Omega_G \tau_2/2\right) \nonumber 
\label{eq934}
\end{eqnarray}
and where $\theta_{1,2}=\Omega_G\tau_{1,2}$ are the so-called pulse areas. 
$E_{c_{1,2}}$ are as in the previous Sections whilst now: 
$D_{1,2}\left(\varphi\right)=\sqrt{E^2-D^2_{0max,min}\varphi^2}$ 
and $E={\hbar \omega }_0$.

In going from Eq.~(\ref{eq930}) to Eq.~(\ref{eq933}) we have tacitly made some 
assumptions on the (to be fully averaged) spectral diffusion term,  
$e^{-i\int^{2{\overline{\tau }}_{12}}_0{\Delta \omega \left(t'\right)s\left(t'\right)dt'}}$
(where $\hbar\Delta \omega(t)=E(t)-E$ represents the time fluctuation of the ATS's 
energy gap which is due to local strain and/or electric field fluctuations \cite{Bla1977}). 
The theory of spectral diffusion (SD) \cite{Bla1977} for the ATSs is still to be 
accomplished, but we can safely assume that what was found by many Authors for 
NMR's spin-echoes as well as for the 2LS polarization echoes in glasses holds good 
for the ATSs too. Namely: there is a wide range of waiting times ${\tau }_{12}$ 
values where the decay of the echo amplitude well approximates a simple exponential 
form, so that one can replace the SD term with 
$e^{-{2{\overline{\tau }}_{12}}/{{\tau }_\phi}}$ , ${\tau }_\phi(T)$ 
being  a SD characteristic time depending only on temperature. There should be a 
SD time ${\tau }_{\phi(3)}$ for the ATSs just like there is a SD time 
${\tau }_{\phi(2)}$ for the standard 2LSs' ensemble. For the latter, STM theory 
has shown \cite{Gal1988,Bla1977} that the latter parameter is independent of $E$ 
and thus for the ATS we shall assume the very same and, moreover, that (just as for 
the phonon damping rate and for Rabi frequency) its own dependence on $B$ is 
weak or absent. This allows us to lump the SD problem together with phonon 
damping, yielding - in essence - an overall exponential relaxation rate:
\begin{equation}
w(E,D)={{\tau }_\phi}^{-1}+{\tau }^{-1}(E,D)
\label{eq935}
\end{equation}
where the SD-time ${\tau }_{\phi(3)}$ is typically much shorter than the 
phonon-damping time $\tau $ and depends on temperature only, through:
\begin{equation}
{{\tau }_{\phi(3)}}^{-1}=c_{ATS}T
\end{equation}
where $c_{ATS}$ is an appropriate constant. The assumption of an overall 
simple-exponential decay of the echo amplitude with waiting time 
${\tau }_{12}$ and characteristic time ${\tau}_{\phi(2)}=1/c_{2LS}T$ seems 
to be well verified experimentally \cite{Ens1996} for single-component glasses 
(uncontaminated a-B$_2$O$_3$, a-SiO$_2$ etc.). Clearly, a better theory for SD
in multi-component glasses is however in order.

\begin{figure}[htbp]
  \centering
  \includegraphics[scale=0.30] {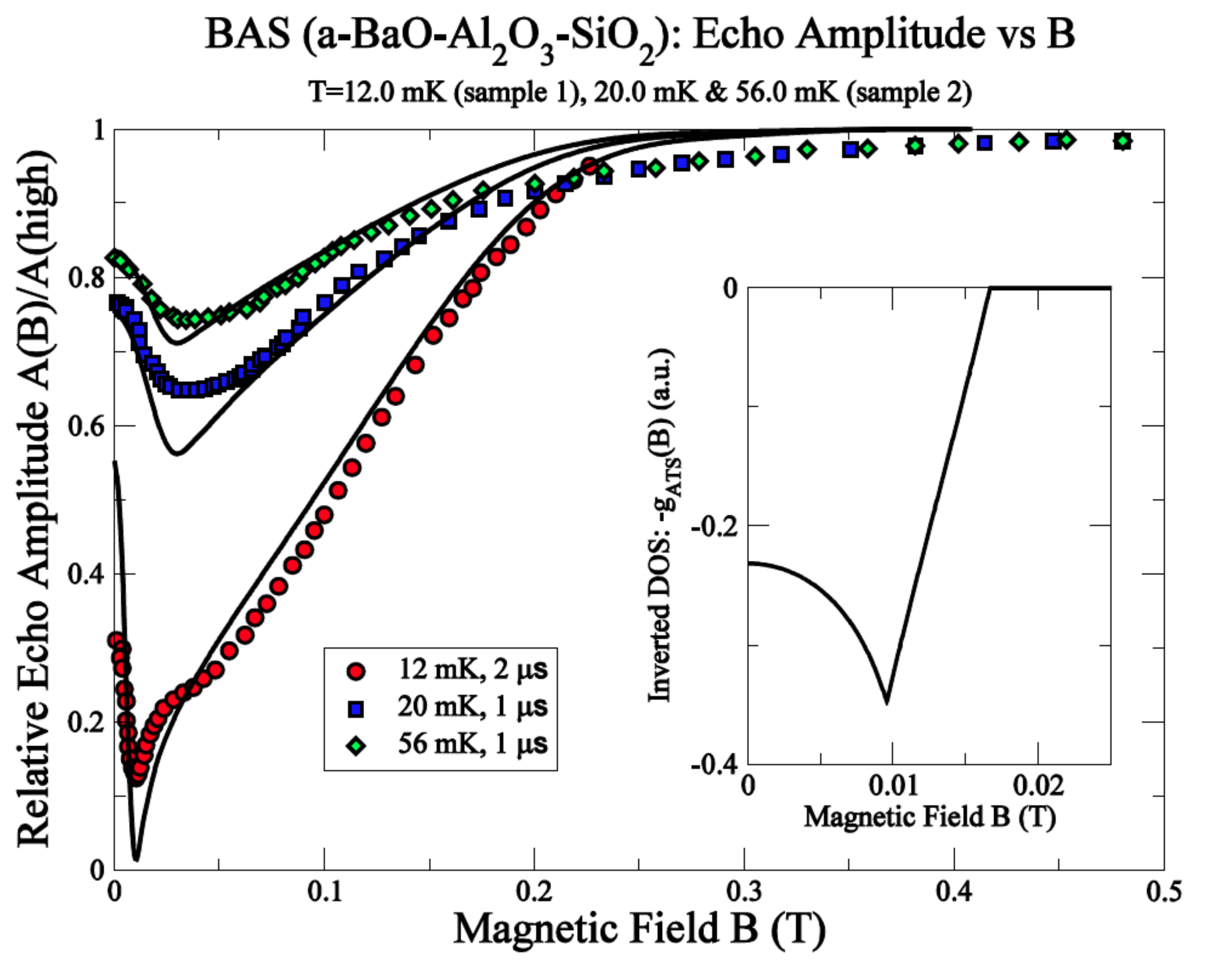}
\caption{ Magnetic field dependence of the polarization echo {\it relative}
amplitude (relative to its value at ``high'' fields, where saturation occurs) 
for the AlBaSiO glass \cite{Lud2002} (also referred to as BAS glass) at given 
experimental conditions. We believe that two separate samples have been used. 
The continuous curves are from our theory. Nominal frequency 1 GHz, 
$\tau_2=2\tau_1$=0.2 $\mu$s. Inset: behaviour of the ATS DOS for the same 
parameters (this is the physical origin of the effect).}
\label{image7e}
\end{figure}

\begin{figure}[htbp]
  \centering
  \includegraphics[scale=0.70] {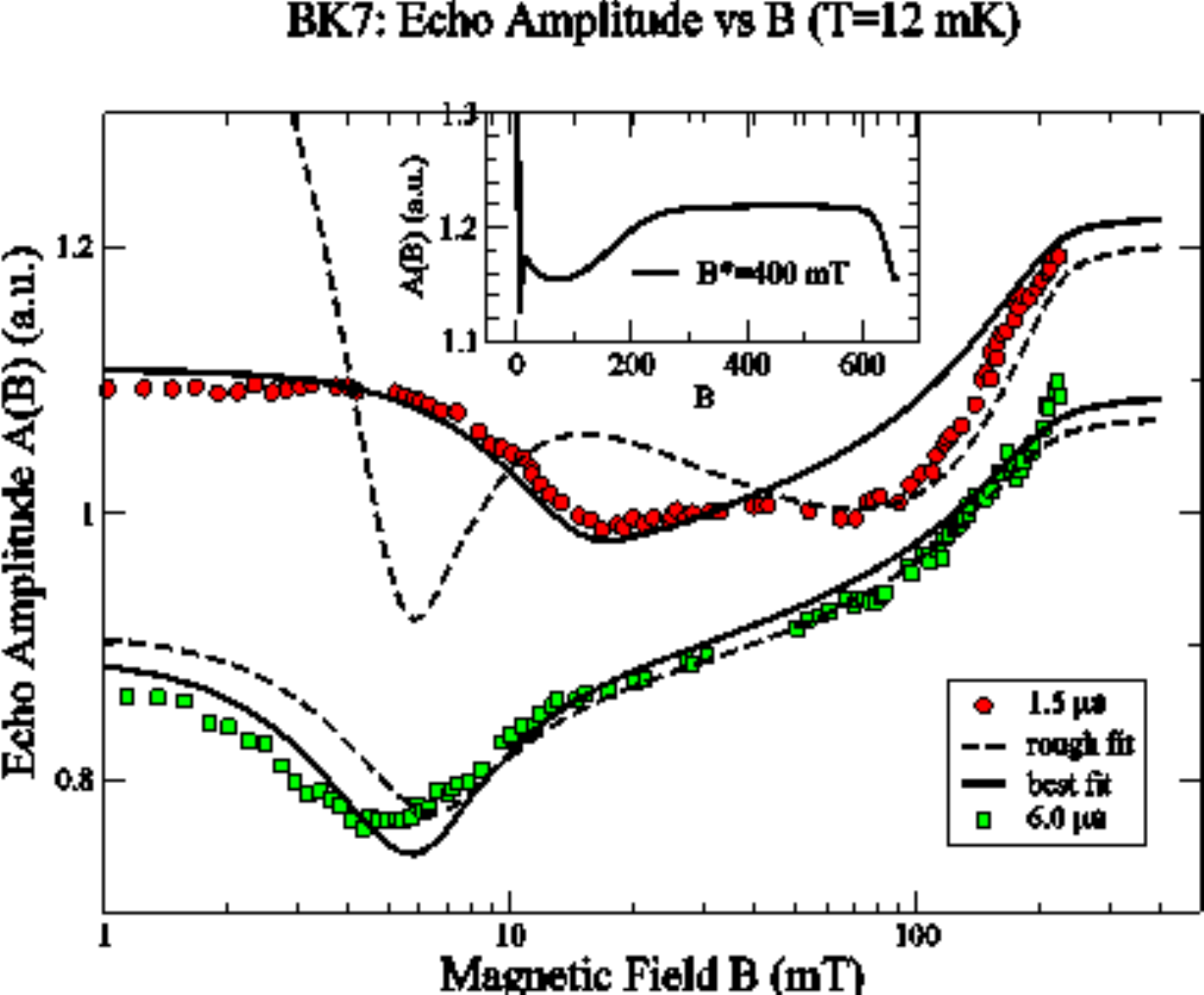}
\caption{ Magnetic-field dependence of the polarization echo amplitude for 
borosilicate BK7 glass \cite{Lud2002} at given experimental conditions. Dashed 
curves (preliminary rough fit) and continuous curves from our theory: there are no 
more than two observable maxima or minima (so, no true oscillations). Nominal 
frequency 0.9 GHz, $\tau_2=2\tau_1$=0.2 $\mu$s. Inset: our prediction for the 
higher magnetic field regime ($B^{*}$ as defined in Section 5, Ref. \cite{JBK2016}).}
\label{image8e}
\end{figure}

We now make use of Eq.~(\ref{eq933}) to fit some of the available experimental 
data for the multi-silicates, the idea being that the total amplitude is given by a
sovrapposition of 2LS and ATS contributions:
$A\left(\varphi\right)=A_{2LS}+\Delta A\left(\varphi\right)$ (which must still 
be averaged with respect to the ATS magnetic-orientation angle $\beta$).
Fig.~\ref{image7e} shows the experimental results for the {\it relative} echo 
amplitude in AlBaSiO (or BAS glass) as a function of $B$. Values of $B$ up to 
0.6 T have been explored, and for three distinct temperatures. The 
data are then fitted with our theory (full curves) with the parameters reported in 
Table~\ref{table91}. The agreement between theory and experiment is indeed 
highly satisfactory, given the discussed simplifications and assumptions that have 
been used in the theory. Only one minimum in $A(B)$ is found and the inset in 
Fig. \ref{image7e} shows that, again, it is the ATS magnetic DOS that is causing 
the magnetic effect (Section 3). In fact, by enforcing the strict-resonance condition 
$\sigma (E)\to \delta (E-\hbar \omega )$ Eq.~(\ref{eq933}) would collapse to a 
quantity very much like the DOS (convoluted with slow-varying, in $E$, 
corrections) and with the very same DOS behaviour, reproducing in this way 
the qualitative shape of $\Delta A(B)$. It is however the non-resonant convolution 
of this quasi-DOS with other (smooth) $E$-dependent functions that produces the 
rounding of the minimum in $\Delta A(B)$ and the $B^{-2}$ saturation that is
always observed. Interestingly enough, now $\tau$ (though $\tau_\phi\ll \tau$) 
the phonon-damping term plays a main role in the rounding of the high-$B$ tail 
of $\Delta A(B)$ to the $B^{-2}$ (as observed) saturation. The ATS approach 
is the only theory that predicts also a linear-in-$B$ intermediate decay regime of the 
echo amplitude, and this is often experimentally observed. Details of the ETM theory
for the electric echo are interesting and will be published elsewhere.
Next, in Fig.~\ref{image8e} we produce the comparison of theory and experiment 
for data for the echo amplitude in BK7 (this is good optical glass, hence devoid of 
true microcrystals, but nevertheless containing the (large) RERs) at two different 
values of the waiting time $\tau_{12}$. It is truly remarkable how our ETM theory, 
despite the many simplifications and assumptions and the total absence of multi-level 
quantum physics (as advocated by the NEQM approach), can reproduce all the 
features qualitatively characterizing the experimenatl data, including every change in curvature of $A(B)$ vs. $B$. A preliminary rough fit, reported, not aiming 
at high ${\chi }^2$ agreement, reproduces also two maxima (and minima) that 
the NEQM approach takes as indication of the multiple (rapid) oscillations ensuing 
from the quantum beatings ascribed to the Zeeman- and NEQM-splitting of the 
generic 2LS \cite{Wur2002} and NEQM-carrying tunneling particle. There are in fact
never more than two observed minima, in the experimental data, and these can be 
reproduced by the present, simpler ATS-ETM approach.
Finally, in the inset of Fig.~\ref{image8e} we show what the experimentalists 
overlooked, by not exploring higher magnetic-field values. Using a simple-minded 
low-$\varphi$ correction for the lower energy gap at higher fields, we plot the 
expected behaviour of $A(B)$ for intermediate fields. After the two observed 
minima, there is only an apparent saturation, yet new interesting features ought to
characterise $A(B)$ at higher fields ($B>$600 mT). This, just as it happens for the 
dielectric constant (Section 5). A full description of the high field effects, however, 
requires a calculation involving all three ATS energy levels (this was never done).
\begin{table}[htbp]
\begin{center}
\begin{tabular}{ |c|c|c|c|c|c|c|c| }
\hline
Glass type & $D_{min}$ & $D_{0min}\left|\frac{q}{e}\right|S_{\Delta }$ & $D_{0max}\left|\frac{q}{e}\right|S_{\Delta }$ & ${\Gamma }^{-1}$ & $c_{ATS}-c_{2LS}$ & ${{\rm p}}_{{\rm 1}}{{\rm F}}_0$ & ${\tan  \omega \Delta \tau }$ \\
 & ${\rm (mK)}$ & $({\rm K}\AA^2)$ & $({\rm K}\AA^2)$ & ${\left(\mu {\rm s}{{\rm K}}^5\right)}^{-1}$ & ${\left(\mu {\rm s} {\rm K}\right)}^{-1}$ & ${\rm D}~{\rm kV}~{{\rm m}}^{-1}$ &  \\
\hline
AlBaSiO & 17.74 & 0.95$\times$10${}^{3}$ & 2.13$\times$10${}^{4}$ & 9.22$\times$10${}^{6}$ & 5.008 & 0.461 & 0.247 \\
(sample 1) & & & & & & & \\
\hline
AlBaSiO & 27.20 & 1.14$\times$10${}^{3}$ & 8.96$\times$10${}^{3}$ & 2.57$\times$10${}^{5}$ & 3.825 & 0.450 & 0.245 \\
(sample 2) & & & & & & & \\
\hline
BK7\newline & 16.76 & 0.92$\times$10${}^{3}$ & 1.34$\times$10${}^{4}$ & 8.91$\times$10${}^{6}$ & 1.03 (*) & 0.60 & 0.207 \\ 
(1.5 $\mu $s) & & & & & & & \\
\hline
BK7\newline & 15.94 & 0.89$\times$10${}^{3}$ & 3.31$\times$10${}^{4}$ & 3.25$\times$10${}^{6}$ & 5.72 (*) & 0.98 & 0.204 \\
(6 $\mu $s) & & & & & & & \\
\hline
\end{tabular}
\caption{Fitting parameters for the echo amplitude's magnetic-field dependence. 
(*) For BK7 (best-fit parameters only), only $c_{ATS}$ is involved.}
 \label{table91}
\end{center}
\end{table}

This concludes this very short survey of the main results obtained within the ETM 
model for the low-temperature magnetic-field effects. The treatment shows that all the 
magnetic effects (basically: an enhancement at low fields, followed by a decrease in 
each probed quantity at the higher fields) appear to be the consequences of the very
same magnetic behaviour of the ATS DOS (described in Section 2). In Fig. 
\ref{alltogether} we have re-drawn the magnetic-field dependence of the relative 
dielectric constant's chenge and (different scale) of the {\it inverted} relative echo 
amplitude variation. The variation of the heat capacity $C_p$ has not been inserted 
for clarity, nevertheless it follows exactly the very same trend as for 
$-A(B)/A({\rm high})$ (see Figs. \ref{heat_capa_fit}(a) and (b)). The schematic 
change of the ATS DOS, $g_{ATS}(B)$ is also shown (not to scale) and it appears to 
determine the trend observed in all three experiments. 
\begin{figure}[h]
\centering
{
   {\includegraphics[scale=0.48] {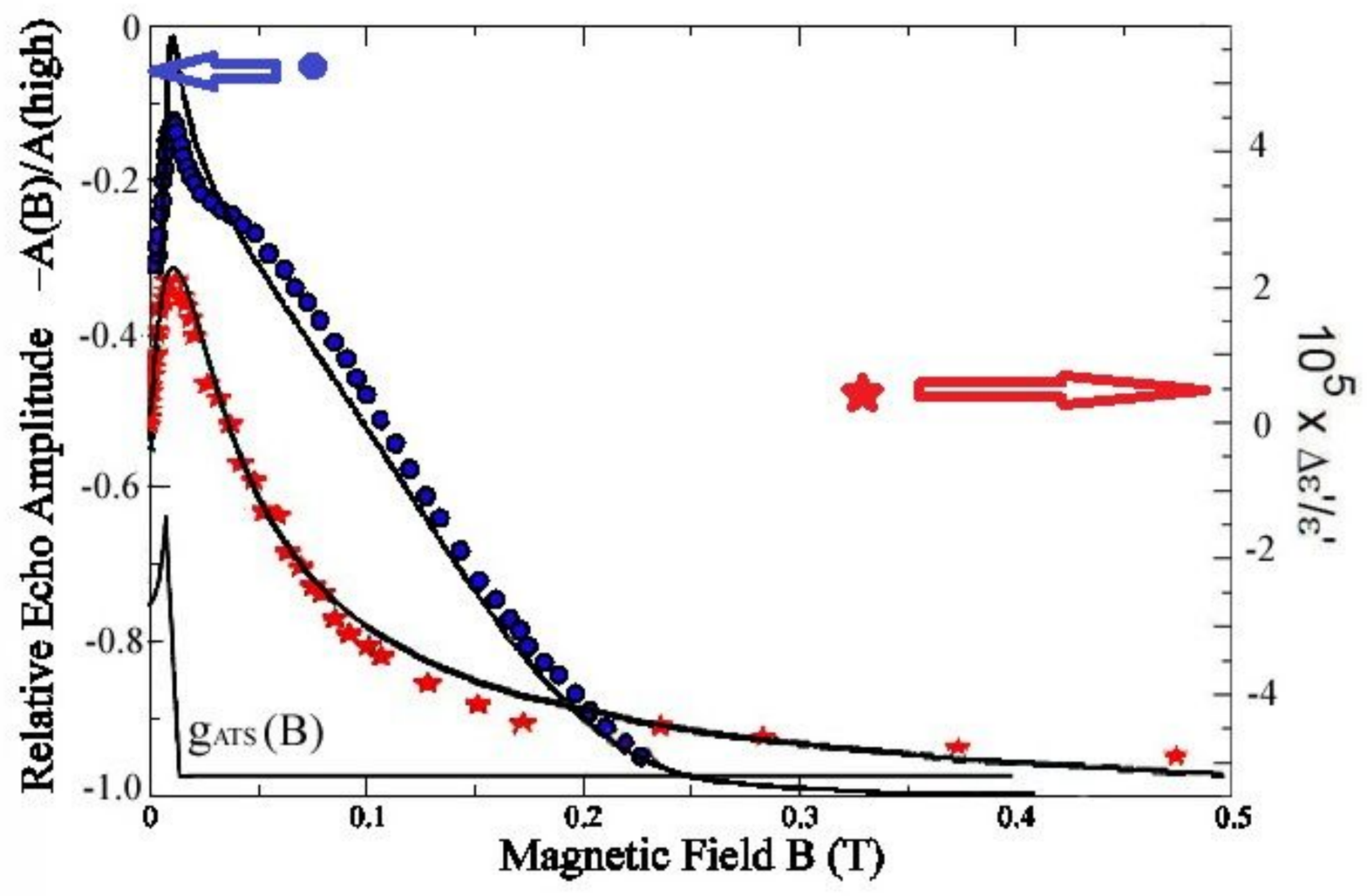} }
\vskip -3mm
}
\caption{ Some data (for the same glass type, BAS) showing the 
magnetic-field variation of (minus) the echo amplitude $-A(B)/A({\rm high})$ 
(blue dots) and of the dielectric constant variation (different scale, red stars). The 
qualitative variation (which mimicks that of the heat capacity $C_p$, see Figs. 
\ref{heat_capa_fit}(a) and (b)) derives from that of the magnetic DOS, 
$g_{ATS}(B)$, drawn not to scale. The black lines are all theory predictions.}
\label{alltogether}
\end{figure}

\section{Consequences, Conclusions and Outlook}

\subsection{A new scenario for the glass transition?}
Having conceived a heterogeneously disordered, cellular make-up of the glassy 
state, it seems natural to ask what the implications for the onset of the glassy
state would be. The crucial question is the definition of the glass transition, which 
can only be kinetic in nature given the continuity that is being advocated between
the dynamical heterogeneities of the supercooled state and those of the glassy
state. Namely, it is envisaged that the better-ordered regions, or RERs or solid-like
clusters in the supercooled state grow in size with decreasing temperature till a
limit size $\xi_0$ is reached. Thereafter a mutually hindering state sets in for the
maximally grown RERs and they must grow through a completely different (and 
much slower) mechanism for consolidation to take place at the expense of the 
material (if any) in the interstitial spaces. Naively we can take the temperature where
$\xi_0$ is attained as the glass transition temperature $T_g$. At this temperature 
the substance is made up of close-packed RERs of maximum size $\xi_0$ which are 
better ordered, though not crystalline, and - depending on composition - of fluid-like
atomic species in the interstitial spaces (or cages) between the RERs. The true 
crystals have not had the chance to grow, either because kinetically disfavoured or 
because their size does not allow for true order (thus RERs have formed instead). 
The time necessary for the coalescence and grow of true crystals has not been 
made available in the quench.

We can describe the dependence of $T_g$, the temperature where the polycluster
forms, on the cooling rate $\kappa$ by envisaging a characteristic microscopic 
time $\tau_0$ for the cells' rearrangement and an entropy (per atomic species) for 
the many ways the polycluster can be formed from the cooperative nucleation of 
such cells:
\begin{equation}
s(T)=k_B \ln\left[ \frac{T_c-T}{\kappa\tau_0} \right]
\end{equation}
much in the same spirit as in Adam-Gibbs' treatment for the onset of the glassy
state \cite{AG1965}. On dimensional grounds, and treating this entropy  as a 
response susceptibility for the glass transition, one then writes (if $w_0$ is a 
characteristic energy, per atomic species, for the polycluster formation):
\begin{equation}
s(T_g)=k_B \ln\left[ \frac{T_c-T_g}{\kappa\tau_0} \right]=\frac{w_0}{T_g}
\end{equation}
If the written form of the above susceptibility is regarded as a Curie-like approximation
in which the nucleating cells are not yet interacting, then the molecular-field 
improved form of the Curie-Weiss type would lead us to write
\begin{equation}
k_B \ln\left[ \frac{T_c-T_g}{\kappa\tau_0} \right]=\frac{w_0}{T_g-\Theta}
\label{glasstrans}
\end{equation}
where $\Theta$ is a characteristic temperature below $T_g$ that takes nucleating 
cell-cell interactions into account. A graphical study of the above Eq. (\ref{glasstrans})
for $T_g(\kappa)$ shows that $T_g$ increases logarithmically with $\kappa$, as is 
known experimentally (see e.g. \cite{Sim1996}) and from computer simulations (see
e.g. \cite{Buc2002}). A full derivation from nucleation theory of the above reasoning 
will be reported elsewhere.

\subsection{Some conclusions from the magnetic effects: estimate of cell size}
Qualitatively at least, our cellular-structure based ETM explains all low-temperature 
experimental observations so far [for a more complete discussion see \cite{JBK2016}]. 
Looking at the parameters which have been used for the best fits, one cannot fail to 
recognize (see e.g. Table \ref{table91}) that for the echo experiments the extracted 
values of the cutoffs for the parameter combinations $D_0\frac{q}{e}S_{\triangle}$ 
are approximately one order of magnitude smaller than those used for the other 
experiments, namely for $C_p$ and $\epsilon$. The latter have been carried out 
inside higher temperature ranges, as it turns out. This could be explained through 
a mechanism where the number of elementary atomic tunneling systems $N$ within 
each single one ATS (hence inside each interstice, or cage, between the RERs or 
mosaic cells or grains) gets to be characterised by a temperature dependence 
$N(T)=N_0\exp\{-E_0/(k_BT)\}$. Namely: there is a consolidation mechanism 
especially important at the lowest temperatures where ions from the interstices get to 
be absorbed in the cells (see Fig. \ref{lowTstructure}) and the resulting number of 
coherently tunneling particles making up each 
ATS diminishes with diminishing temperature. We have conducted an analysis of the
paramagnetic magnetization of samples of Duran, BAS and BK7 glass reported in the
literature as a function of temperature using the idea of the cell model and ATS 
tunneling in the interstices with a temperature-dependent $N(T)$ \cite{Bon2015}. 
We have obtained in this way good fits to the data and an estimate of the Fe impurity 
concentrations that are in agreement with the concentrations extracted from the 
low-temperature $C_p$ data (Tables \ref{tab_imp_extr} and \ref{tab_imp_extr_dur}).
The number $N$ of coherently tunneling ions making up each ATS then enters the 
parameter combination $D_0\frac{q}{e}S_{\triangle}$ as $[N(T)]^3$ times a 
combination of factors specific for a single atomic tunneling particle and the $T$-
dependence of the extracted combination of cutoff and other tunneling parameters
receives its rationale. Though it might seem surprising that the tunneling parameter 
$D_0$ of a collection of $N$ coherently tunneling atomic particles gets to scale like 
$N$ times a microscopic tunneling parameter, we remark that this is similar to what 
happens in the theory and experiments of a drop of coherent atoms in a Bose-Einstein 
condensate trapped and subject to a double-welled tunneling potential \cite{BE1,BE2}

At this point one could ask if the low temperature experiments hold some information
on the cell size for the proposed polycluster structure of glass, given that we have 
extracted values of the ATS concentration in the form of the quantity $n_{ATS}P^*$
(where $n_{ATS}$ is the ATS number (mass) density) (see Tables). 
It is reasonable to assume, in fact, that on average four ATSs sit in each interstice 
between four tetrahedrally close-packed cells and this allows for a determination of 
the cell size $\xi$. If $\xi$ is the cell's radius, the volume of the interstitial space is 
$2\sqrt{2}\xi^3/3$ and therefore we have, on average:
\begin{equation}
\frac{4}{\frac{2\sqrt{2}}{3}\xi^3}=x_{ATS}=n_{ATS}\rho=\frac{n_{ATS}P^*\rho}{P^*}
\end{equation} 
where $\rho$ is the solid's mass density. The parameter $P^*$ could be determined, 
in principle, from the normalization condition for the ATS parameter distribution:
\begin{equation}
2\pi P^* \ln\left( \frac{D_{max}}{D_{min}} \right) 
\ln\left( \frac{D_{0max}}{D_{0min}} \right) =1. 
\end{equation}
However $D_{max}$ remains unknown from the fits to the data, so we can only 
make the reasonable guess that the quantity  
$\ln\left( \frac{D_{max}}{D_{min}} \right) \ln\left( \frac{D_{0max}}{D_{0min}} \right)$
is of order 1, to estimate $P^*\approx 1/(2\pi)$. We then get the estimating 
formula for the average cell radius
\begin{equation}
\xi\approx \left[ \frac{3}{\pi\sqrt{2}(n_{ATS}P^*)\rho} \right]^{1/3}
\label{estimate}
\end{equation}
so that at this point we can use the values of $n_{ATS}P^*$ obtained in Section 3
to give cell size estimates. We use, for the silicates:
$n_{ATS}P^*\approx$ 5 $\times$ 10$^{16}$ g$^{-1}$ (BAS glass),
9  $\times$ 10$^{16}$ g$^{-1}$ (Duran) and [as obtained in \cite{Bon2015}]
1 $\times$ 10$^{16}$ g$^{-1}$ (BK7 glass). Then, from the literature \cite{Sie2001}
we get: 
$\rho\simeq$ 3.1 g cm$^{-3}$ (BAS glass),
2.3 g cm$^{-3}$ (Duran) and
2.5 g cm$^{-3}$ (BK7 glass). Using the above estimating formula 
Eq. (\ref{estimate}) we arrive at the size of the cells in terms of their radius:
$\xi\approx$ 1.63 $\times$ 10$^{-6}$ cm or 163 $\AA$ (BAS glass),
1.54 $\times$ 10$^{-6}$ cm or 154 $\AA$ (Duran) and finally
3.00 $\times$ 10$^{-6}$ cm or 300 $\AA$ (BK7 glass). Thus, from the low
temperature experiments we get to estimate that for these silicates the cell size 
should be some 300 to 600 $\AA$ in diameter (2$\xi$). 
Is this a reasonable estimate?

While high-resolution electron microscopy (HREM) images for the mentioned
silicate glasses are not available in the literature, some HREM images of (inevitably)
very thin samples of related glasses can be found. These are presented below, 
for the case of amorphous SiO$_2$ (Fig. \ref{cellfig}(a)), amorphous 
(B$_2$O$_3$)$_{0.75}$-(PbO)$_{0.25}$ (Fig. \ref{cellfig}(b)) and amorphous
LiO$_2$$\cdot$SiO$_2$ (equimolar mixture, Fig. \ref{cellfig}(c)). The cellular 
structure of these thin glass samples is clearly visible in these images, with the 
estimates for the diameter size 2$\xi\approx$ 500 $\AA$, 600 $\AA$ and, 
respectively, 500 $\AA$. While the second glass is not a silicate, the size of the 
cells as seen in HREM imaging for the two other silicates compares very favourably
with the estimates for other silicates obtained from the low temperature work.
It is therefore tempting to conclude that the estimate from the tunneling data at
low temperatures lead to cellular sizes that are compatible with HREM imaging.
\begin{figure}[h!]
\centering
   \subfigure[]{\includegraphics[scale=0.30] {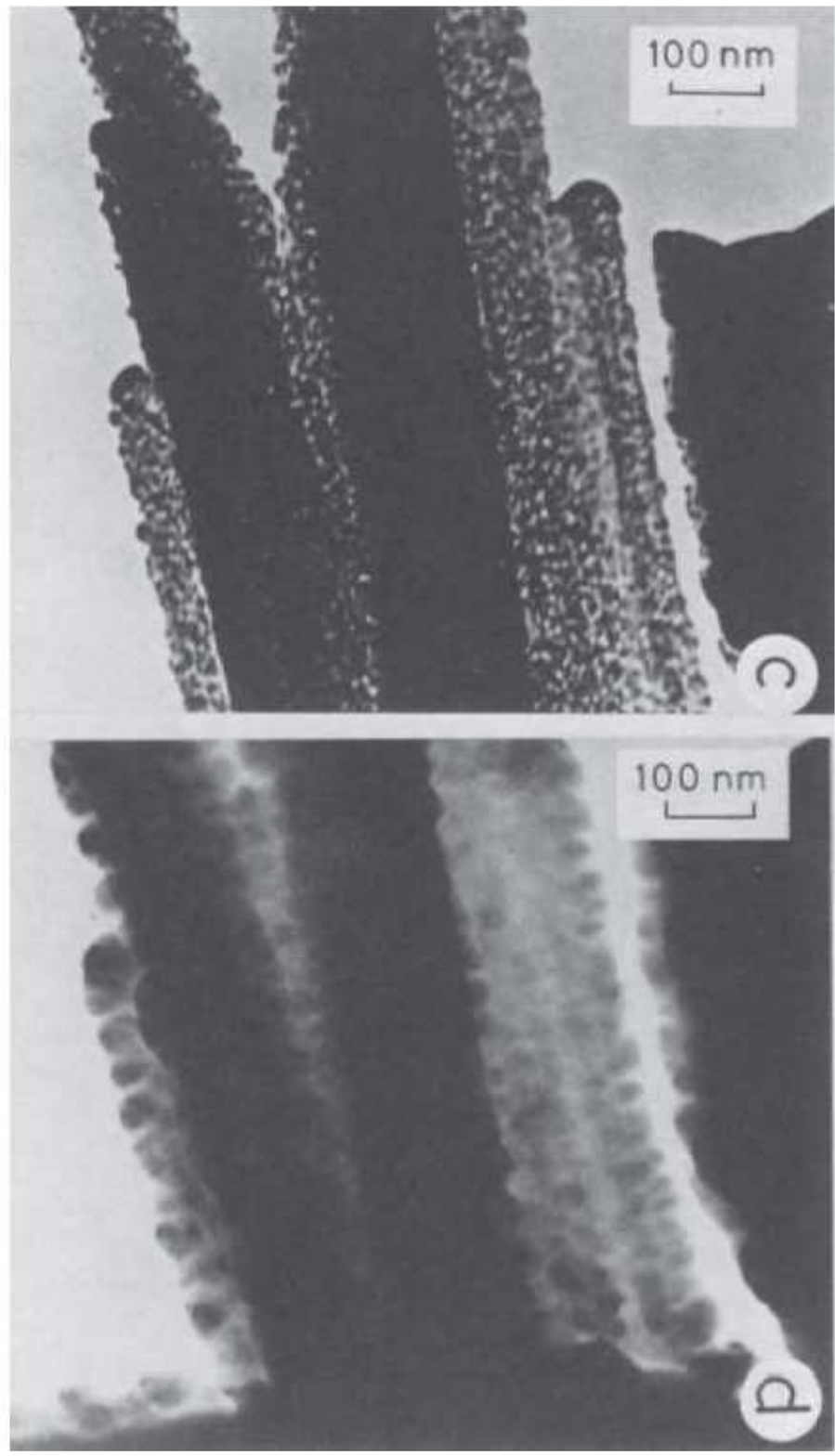}}
     \subfigure[]{\includegraphics[scale=0.25] {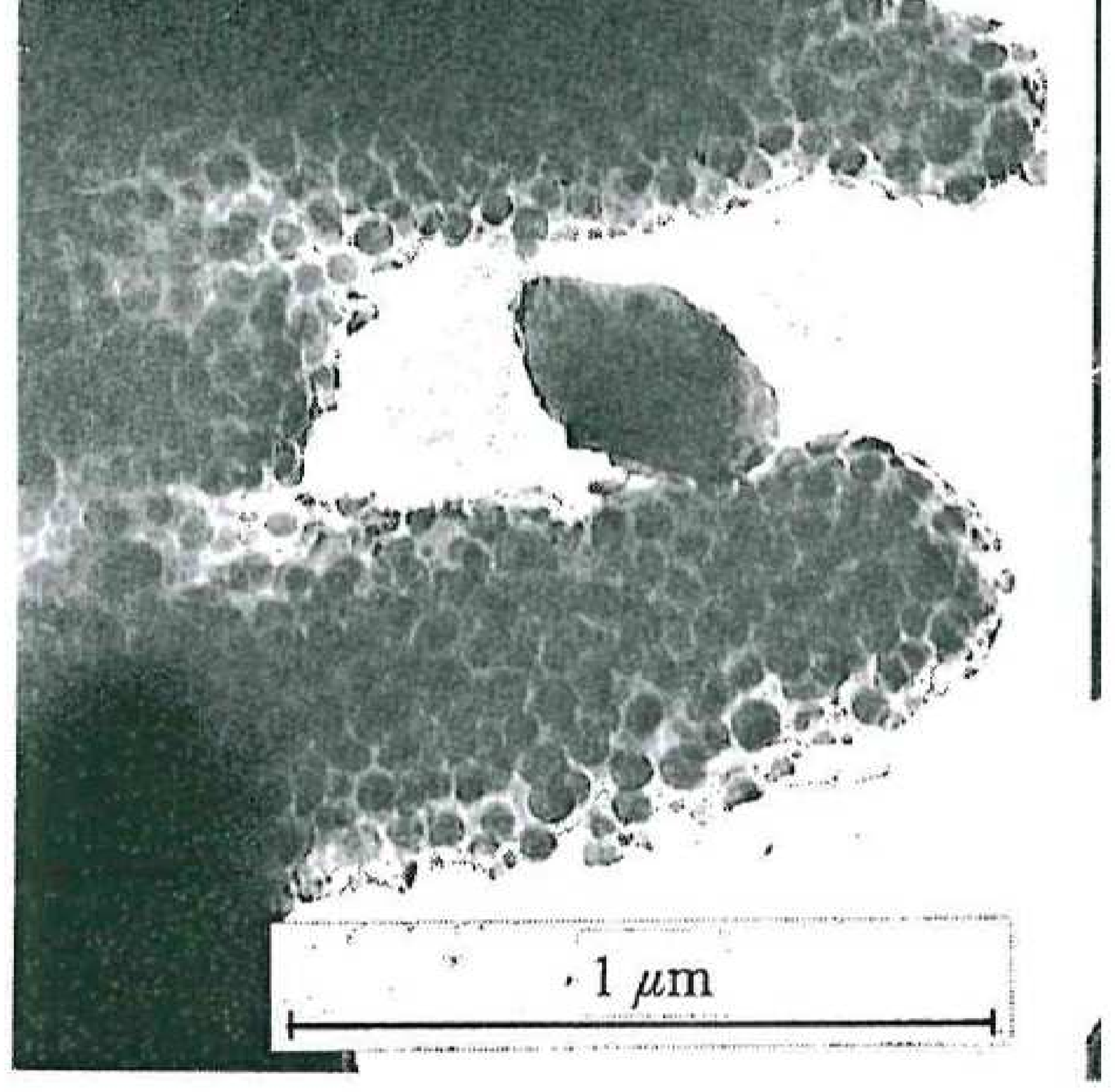}} \vskip 20mm
      \subfigure[]{\includegraphics[scale=0.40] {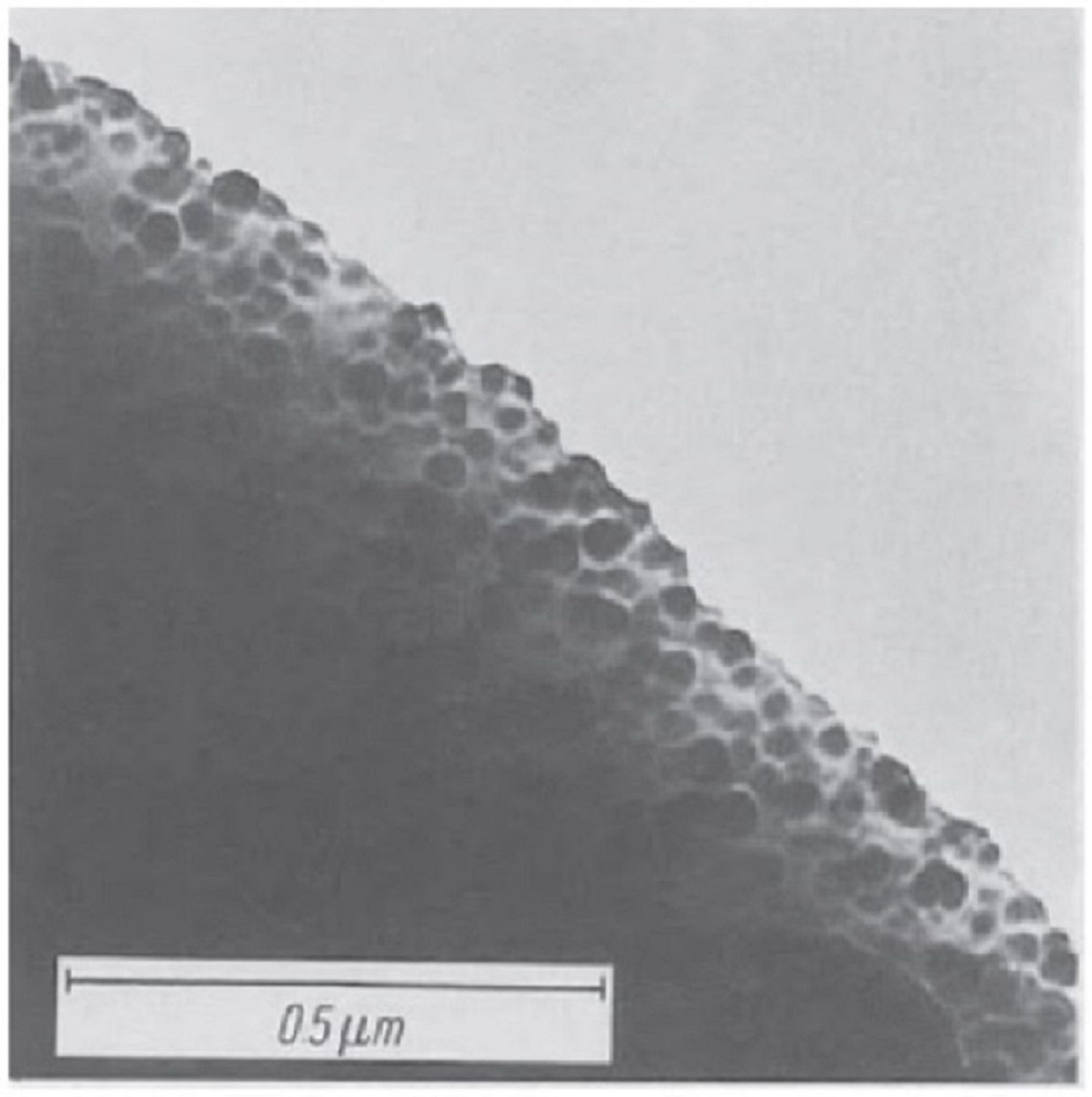}} 
\caption{HREM images of the cellular structure of thin glass samples. (a) Amorphous
SiO$_2$ (especially panel d, from \cite{Zar1}, see also \cite{Zar2}), 
(b) amorphous B$_2$O$_3$-PbO (from \cite{Zar2}), (c) amorphous Li$_2$O-SiO$_2$ 
(from \cite{Vogel}). }
\label{cellfig}
\end{figure}
This is also consistent with estimates for the number $N$ of coherently tunneling
atomic particles that make up each ATS. While $N$ depends on temperature as
stated, estimates \cite{Pal2011} range from 10 to 10$^2$ and while the nature of 
the microscopic tunneling entities is still unknown and should depend on composition 
(in the case of pure SiO$_2$ this $N$ should be close to 1) these values are not 
incompatible with a cell size of some 500 $\AA$ in diameter. The question of the 
nature of the atomic coherently tunneling particles making up each ATS in each 
cellular interstice remains completely unanswered.

\subsection{Conclusions}
In conclusion, the cellular glass-structure backed ATS ETM for the magnetic effects 
in multi-component glasses (the multi-silicates BAS (or AlBaSiO), Duran and BK7) 
and contaminated mono-component vitreous glycerol has been fully justified in 
terms of a (not entirely) new vision for the intermediate-range structure of 
real glasses. In this scheme the particles are organized in regions of enhanced
regularity (RERs) and more mobile {\em charged} particles trapped in the 
interstices (or cages) between the close-packed RERs. These are coherent
(owing to proximity and strong Coulomb forces) atomic tunnelers that can be
modeled in terms of single quasi-particles with highly renormalized tunneling 
parameters. This model explains a large number of experimental data and facts 
with remarkable consistency also in terms of cross-checks like the 
determination of the concentration of trace paramagnetic impurities \cite{Bon2015}. 

The fact that pure amorphous silica (Spectrosil-I, for example \cite{Lud2003})
shows no detectable magnetic effects is a consequence of the extremely small
size of the RERs for a-SiO${}_2$, deprived of almost any nucleation centres
for both mosaic cells (RERs) and polycluster formation. These RERs will 
therefore trap a very small number $N$ of dangling-bond ionic particles, or
none at all given the covalent nature of the Si-O bonds. Hence, no magnetic
effects are observable in this purest, single-component SiO$_2$ glass. 

We remark at this point that the ETM with two types of TSs is not the only 
theoretical explanation that has been put forth for the explanation of the magnetic
effects at low temperatures in the multi-silicate glasses. It is however the only 
theory that successfully explains all of the experimental data so far, including those
for the compositional effects, and that links the deviations from the STM to the 
real structure of glasses at the intermediate atomic range. Two other approaches 
have been in fact proposed to date, the one already mentioned based on the 
coupling of the 2LSs to nuclei in the sample carrying NEQMs \cite{Wur2002} and 
the one based on the coupling of the 2LSs to paramagnetic Fe$^{3+}$ impurities 
\cite{Bor2007,Bor2011}. These two other approaches have shortcomings that will 
not be discussed here, since their model justification has no bearing on the real 
atomic structure of glasses and on the glass-forming process from glass-forming 
liquids (the Zachariasen-Warren picture being -- in fact -- always tacitly assumed). 
Polymeric glasses should also be well described by the present cellular model 
(possibly also their surfaces \cite{Koi2017}).   

The present theory on the one hand bears heavily on the true structure of real 
glasses and it implies that the amorphous state should no longer be regarded as a
dynamically arrested liquid, but rather as a new type of solid. It also shows on the
other hand that the magnetic (and compositional) effects are a mere manifestation 
of the heterogeneous, cellular-type intermediate atomic structure of real glasses 
which abandons the Zachariasen-Warren picture for good. A cellular-type structure 
that has been advocated for by scientists, especially (but not only) in the ex-USSR, 
now for almost a century. It is not impossible that with this vision in mind the TSs 
could become in the near future the right probes with which to study the structure 
of real glasses in the laboratory.

\section*{Acknowledgements}
The Author is very grateful to Maksym Paliienko and Silvia Bonfanti for their 
help with data fitting. He gratefully acknowledges stimulating discussions with 
A.S. Bakai. Part of this work was carried out whilst visiting
the Physics Department of McGill University in Montreal (CA). The Author is
grateful to Hong Guo for support and to him, to Martin Grant and Mark Sutton 
for useful discussions. On-going support from INFN-Pavia through Iniziativa
Specifica GEOSYM-QFT is gratefully acknowledged.


\newpage
\begin{thebibliography}{99}

\bibitem[*]{J} email: giancarlo.jug@uninsubria.it (corresponding author)


\bibitem{Jug2004} G. Jug:
\newblock Theory of the Thermal Magnetocapacitance of Multi-component
Silicate Glasses at Low Temperature,
\newblock {\em Phil. Mag.} {\bf 84}(33), 3599--3615 (2004).

\bibitem{Phi1981} W.A. Phillips (Ed.):
\newblock {\it Amorphous Solids: Low Temperature Properties},
\newblock  (Springer Verlag, Berlin 1981).

\bibitem{Esq1998} P. Esquinazi (Ed.):
\newblock {\it Tunneling Systems in Amorphous and Crystalline Solids}
\newblock (Springer, Berlin, 1998).

\bibitem{Zac1932} W.H. Zachariasen:
\newblock The Atomic Arrangement in Glass,
\newblock {\em J. Am. Chem. Soc.} {\bf 54}, 3841--3851 (1932); 
\newblock {\it ibid.}: 
\newblock The Vitreous State,
\newblock {\em J. Chem. Phys.} {\bf 3}, 162--163 (1935).

\bibitem{War1934} B.E. Warren:
\newblock The Diffraction of X-Rays in Glass,
\newblock {\em Phys. Rev.} {\bf 45}, 657--661 (1934).

\bibitem{Mac1985} W.M. MacDonald, A.C. Anderson and J. Schr\"oder:
\newblock Low-temperature Behavior of Potassium and Sodium Silicate Glasses,
\newblock {\em Phys. Rev. B} {\bf 31}, 1090--1101 (1985).

\bibitem{Jug2010}
G. Jug and M. Paliienko:
\newblock Evidence for a Two-component Tunnelling Mechanism in the
Multicomponent Glasses at low Temperatures,
\newblock {\em Europhys. Lett.} {\bf 90} 36002 (2010).

\bibitem{Ens2002} C. Enss:
\newblock Anomalous Behavior of Insulating Glasses at Ultra-Low Temperatures,
\newblock {\em Adv. in Solid State Phys.} {\bf 42}, 335--346 (2002).

\bibitem{Str2000} P. Strehlow, M. Wohlfahrt, A.G.M. Jansen, R. Haueisen, 
G. Weiss, C. Enss and S. Hunklinger: 
\newblock Magnetic Field Dependent Tunneling in Glasses,
\newblock {\em Phys. Rev. Lett.} {\bf 84}, 1938--1941 (2000).

\bibitem{Woh2001}
M. Wohlfahrt, P. Strehlow, C. Enss, and S. Hunklinger:
\newblock Magnetic-Field Effects in Non-Magnetic Glasses,
\newblock {\em Europhys. Lett.} {\bf 56}, 690--694 (2001);
\newblock M. Wohlfahrt: Ph.D. Thesis (Heidelberg 2001,
www.ub.uni-heidelberg.de/archiv/1587).

\bibitem{Nag2004} P. Nagel, A. Fleischmann, S. Hunklinger and C. Enns:
\newblock Novel Isotope Effects Observed in Polarization Echo Experiments,
\newblock {\em Phys. Rev. Lett.} {\bf 92}, 245511 (2004).

\bibitem{Leb1921} A.A. Lebedev:
\newblock O Polimorfizme i Otzhige Stekla,
\newblock {\em Trud'i Gos. Opt. Inst.} {\bf 2} 1-20 (1921) (in Russian);
\newblock {\it ibid.}, {\em Izv. Akad. Nauk SSSR, Otd. Mat. Estestv. Nauk,
Ser. Fiz.} {\bf 3}, 381 (1937).

\bibitem{Ran1930} J.T. Randall, H.P. Rooksby and B.S. Cooper:
\newblock The Diffraction of X-Rays by Vitreous Solids and its Bearing on
their Constitution,
\newblock {\em Nature} {\bf 125}, 438 (1930);
\newblock {\it ibid}: X-Ray Diffraction and the Structure of Vitreous
Solids -- I,
\newblock {\em Z. Kristallogr.} {\bf 75}, 196--214 (1930).

\bibitem{PK1990} E.A. Porai-Koshits:
\newblock Genesis of Concepts on Structure of Inorganic Glasses,
\newblock {\em J. Non-cryst. Sol.} {\bf 123}, 1--13 (1990).

\bibitem{Wri2014} A.C. Wright:
\newblock Crystalline-like Ordering in Melt-quenched Network Glasses?
\newblock {\em J. Non-cryst. Solids}, {\bf 401} 4--26 (2014);
\newblock {\it ibid.}:
\newblock The Great Crystallite versus Random Network Controversy: A Personal
Perspective,
\newblock {\em Int. J. Appl. Glass Sci.} {\bf 5}, 31--56 (2014).

\bibitem{Bak1994} A.S. Bakai:
\newblock The Polycluster Concept of Amorphous Solids,
\newblock Beck/G\"unterodt (Eds.), {\em Topics in Applied Physics} {\bf 72},
209--255 (Springer-Verlag, Berlin Heidelberg 1994).

\bibitem{Bak2013} A.S. Bakai:
\newblock {\em Poliklastern'ie Amorfn'ie Tela},
\newblock Khar'kov ``Synteks'' (Khar'kov, Ukraine 2013) (in Russian).

\bibitem{AG1965} 
G. Adam and J.H. Gibbs:
\newblock On the Temperature Dependence of Cooperative Relaxation Properties
in Glass-Forming Liquids,
\newblock{\em J. Chem. Phys.} {\bf 43}, 139--146 (1965).

\bibitem{Ber2011} L. Berthier and G. Biroli:
\newblock Theoretical perspective on the glass transition and amorphous
materials,
\newblock {\em Rev. Mod. Phys.} {\bf 83}, 587--645 (2011).

\bibitem{Ang1988} C.A. Angell:
\newblock Perspective on the Glass Transition,
\newblock {\em J. Phys. Cem. Solids} {\bf 49}, 863--871 (1988).

\bibitem{Lub2007} V. Lubchenko and P.G. Wolynes:
\newblock Theory of Structural Glasses and Supercooled Liquids,
\newblock {\em Annu. Rev. Phys. Chem.} {\bf 58}, 235--266 (2007).

\bibitem{Sim2009} S.L. Simon and G.B. McKenna:
\newblock Experimental Evidence Against the Existence of an Ideal Glass
Transition,
\newblock {\em J. Non-Cryst. Solids} {\bf 355}, 672--675 (2009).


\bibitem{Hag1935} G. H\"agg:
\newblock The Vitreous State,
\newblock {\em J. Chem. Phys.} {\bf 3}, 284--49 (2016).

\bibitem{Koi2016} U. Satoshi and H. Koibuchi:
\newblock Finsler Geometry Modeling of Phase Separation in Multi-Component 
Membranes,
\newblock {\em Polymers} {\bf 8}, 284 (2016).


\bibitem{Hwa2012} J. Hwang, Z.H. Melgarejo, Y.E. Kalay, I. Kalay, M.J. Kramer, 
D.S. Stone, P.M. Voyles:
\newblock Nanoscale Structure and Structural Relaxation in Zr$_50$Cu$_45$Al$_5$
Bulk Metallic Glass,
\newblock {\em Phys. Rev. Lett.} {\bf 108}, 195505 (2012).

\bibitem{Tre2012} M.M.J. Treacy and K.B. Borisenko:
\newblock The Local Structure of Amorphous Silicon,
\newblock {\em Science} {\bf 335}, 950--953 (2012).

\bibitem{Phi1983} J.C. Phillips:
\newblock Realization of a Zachariasen Glass,
\newblock {\em Solid State Comm.} {\bf 47}, 203--206 (1983).


\bibitem{Hur1995} M.M. Hurley and P. Harrowell:
\newblock Kinetic Structure of a Two-dimensional Liquid,
\newblock {\em Phys. Rev. E} {\bf 52}, 1694--1698 (1995).

\bibitem{Sil1999} H. Sillescu:
\newblock Heterogeneity at the Glass Transition: a Review,
\newblock {\em J. Non-Cryst. Solids} {\bf 243}, 81-108 (1999).

\bibitem{Edi2000} M.D. Ediger:
\newblock Spatially Heterogeneous Dynamics in Supercooled Liquids,
\newblock {\em Annu. Rev. Phys. Chem.} {\bf 51}, 99--128 (2000).

\bibitem{Vol2005} K. Vollmayr-Lee and A. Zippelius:
\newblock Heterogeneities in the Glassy State,
\newblock {\em Phys. Rev. B} {\bf 72}, 041507 (2005);
\newblock K. Vollmayr-Lee, W. Kob, K. Binder and A. Zippelius:
\newblock Dynamical heterogeneities below the glass transition,
\newblock {\em J. Chem. Phys.} {\bf 116}, 5158--5166 (2002).

\bibitem{Don1999} C. Donati, S.C. Glotzer, P.H. Poole, W. Kob and 
S. Plimpton:
\newblock Spatial Correlations of Mobility and Immobility in a Glass-forming 
Lennard-Jones Liquid,
\newblock {\em Phys. Rev. E} {\bf 60}, 3107--3119 (1999). 

\bibitem{deG2002} P.-G. de Gennes:
\newblock A Simple Picture for Structural Glasses,
\newblock {\em Comptes Rendus - Physique} {\bf 3}, 1263--1268 (2002).

\bibitem{Bal1973} H.P. Baltes:
A Cellular Model for the Specific Heat of Amorphous Solids at Low Temperatures,
\newblock {\em Solid State Commun.} {\bf 13}, 225--228 (1973).

\bibitem{Phi1987} W.A. Phillips:
\newblock Two-level States in Glasses,
\newblock {\em Rep. Prog. Phys.} {\bf 50}, 1657--1708 (1987).

\bibitem{Sus1962} J. A. Sussmann:
\newblock Electric Dipoles due to Trapped Electrons,
\newblock {\em Proc. Phys. Soc. (London)} {\bf 79}, 758--774 (1962).

\bibitem{Jug2013} G. Jug and M. Paliienko:
\newblock Multilevel Tunneling Systems and Fractal Clusters in the 
Low-Temperature Mixed Alkali-Silicate Glasses,
\newblock {\em Sci. World J.} {\bf 2013}, 1--20 (2013).

\bibitem{Jug2009} G. Jug:
\newblock Multiple-well Tunneling Model for the Magnetic-field Effect in
Ultracold Glasses
\newblock {\em Phys. Rev. B} {\bf 79}, 180201 (2009).

\bibitem{Jug2014} G. Jug, M. Paliienko and S. Bonfanti:
\newblock The Glassy State — Magnetically Viewed from the Frozen End,
\newblock {\em J. Non-Crys. Solids} {\bf 401},  66--72 (2014).

\bibitem{CYu1988} C. C. Yu and A. J. Leggett: 
\newblock Low Temperature Properties of Amorphous Materials: Through a Glass 
Darkly, 
\newblock {\em Comm. Cond. Mat. Phys.}, {\bf 14}, 231--? (1988).

\bibitem{JBK2016} G. Jug, S. Bonfanti and W. Kob:
\newblock Realistic Tunneling Systems for the Magnetic Effects in non-metallic
Real Glasses,
\newblock {\em Phil. Mag.} {\bf 96}, 648--703 (2016).



\bibitem{Bon2015} S. Bonfanti and G. Jug:
\newblock On the Paramagnetic Impurity Concentration of Silicate Glasses 
from Low-Temperature Physics
\newblock {\em J. Low Temp. Phys.} {\bf 180}, 214–-237 (2015).

\bibitem{Sie2001} L. Siebert:
\newblock Ph.D. Thesis Heidelberg University (2001),
www.ub.uni-heidelberg.de/archiv/1601


\bibitem{Car1994} H.M. Carruzzo, E.R. Grannan and C.C. Yu:
\newblock Non-Equilibrium Dielectric Behavior in Glasses at Low Temperatures:
Evidence for Interacting Defects,
\newblock {\em Phys. Rev. B} {\bf 50}, 6685--6695 (1994).


\bibitem{Lud2003} S. Ludwig, P. Nagel, S. Hunklinger and C. Enss:
\newblock Magnetic Field Dependent Coherent Polarization Echoes in Glasses,
\newblock {\em J. Low Temp. Phys.} {\bf 131}, 89--111 (2003).

\bibitem{Pal2011} M. Paliienko:
\newblock Multiple-welled Tunnelling Systems in Glasses at low
Temperatures
\newblock (Ph.D. Thesis, Universit\`a degli Studi dell'Insubria, 2011)
http://insubriaspace.cineca.it/handle/10277/420

\bibitem{LeC2002} F. LeCochec, F. Ladieu and P. Pari:
\newblock Magnetic field effect on the dielectric constant of glasses: 
Evidence of disorder within tunneling barriers,
\newblock {\em Phys. Rev. B} {\bf 66}, 064203 (2002).

\bibitem{Smo1979} B.P. Smolyakov and E.P. Khaimovich, 
(courtesy A. Borisenko)
\newblock {\em Pis'ma Zh. Eksp. Teor. Fiz.} {\bf 29}, 464 (1979) (in Russian);
\newblock {\em ibid.}: 
\newblock Dynamic processes in dielectric glasses at low temperatures,
\newblock {\em Sov. Phys. Uspekhi}, {\bf 25}, 102–115 (1982).  


\bibitem{Lud2002} S. Ludwig, P. Nagel, S. Hunklinger and C. Enss:
\newblock Direct Coupling of Magnetic Fields to Tunneling Systems in Glasses,
\newblock {\em Phys. Rev. Lett.} {\bf 88}, 075501 (2002). 

\bibitem{Wur2002} A. W\"urger, A. Fleischmann and C. Enss:
\newblock Dephasing of Atomic Tunneling by Nuclear Quadrupoles,
\newblock {\em Phys. Rev. Lett.} {\bf 89}, 237601 (2002).

\bibitem{Bla1977} J.L. Black and B.I. Halperin:
\newblock Spectral Diffusion, Phonon Echoes and Saturation Recovery in Glasses at Low Temperatures,
\newblock {\em Phys. Rev. B} {\bf 16}, 2879--2895 (1977).

\bibitem{Gur1990} V.L. Gurevich, M.I. Muradov and D.A. Parshin:
\newblock Electric Dipole Echo in Glasses,
\newblock {\em Sov. Phys. JETP} {\bf 70}, 928 (1990).

\bibitem{Gal1988} Yu.M. Galperin, V.L. Gurevich and D.A. Parshin:
\newblock Nonlinear Resonant Attenuation in Glasses and Spectral Diffusion,
\newblock {\em Phys. Rev. B} {\bf 37}, 10339--10349 (1988). 

\bibitem{Ens1996} C. Enss, S.  Ludwig, R. Weis and S. Hunklinger:
\newblock Decay of Spontaneous Echoes in Glasses,
\newblock {\em Czechoslovak J. Phys.} {\bf 46}, 2247--2248 (1996). 


\bibitem{BE1} A. Smerzi, S. Fantoni, S. Giovanazzi, and S. R. Shenoy:
\newblock Quantum Coherent Atomic Tunneling between Two Trapped
Bose-Einstein Condensates,
\newblock {\em Phys. Rev. Lett. } {\bf 79}, 4950 (1997).  

\bibitem{BE2} 
M. Albiez, R. Gati, J. F\"olling, S. Hunsmann, M. Cristiani and M.K. Oberthaler:
\newblock Direct Observation of Tunneling and Nonlinear Self-Trapping
in a Single Bosonic Josephson Junction,
\newblock {\em Phys. Rev. Lett. } {\bf 95}, 010402 (2005).  

\bibitem{Sim1996} D. Simatos, G. Blond, R. Roudaut, D. Champion, J. Perez and
A.L. Faivre:
\newblock Influence of Heating and Cooling Rates on the Glass Transition 
Temperature and the Fragility Parameter of Sorbitol and Fructose as measured 
by DSC,
\newblock {\em J. Thermal Analysis} {\bf 47}, 1419--1436 (1996). 

\bibitem{Buc2002} J. Buchholz, W. Paul, F. Varnik and K. Binder:
\newblock Cooling Rate Dependence of the Glass Transition Temperature of
Polymer Melts: Molecular Dynamics Study,
\newblock {\em J. Chem. Phys.} {\bf 117}, 7364--7372 (2002).

\bibitem{Zar1} J. Zarzycki:
\newblock {\em Proc. X Intern. Congress on Glass}, Kyoto, Japan, No. 12, p. 28
(1974).

\bibitem{Zar2} J. Zarzycki:
\newblock {\it Glasses and the Vitreous State} (Cambridge University Press, 
Cambridge 1991), p. 172.

\bibitem{Vogel} W. Vogel:
\newblock {\it Glasses Chemistry} (Springer-Verlag, Berlin 1992) 2nd edition, p. 74.

\bibitem{Bor2007} A. Borisenko:
\newblock Hole-compensated Fe$^{3+}$ Impurities in Quartz Glasses:
a Contribution to Subkelvin Thermodynamics,
\newblock {\em  J. Phys.: Condens. Matter}, {\bf 19}, 416102 (2007) .

\bibitem{Bor2011} A. Borisenko and G. Jug:
\newblock Paramagnetic Tunneling Systems and Their Contribution to the 
Polarization Echo in Glasses,
\newblock {\em  Phys. Rev. Lett.} {\bf 107}, 075501 (2011).

\bibitem{Koi2017} E. Proutorov and H. Koibuchi:
\newblock Orientation Asymmetric Surface Model for Membranes: 
Finsler Geometry Modeling,
\newblock {\em  Axioms} {\bf 6}, 10 (2017).


\end{thebibliography}
\end{document}